\documentclass[aps,prb,twocolumn,longbibliography,superscriptaddress]{revtex4-2}
\usepackage{amsmath,amssymb,bm,graphicx,color,gensymb,bbold,hyperref,keyval,url,latexsym}
\usepackage[dvipsnames,svgnames,table]{xcolor}
\usepackage{enumitem}
\usepackage[normalem]{ulem}

\usepackage{hyperref}
\hypersetup{
    pdfstartview={FitH},
    colorlinks=true,    
    linkcolor=NavyBlue, 
    citecolor=Maroon,   
    filecolor=NavyBlue, 
    urlcolor=NavyBlue   
}

\begin{document}

\title{Revisiting magnon bound states:  ferro-antiferromagnetic $J_1$--$J_2$ square-lattice model}

\author{Cesar A. Gallegos}
\affiliation{Department of Physics and Astronomy, University of California, Irvine, California 92697, USA}
\author{A. L. Chernyshev}
\affiliation{Department of Physics and Astronomy, University of California, Irvine, California 92697, USA}
\date{\today}
\begin{abstract}
A comprehensive analysis of pairing and bound states of magnons in the Heisenberg $S=1/2$ square-lattice $J_1$--$J_2$ ferro-antiferromagnetic model is presented.  We highlight the similarities and differences between the bound states of spin flips on a lattice and those of particles in the continuum. Magnon bound states at finite pair momentum are studied throughout the Brillouin zone and a convenient lattice partial-wave nomenclature is advocated. The mechanisms of enhanced stability or fragility of these bound states for the high-symmetry pair momenta are identified and quantified as relating to the effective dimensional reduction or enhancement, respectively. These effects are also shown to control the evolution of the bound states with the model parameters, providing  a transparent framework for understanding magnon pairing in a more general setting. A method to determine the bound state phase boundaries is presented.
\end{abstract}
\maketitle

\section{Introduction}
\label{Sec:Intro}

The two-body problem is a recurrent theme across physics, from nucleons, mesons, and positronium in nuclear and particle physics~\cite{PairingNuclearRMP2003, PairingParticlePR1950, Fermi1949, QED}, to excitons, Cooper pairs, and magnon bound states (BSs) in condensed matter~\cite{Wannier1937, Cooper1956, Combescot2015, Hanus1963, Wortis63, Rastelli2013}. 
Quantum magnets provide a controlled platform for multi-magnon physics, where magnon BSs occur as direct consequences of magnon pairing~\cite{WoodlandLovas2023CoNb2O6, Armitage2021MagBoundStates, Ghioldi2022,Zhitomisky2025} and possibly lead to long-range spin-multipolar phases~\cite{Mila_17,exp2017licuvo4,exp2019sq,zheludev2}. 
In particular, a long-sought spin-nematic phase~\cite{1969nm,1984nm,Papanicolaou1988}---a quantum analogue of a classical liquid crystal, sometimes viewed as an intermediary between a conventional dipolar magnetic order and a spin liquid~\cite{Savary2016QSL}---originates from the Bose-Einstein condensation (BEC) of two-magnon BSs in a fully polarized spin state~\cite{Chubukov1991Nematic, Penc2011Nematic, NematicShengtao2023}, preempting a single-magnon BEC that leads to a conventional magnetic order~\cite{Batyev,Batista_14}. 
Therefore, a clear understanding of the mechanisms underlying magnon pairing and the rationalization of various puzzling behaviors of magnon BSs are of central importance, as it will provide insights into the fragility of spin-nematics~\cite{NematicShengtao2023,ueda_phasesep} and may lead to a more robust realization of these elusive phases.

However, some obstacles complicate this path. 
Pairing in spin systems carries an aura of being qualitatively different from the other pairing problems on the lattice~\cite{Mattis1986RMP, Kornilovitch2024,Reiter1968,RodriguezDemlerPRB2022} in several interconnected ways.
From the earliest studies~\cite{Bethe1931, Dyson1956, Feynman1998SM,Akhiezer1968SW}, it was made clear that, contrary to the complete and orthogonal basis of the single-spin-flips, the two-spin-flip wavefunctions in the ferromagnetic Heisenberg model do not form an orthogonal basis. 
Nor can this basis be constructed from the single-spin-flip ones in a direct product fashion due to the on-site spin algebra. In one dimension (1D), the resolution of this difficulty has led to the development of the Bethe ansatz~\cite{Bethe1931}, while in higher dimensions the prevalent approach to the two-body problem has been to work directly within the formally exact non-orthogonal basis~\cite{Hanus1963,Wortis63, Rastelli2013, Akhiezer1968SW, Mattis2006book, Fukuda1963,Callaway1965}. 
By contrast, the $1/S$-formulation of spin-wave theory, which operates in an orthogonal and commutative but  overcomplete basis of bosonized spin-flips known as magnons, is regarded as an approximate approach to the pairing problem. Some works have argued that, because of the overcompleteness, the bosonic representation of spin-flips is inherently incapable of describing magnon pairing and BSs, thereby also undermining the applicability of spin-wave theory as a whole~\cite{Mattis2006book}. 

Concomitantly, the on-site hard-core repulsion of spin flips due to the same spin algebra leads to the magnon interaction  that betrays the standard potential-like form~\cite{Reiter1968,RodriguezDemlerPRB2022}. 
Instead of depending only on the transferred momenta in the scattering process, magnon interaction also depends on the incident particles' momenta, rendering the problem neither directly nor uniquely mappable onto the continuum problems, a feature also familiar in other hard-core repulsion contexts~\cite{Kornilovitch2024}.

Taken together, these aspects have obscured direct analogies with familiar textbook two-body problems of Fermi or Bose particles in the continuum and contributed to a widespread impression that magnon pairing is fundamentally different from them. 
In this work, we show that this impression is demonstrably incorrect, that the overcompleteness of the bosonic basis has no bearing on the magnon BS problem, and that the hard-core repulsion does not prohibit the use of the $1/S$-approach. 
In fact, the $1/S$-approach to the magnon pairing turns out to be consistent in every respect, leading to the bound-state Schr\"odinger equation that is {\it identical} to the one following from the more cumbersome formally exact approaches which utilize the non-orthogonal basis.

Formalism aside, puzzling aspects of magnon pairing abound. There are significant departures from intuitive expectations for magnon BSs even in simple models, adding to the aura of magnon BSs being special and leading to various misinterpretations.

An attractive interaction among excitations is the primary prerequisite for pairing. Because spin-flips in ferromagnets (FMs) attract by minimizing energy at neighboring sites, FM spin models on several lattices have provided a natural playground for magnon BS studies~\cite{Hanus1963,Wortis63, Rastelli2013, Mattis2006book, Fukuda1963,Callaway1965,Oguchi1971, Zhitomirsky2010, Paulson1975, Loly1976, Oguchi1975, Tonegawa1970, Torrance1970, Loly1986, Mogilner89, Rastelli1991, Rastelli1992,Penc_17, Mook2023}.
With the local attraction definitely present, one might na\"ively expect a condensate of magnon BSs to be the true ground state in a FM in dimensions two (2D) or lower~\cite{firstBEC_FM1950}, in a broad analogy with the Cooper problem~\cite{Cooper1956, Combescot2015}. Contrary to this intuition, pure FMs fail to produce a BS-condensed ground state in {\it any} dimension, an outcome recognized as being due to hard-core repulsion of spin flips in the $s$-wave channel~\cite{Zhitomirsky2010}. 

The key finding of the earlier works is that magnon BSs {\it do} occur in the spectra of simple FMs, but at a {\it finite} momentum of the pair, not as a global energy minimum~\cite{Wortis63, Fukuda1963,Callaway1965,Oguchi1971}.
However, even for these simple models, clarity is difficult to reach. For instance, for the Heisenberg FMs on 1D-chain, 2D-square, and 3D-simple-cubic lattices, the total number of distinct BS branches throughout their Brillouin zones (BZs) was found to be one, two, and three, respectively~\cite{Wortis63}, with the conventional explanation associating this trend with the dimensionality of the problem~\cite{Wortis63,Akhiezer1968SW, Mattis2006book}.  

For the Heisenberg nearest-neighbor ($J_1$-only) FM model on the 2D square lattice, one of its two BS branches  is found to exist at {\it any} momentum  ${\bf K}$ of the pair throughout the BZ, except at ${\bf K}\!=\!0$---the global band minimum~\cite{Rastelli2013, Wortis63}. 
Near that minimum, the BS is not merely exponentially shallow, as an analogy with the Cooper problem would imply~\cite{Cooper1956}, but super-shallow, with the coupling constant in the exponent being $\propto\!{\bf K}^2$. 
This renders such a super-shallow BS physically irrelevant up to a large momentum, as its size exceeds any reasonable length scale of a real system~\cite{Wortis63}. 
Such a trend should be related to the same hard-core repulsion effects projecting out the $s$-wave BS for ${\bf K}\!=\!0$.
Instead of explicating this connection, previous studies have declared that one BS {\it must} exist for every ${\bf K}$ in {\it all} 2D Heisenberg FMs~\cite{Rastelli1992,Mattis2006book}, also conjecturing a relation between such super-shallow BSs and the Mermin-Wagner theorem~\cite{Rastelli1991, Rastelli1992}. 

The most puzzling behavior of the BSs in this model concerns large momenta of the pair in the vicinity of the high-symmetry points or directions where the BSs appear unnaturally deep, at odds with the  dimensionality of the pairing problem and the expected shallow binding energies in 2D~\cite{Cooper1956}. 
The existing studies address this observation primarily at a technical level, leaving its physical origin mysterious~\cite{Mattis2006book, Rastelli2013}. 
In addition, few works categorize, reference, or take advantage of the symmetries and symmetry classifications of the BSs~\cite{Loly1976,Callaway1965,Paulson1975}.

In this work, we attempt to demystify most of these findings by relating the unusually deep BSs to the high symmetries and degeneracies of the two-magnon continua in simple FM models, which lead to the effective dimensional reduction of the magnon pairing problem from 2D to one-dimensional (1D) and even zero-dimensional (0D) universalities. 
The one-two-three BS counting paradigm and the statement of existence of one BS for all momenta in 2D are easily falsified by considering models with more complicated couplings, such as the $J_1$--$J_2$ ferro-antiferromagnetic model discussed below and the FM-only model on the non-Bravais lattices~\cite{Oguchi1975}. 
We also employ the partial-wave nomenclature for the classification of the BSs, which, aside from mere convenience, provides one with a deeper and more systematic understanding of their orthogonalities and mixings. 

The recent interest in the spin-nematic states in models and materials containing a mix of FM and antiferromagnetic (AFM) or multiple-spin-exchange interactions~\cite{Chubukov1991Nematic, Lauchli09, Penc2011Nematic,Penc_26} has brought  these and other unresolved issues with magnon BSs back into focus. 
For the AFM phases, a uniform field is also necessary to  reach the fully polarized FM state, from which a condensation of the two-spin-flip BSs leads to the spin-nematics~\cite{Zhitomirsky2010}. 

The Heisenberg $J_1$--$J_2$ FM-AFM model on the 2D square lattice has been explored in this context~\cite{Shannon_2004,Richter2010,Iqbal2016, NematicShengtao2023, Shannon2006, Penc2011Nematic, Shannon2015}.  
The  $d$-wave spin-nematic phase has been unequivocally established for this model using analytical and numerical insights, and the nature of this state as that of the BEC of the $d$-wave magnon pairs has been thoroughly discussed~\cite{Shannon2006,NematicShengtao2023}. 
The spin-nematic state of the condensed magnon BSs turned out to be stable only at finite fields, in a narrow field region near the fully polarized FM state, and only for $0.6\!\alt\!J_2\!<\!\infty$. 
This fragility is due to the more complex many-body effects, such as the pair-attraction leading to the collapse of the condensate into an ordered AFM state at smaller $J_2$ and a strong suppression of the single-magnon gap in the presence of the BS condensate, which leads to a much narrower spin-nematic region than originally expected~\cite{NematicShengtao2023}. 
 
For the spectrum of the magnon BSs and its partial-wave composition as a function of $J_2$, the transformation from the much-scrutinized $J_1$-only FM model to the limit that hosts a nematic condensate is both dramatic and fundamental, calling for deeper insights.
The BS spectrum evolves from a set of two branches, one limited to a corner of the BZ and the other, however shallow, spanning its entirety, to a single branch, which {\it does not} span the entire BZ, but attains a global energy minimum. 
This latter BS branch is also of $d$-wave character, demanding an explanation of how a higher partial wave can be favored in the system with only local attraction. 
In addition, some of the unnaturally deep BSs of the original $J_1$-only model near the high-symmetry pair momenta discussed above also appear surprisingly resilient to the introduction of the $J_2$ term, while others disappear more readily, providing further enigmas. 

In this work, we revisit the two-magnon BS problem in the $J_1$--$J_2$ FM-AFM square-lattice model. 
With a discussion of the reasons for the appearance of the higher partial wave BS as a ground state of this model at a finite $J_2$  provided in Ref.~\cite{NematicShengtao2023}, we show that the introduction of $J_2$ leads to a more generic magnon band structure, which removes the highly degenerate composition of the continuum of the $J_1$-only model and stabilizes a BS with $d$-wave symmetry as the true ground state. 

We also demonstrate that the evolution of magnon BSs can be understood either through the prism of effective dimensional reduction and enhancement, or through the switching of the magnon-magnon interaction in a given channel from attractive to repulsive. These mechanisms, arising either from the structure of the two-magnon continuum or from that of the interaction, control the robustness or fragility of the BSs across momentum space. They are also responsible for the complete disappearance of the BSs from parts of momentum space as a function of $J_2$. With these insights, we study the emergence and disappearance of the BSs as a function of $J_2$ and derive asymptotic expressions for their binding energies at the high-symmetry momenta. The BS problems are solved using lattice partial-wave nomenclature, which is a technically  convenient and consistent framework for the lattice problem. We also present an important methodological development that enables the precise determination of the BS phase boundaries.

The paper is organized as follows. 
In Section~\ref{Sec:MagnonBSEq}, we derive the Schr\"odinger equation for the two-particle wavefunction of spin-flip excitations on a lattice, review the BS problem of particles in the continuum, and introduce lattice partial waves. In Section~\ref{Sec:BS_FM}, we revisit the magnon BS problem in the square-lattice FM and highlight its similarities to and differences from the BS problem for particles in the continuum. In Section~\ref{Sec:BS_FM_AFM}, we extend this analysis to the $J_1$--$J_2$ FM-AFM model and provide a detailed study of the evolution of the BSs as a function of $J_2$. Section~\ref{Sec:Summary} summarizes our results and their implications. The appendixes provide technical details and the Supplemental Material~\cite{SM} presents an animation of the evolution of the BSs as a function of $J_2$ along the high-symmetry contour in the BZ.

\section{Magnon pairing problem}
\label{Sec:MagnonBSEq}

In this Section, we demonstrate the equivalence of two approaches to deriving the Schr\"odinger equation (SE) for the two-magnon BS wavefunction in a FM state, first using the non-orthogonal basis of spin flips and second using the $1/S$-expansion in the spin Hamiltonian. 

To lay out the expectations for the pairing by a short-range attraction in the continuum, we also briefly review the textbook BS problems with a contact potential, introduce partial-wave decomposition, and discuss the notions of the dimensional enhancement and reduction. Finally, we discuss differences of the magnon pairing problem and the rendering of the lattice SE for two-magnon BS to a system of algebraic equations using partial waves.

\subsection{Magnon BS Schr\"odinger equation}

For the Fermi or Bose particles, the Bethe-Salpeter equation on the two-particle Green's function (GF) provides the most general path to the BS problem~\cite{QED, FetterWalecka2012} as the additional poles of the two-particle GF correspond to the paired states~\footnote{We also note that a reduced form of the Bethe-Salpeter equation has been utilized in the related hard-core boson problems under the same name~\cite{ueda_phasesep, JackeliZhitomirsky2004}.}. In the non-relativistic limit and with no retardation in the pairing interaction, the Bethe-Salpeter equation can be consistently reduced to the Schr\"odinger equation---a single-particle equation for the BS of a pair~\cite{QED}. Because of the spin's non-commutative algebra, the diagrammatic rules for spin GFs are non-trivial~\cite{Izyumov1988book} and no equivalent reductionist path to the SE can be taken without approximations. 

However, for a narrower problem of the BSs of spin flips in the FM background, the Schr\"odinger equation for the BS pair can still be derived exactly using significantly more pedestrian approaches~\cite{Fukuda1963, Wortis63}. Here we replicate one of them: the real-space formulation of the two-particle wavefunction in the non-orthogonal basis to obtain the {\it exact} SE for the magnon BS~\cite{Fukuda1963, Callaway1965, Paulson1975, Zhitomirsky2010}. We complement it with the standard $1/S$ expansion of the bosonized spin Hamiltonian,  demonstrate the equivalence of their results for the two-magnon SE in the case of the Heisenberg FM on a Bravais lattice, and discuss the origin of such an equivalence.  

We note that while the former approach is formally exact, its extensions to the AFM~\cite{Rastelli1992} or other fluctuating ground states are not available, while the latter not only is easier because of its orthogonal bosonic basis, but also suffers no such difficulties for the fluctuating problems.

Keeping in mind a generalization of the magnon pairing problem to the polarized FM-like state of the models with mixed interactions, we demonstrate two approaches to the derivation of the magnon BS SE for the spin-$S$ Heisenberg model on a Bravais lattice and in a field
\begin{align}
\hat{\cal H}=\frac{1}{2}\sum_{i,{\bm \delta}}J_{\bm \delta}\,{\bf S}_i\cdot {\bf S}_{i'}-H\sum_i S_i^z, 
\label{eq:HH}
\end{align}
where $J_{\bm\delta}$ are the exchanges between sites $i$ and $i'$, with $\bm \delta\!=\!{\bf r}_{i'}-{\bf r}_i$. For the case of the FM ground state, the field is zero, and for the AFM ground state, the applied field is assumed to be large enough to fully coalign the spins in a state $|0 \rangle \!=\! |\!\uparrow\uparrow\uparrow\dots\rangle$ with the energy
\begin{align}
\frac{E_0}{N}=\frac12 S^2{\cal J}_{0}-HS
\label{eq:E_0}
\end{align}
where  $N$ is the number of lattice sites, and ${\cal J}_{0}\!=\!\sum_{\bm \delta} J_{\bm \delta}$ is the ${\bf k}\!=\!0$ component of the Fourier transform of the exchanges, ${\cal J}_{\bf k}\!=\!\sum_{\bm \delta} J_{\bm \delta}e^{-i {\bf k}{\bm \delta}}$.

\subsubsection{The SE in  the non-orthogonal basis of spin flips}

The single-spin-flip excitations in the polarized state,
\begin{align}
|{\bf k}\rangle = \frac{1}{\sqrt{2SN}}\sum_i e^{-i{\bf kr}_i}S_i^-|0\rangle,
\label{eq:1mag}
\end{align}
are exact and fully {\it orthogonal} eigenstates of the model~\eqref{eq:HH}, with the energy relative to $E_0$ in Eq.~(\ref{eq:E_0}), 
\begin{align}
\varepsilon_{{\bf k}}=H-S\big({\cal J}_{0}-{\cal J}_{\bf k}\big), 
\label{eq:ek_Jq}
\end{align}
where ${\cal J}_{\bf k}\!=\!{\cal J}_{-\bf k}$, as $J_{\bm \delta}\!=\!J_{-\bm \delta}$ for the Bravais lattices. 

One can be tempted, and rightfully so, to build the two-spin-flip states in the same plane-wave-like fashion,
\begin{align}
|{\bf k}_1,{\bf k}_2\rangle = \frac{1}{2SN}\sum_{i,j} e^{-i({\bf k}_1{\bf r}_i+{\bf k}_2{\bf r}_j)}S_i^-S_j^-|0\rangle.
\label{eq:2mag}
\end{align}
However, this basis has a slight problem. Because of the on-site spin algebra, these states are not fully orthogonal,  
\begin{align}
\langle {\bf 4},{\bf 3}|{\bf 1},{\bf 2}\rangle = \delta_{{\bf 1},{\bf 3}}\delta_{{\bf 2},{\bf 4}} + \delta_{{\bf 1},{\bf 4}}\delta_{{\bf 2},{\bf 3}}-\frac{1}{SN}\delta_{{\bf 1}+{\bf 2}, {\bf 3}+{\bf 4}},
\label{eq:2mag_north}
\end{align}
where we use ${\bf i}\!=\!{\bf k}_i$ for brevity. Although the non-orthogonality correction vanishes in the thermodynamic limit as $1/N$ and does not affect the states of the two-particle continuum, it signals that the BS problem needs additional care, since pairing is also a $1/N$ effect. 

While the pairing problem in nuclear, polaronic, and other contexts routinely uses the non-orthogonal basis of states as an approximation~\cite{FetterWalecka2012,tJ94}, in spin systems such as the fully polarized FM-like model considered here,  the BS problem can be solved exactly. This is because the conservation of total $S^z$ ensures that there are no matrix elements connecting the two-spin-flip states to states outside that subspace. Our discussion of this route to the magnon BS SE for the model~\eqref{eq:HH} closely follows Refs.~\cite{Fukuda1963, Callaway1964, Paulson1975, Zhitomirsky2010, Rastelli2013}. 

Using the ansatz for the paired two-spin-flip state
\begin{align}
|\psi_{2}\rangle=\sum_{i,j}  \psi_{ij} |ij\rangle,\quad  |ij\rangle =S^-_i S^-_j|0\rangle,
\label{eq:psi2sf}
\end{align}
and imposing the eigenvalue condition, $\hat{\cal H}|\psi_{2}\rangle = E_2|\psi_{2}\rangle$, yields the first step toward the magnon BS SE
\begin{align}
\sum_{i,j}  \psi_{ij} \big[\hat{\cal H}, S^-_i S^-_j \big]|0\rangle=E|\psi_{2}\rangle,
\label{eq:Hpsi2}
\end{align}
where $E\!=\!E_2-E_0$ is the two-spin-flip energy relative to the ground-state energy~(\ref{eq:E_0}). 

Evaluating the single-spin-flip commutator,
\begin{align}
[\hat{\cal H},S_i^-] = H S_i^- + \sum_{\bm \delta} J_{\bm \delta} \big( S_i^z S_{i+\delta}^- - S_{i+\delta}^z S_i^- \big),    
\end{align}
and using $[\hat{\cal H},S_i^-S_j^-] \!= \![\hat{\cal H},S_i^-]S_j^- + S_i^-[\hat{\cal H},S_j^-]$ together with the redundancy of $|ij\rangle\!=\!|ji\rangle$ in Eq.~\eqref{eq:Hpsi2}, one obtains the real-space SE for the $\psi$ amplitudes in Eq.~\eqref{eq:psi2sf},
\begin{align}
E\psi_{ij} &= 
2(H-S{\cal J}_0)\psi_{ij} 
+ S\sum_{\bm \delta} J_{\bm \delta} \big( \psi_{i+\delta,j} + \psi_{i,j+\delta} \big)
\nonumber \\
&+\frac12 J_{ij}\big(\psi_{ij}+\psi_{ji}- \psi_{ii} - \psi_{jj} \big),
\label{eq:RealSpaceHpsi2}
\end{align}
where $J_{ij}=J_{{\bf r}_j-{\bf r}_i}$. The first line on the right-hand side corresponds to the energy of independent magnons, while the second line represents the interaction between them.

The $\psi_{ij}$ ($\psi_{ji}$) terms in the latter shift the energy of the two spin-flips coupled by $J_{ij}$, and are the lattice version of the potential-like interaction between them.  The terms $\psi_{ii(jj)}$ correspond to the same-site amplitudes, giving rise to the non-potential-like part of the interaction.

We note that for spin $S\!=\!\frac12$, there is an additional constraint, because two spin-flips on the same site are forbidden. However, these formally unphysical amplitudes, $\psi_{ii(jj)}$, are important to retain, because they cancel out their twins from the first line in the magnon-energy part of Eq.~\eqref{eq:RealSpaceHpsi2} for $i\!=\!j\mp\delta$.
They do not affect the physical amplitudes and have no impact on the two-spin-flip wavefunction~\cite{Fukuda1963}, leaving Eq.~\eqref{eq:RealSpaceHpsi2} valid for any spin $S$. 

The two-magnon SE in momentum space is obtained by a double Fourier transform of $\psi_{ij}$ in Eq.~\eqref{eq:RealSpaceHpsi2}, for which it is convenient to introduce the total and relative momenta of the pair, ${\bf K}$ and ${\bf q}$, associated with the center-of-mass and relative coordinates, respectively,
\begin{align}
\psi_{ij}=\frac{1}{N^2}\sum_{{\bf K},{\bf q}}e^{i\frac{\bf K}{2}({\bf r}_i+{\bf r}_j)} e^{i{\bf q}({\bf r}_j-{\bf r}_i)} \psi_{\bf K}({\bf q}) .
\label{eq:2mag_wfnFT}
\end{align}
With that, the algebra outlined in Ref.~\cite{Rastelli2013} transforms Eq.~\eqref{eq:RealSpaceHpsi2} to the two-magnon SE in momentum space,
\begin{align}
\Big(E-E_{\bf K}({\bf q})\Big)\psi_{\bf K}({\bf q})
=\frac{1}{2N}\sum_{\bf p} V_{\bf K}({\bf q},{\bf p})\psi_{\bf K}({\bf p}),
\label{eq:2MagSE}
\end{align}
where $E_{\bf K}({\bf q})\!=\!\varepsilon_{\frac{\bf K}{2}+{\bf q}}+\varepsilon_{\frac{\bf K}{2}-{\bf q}}$ is the two-magnon energy and the interaction is given by
\begin{align}
V_{\bf K}({\bf q},{\bf p}) = {\cal J}_{{\bf q}-{\bf p}}+{\cal J}_{{\bf q}+{\bf p}}-{\cal J}_{\frac{{\bf K}}{2}-{\bf q}}-{\cal J}_{\frac{\bf K}{2}+{\bf q}}.
\label{eq:2MagVKqp}
\end{align}
The solutions of Eq.~\eqref{eq:2MagSE} for a given ${\bf K}$ with $E$ below the minimum of the continuum $E_{\bf K}({\bf q})$, correspond to the two-magnon BSs.

The first two terms in the interaction~\eqref{eq:2MagVKqp} have the usual ``potential-like'' structure, as they depend on the momentum transfer between the initial and final particles in the scattering process, ${\bf q}\pm\bf p$. The remaining two terms depend explicitly on the incident particles' momenta, ${\bf K}/2\pm{\bf q}$, rendering the lattice scattering problem neither directly nor uniquely mappable onto the continuum problem. It is worthwhile to mention that they have the form of the single-particle kinetic energies, which come from the non-orthogonal part of the two-spin-flip states in Eq.~\eqref{eq:2mag_north} that the pairs are built from~\cite{tJ94}.

A minor note concerns the periodicity of the solutions of the SE, Eq.~\eqref{eq:2MagSE}. Although the pair's center of mass formally resides on a different lattice from that of the constituent particles, the solution of the two-particle eigenvalue equation, Eq.~\eqref{eq:2MagSE}, on the Bravais lattice is periodic with respect to ${\bf K}\!\rightarrow\!{\bf K}+{\bf G}$, where ${\bf G}$ is a reciprocal-lattice vector; see App.~\ref{A:BS_Periodicity}.

\subsubsection{The $1/S$ spin-wave approach to the SE}

Within the standard spin-wave theory, bosonization of spin operators~\cite{Dyson1956,Maleev1958,hp1940} provides a natural basis of bosonic magnons, which is orthogonal for any $n$-particle sector due to their full commutativity, so a bosonic equivalent of the two-spin-flip state of Eq.~(\ref{eq:2mag}) does not have the non-orthogonality problem of Eq.~(\ref{eq:2mag_north}) that would necessitate a special treatment of the pairing problem. 

Within this approach, the magnon-magnon interaction appears directly in the $1/S$-expansion of the Hamiltonian, which, in the case of the FM states and isotropic models, conserves the number of particles in scattering due to $U(1)$ symmetry, maintaining a one-to-one correspondence of magnons to spin-flips with $\Delta S^z\!=\!-1$ quantum numbers~\cite{RMP_13}. We also note that the Hermitian Holstein-Primakoff (HP) and non-Hermitian Dyson-Maleev (DM) bosonic spin representations~\cite{Dyson1956,Maleev1958,hp1940} lead to identical results for the two-magnon problem.  

Using the standard HP transformation, $S^z_i\!=\!S-n_i$ and $S^-_i\!=\!a_i^\dagger\big(2S-n_i\big)^{1/2}$  with $n_i\!=\!a_i^\dagger a_i$, for the model~(\ref{eq:HH}) in the polarized state, leads to the expansion in $1/S$ and in powers of bosonic operators, $\hat{\cal H} \!=\!E_0\!+\!\hat{\cal H}^{(2)}\!+\!\hat{\cal H}^{(4)}\!+\!\dots$, with the symmetry allowing only even powers. 

Straightforward algebra yields the same answer for the energy $E_0$ as in Eq.~(\ref{eq:E_0}), and, after a Fourier transform, the linear spin-wave theory (SWT) Hamiltonian,  
\begin{align}
\hat{\cal H}^{(2)} = \sum_{\bf k} \varepsilon_{\bf k} a_{\bf k}^\dagger a^{\phantom{\dagger}}_{\bf k}, 
\label{eq:H2}
\end{align}
with the magnon energy $\varepsilon_{\bf k}$ identical to that of the spin-flip in Eq.~(\ref{eq:1mag}). Equally straightforwardly, the real-space two-magnon interaction is given by, 
\begin{equation}
\hat{\cal H}^{(4)} \!=\! 
\sum_{i, {\bm \delta}} J_{\bm \delta}\Big( n_i n_j -\frac14\Big(a_j^\dagger n_ia_i+ a_i^\dagger n_j a_j + {\rm H.c.}\Big) \Big), \label{eq:H4RealSpace}
\end{equation}
where ${\bf r}_j\!=\!{\bf r}_i+{\bm \delta}$. The first term is the density-density interaction, originating from the $S^z_iS^z_j$ part of the exchange, while the rest of the terms, having the form of a correlated hopping, are from its $S^+_iS^-_j$ part. The elementary  transformation to momentum space yields,
\begin{align}
\hat{\cal H}^{(4)} = 
\frac{1}{4N}\sum_{{\bf 1}+{\bf 2}={\bf 3}+{\bf 4}}  V({\bf 1},{\bf 2};{\bf 3},{\bf 4}) \,
a^\dagger_{{\bf 4}}a_{{\bf 3}}^\dagger a_{{\bf 2}}^{\phantom{\dagger}} a_{{\bf 1}}^{\phantom{\dagger}},
\label{eq:H4}
\end{align}
where ${\bf i}\!=\!{\bf k}_i$. After symmetrization and relabeling the momenta into the total and relative ones before and after the scattering, ${\bf k}_{1,2}\!=\!\frac{\bf K}{2}\pm{\bf q}$ and ${\bf k}_{3,4}\!=\!\frac{\bf K}{2}\pm{\bf p}$, the 4-magnon vertex is given by,
\begin{align}
V({\bf 1},{\bf 2};{\bf 3},{\bf 4}) = {\cal J}_{{\bf q}-{\bf p}}+{\cal J}_{{\bf q}+{\bf p}}-{\cal J}_{\frac{{\bf K}}{2}-{\bf q}}-{\cal J}_{\frac{\bf K}{2}+{\bf q}},
\label{eq:2MagVKqp_1_S}
\end{align}
which is manifestly the {\it same} expression as in Eq.~(\ref{eq:2MagVKqp}), with the first two ``potential-like'' terms coming from the density-density part in Eq.~(\ref{eq:H4RealSpace}) and the two ``non-potential-like'' ones from the correlated hoppings, resulting naturally in their single-particle kinetic-energy form. 

Because there is no retardation in the interaction~(\ref{eq:H4}), one can bypass the full Bethe-Salpeter formalism~\cite{QED} for the BSs, and  formulate the SE for them directly for the two-magnon wavefunction,
\begin{equation}
|{\bf K}\rangle = \frac{1}{\sqrt{2N}}\sum_{\bf q} \psi_{\bf K}({\bf q})\,
a^\dagger_{{\bf K}/2+{\bf q}} a^\dagger_{{\bf K}/2-{\bf q}} |0\rangle,
\label{eq:psi_oguchi}
\end{equation}
with the eigenvalue problem  $\hat{\cal H}|{\bf K}\rangle \!= \!E|{\bf K}\rangle $ resulting in the SE that is identical to Eq.~(\ref{eq:2MagSE}), the route also taken in Ref.~\cite{Oguchi1971} using the DM spin-representation. 

Altogether, with the ease and consistency of the spin-bosonization approach to the BS problem demonstrated here compared with a significantly more cumbersome formally exact treatment in a non-orthogonal basis outlined above, one may wonder what the objection against the SWT formalism was to begin with and why it did not matter. With the two approaches leading to {\it identical} SEs for the magnon BSs problem,  there is certainly no evidence to suggest that the overcompleteness of the bosonic formulation has any projection on the pairing problem, much less to undermine the applicability of SWT as a whole even for the ferromagnets~\cite{Mattis2006book, RodriguezDemlerPRB2022}. 

The objection has been that the spin-bosonization allows for unphysical populations of bosonic spin-flips. But, evidently, for the high symmetry of the FM states within the isotropic models, which translates into the particle-number conservation for their excitations, and for the BS problem involving only one- and two-particle sectors, the resultant SE for the BS is exactly the same, with the overcompleteness having no bearing on the magnon BS problem. An additional insight can be obtained by examining the SE in Eq.~(\ref{eq:2MagSE}), in which the left-hand side with the single-magnon energies is of order ${\cal O}(S)$, while the interaction part on the right-hand side is precisely next order in $1/S$, i.e., ${\cal O}(S^0)$, regardless of the approach. With all the terms for the physical states in each order properly accounted for, 
this leads to an exact SE, underscoring consistency of the $1/S$-approach. 

Beyond its technical ease and conceptual transparency, it is also worth noting that the bosonic framework extends naturally to the AFMs~\cite{Oguchi1971}, while the real-space commutator construction for them or other fluctuating ground states is simply not available~\cite{Rastelli1992}.

\subsection{Pairing in the continuum limit}
\label{sec:continuum}

In order to provide a backdrop of the standard expectations against which the differences and similarities of the magnon pairing problem can be seen, here we recall key results for the pairing of particles by a short-range interaction in continua of different dimensions~\footnote{For the relevant derivations, see Ref.~\cite{Galitskii}:  Problem 2.17 (see also 2.7) for 1D, Problem 4.38 for 2D, and Problems 4.1 and 4.10 for 3D.}. 

For two particles of the same mass $m$ interacting with a potential $U({\bf r}_1-{\bf r}_2)$, the two-body problem reduces to that of a single particle of mass $m/2$ in an external potential. The two-particle kinetic energy 
\begin{align}
\varepsilon_{\frac{\bf K}{2}+{\bf q}}+\varepsilon_{\frac{\bf K}{2}-{\bf q}} = \frac{{\bf K}^2}{4m}+\frac{{\bf q}^2}{m}=E_{\bf K}+\frac{{\bf q}^2}{2m^*},
\label{eq:2pE}
\end{align}
naturally splits into the sum of the energy of their center of mass, $E_{\bf K}$, which is also the energy of the bottom of the two-particle continuum at the momentum ${\bf K}$, and the ``relative'' kinetic energy of an effective mass $m^*\!=\!m/2$. 

Trivially, since the potential $U({\bf r})$ depends only on the relative position ${\bf r}$ of the particles, in momentum space it depends only on the momentum transfer in the scattering process, and not on the total momentum of the pair ${\bf K}$. For the ${{\bf k}_1},{{\bf k}_2}\!\Rightarrow\!{{\bf k}_3},{{\bf k}_4}$ scattering, with ${\bf k}_i\!=\!{\bf K}/2\pm{\bf q}({\bf p})$, and having in mind bosonic symmetrization as in Eq.~(\ref{eq:H4}), this yields $V({\bf q},{\bf p})\!=\!\frac12\big(U({\bf q}-{\bf p})+U({\bf q}+{\bf p})\big)$, leading to the SE for $\psi_{\bf K}({\bf q})$ of the pair from Eq.~(\ref{eq:psi_oguchi}),
\begin{align}
\Big(E-E_{\bf K}-\frac{{\bf q}^2}{m}\Big)\psi_{\bf K}({\bf q})=\frac{1}{2N}\sum_{\bf p} V({\bf q},{\bf p})\psi_{\bf K}({\bf p}),
\label{eq:2pSE}
\end{align}
where we keep the sum instead of the integral for consistency with Eq.~(\ref{eq:2MagSE}). Clearly, the solution of Eq.~(\ref{eq:2pSE}) for the BS energy relative to the bottom of the two-particle continuum, $E-E_{\bf K}$, is independent of ${\bf K}$.

\subsubsection{The simplest case}
\label{simplest}

Consider the simplest case of an attractive contact potential, $U({\bf r})=-\alpha \delta({\bf r})$. It corresponds to a constant in momentum space, $V({\bf q},{\bf p})=-\alpha$, which reduces the SE in (\ref{eq:2pSE}) to an algebraic equation for the wave function, with the solution
\begin{align}
\psi({\bf q})=\frac{\alpha\, C}{2\Delta+{\bf q}^2/m}, \quad
C = \frac{1}{2N}\sum_{{\bf p}} \psi({\bf p}),
\label{eq:psiCk_Cooper}
\end{align}
where we introduced the (positive) binding energy
\begin{equation}
2\Delta = E_{\bf K}-E,
\label{eq:DeltaK}
\end{equation}
relative to the bottom of the two-particle continuum $E_{\bf K}$ and removed the otherwise extraneous index ${\bf K}$.  

The self-consistent condition on the eigenfunction via the auxiliary constant $C$ in Eq.~\eqref{eq:psiCk_Cooper} provides an implicit eigenvalue equation for the pair's binding energy
\begin{equation}
1=\frac{\alpha}{2N}\sum_{\bf p} \frac{1}{2\Delta+{\bf p}^2/m},
\label{eq:SCeq_Cooper}
\end{equation}
such that a solution with $\Delta\!>\!0$ corresponds to a BS.

A direct evaluation of the integral in Eq.~\eqref{eq:SCeq_Cooper} leads to the textbook results and standard expectations for the outcome of the pairing in different dimensions~\cite{Galitskii}. Thus, by examining the small $\Delta$ limit, the existence of a BS can be guaranteed in 1D and 2D cases, in which the integral diverges  as $1/\sqrt{\Delta}$ and $ \ln \Delta$, respectively, yielding a BS solution for an arbitrarily weak attraction strength $\alpha$. 
In 1D, Eq.~(\ref{eq:SCeq_Cooper}) straightforwardly leads to~\cite{Galitskii}
\begin{align}
\Delta^{\rm 1D}\approx \frac{\alpha^2 m}{32}\ .
\label{eq:SE_1D}
\end{align}
For the 2D case, the solution is 
\begin{align}
\Delta^{\rm 2D}\approx \frac{\Lambda}{2m}\, \exp\Big(-\frac{8\pi}{\alpha m}\Big),
\label{eq:SE_2D}
\end{align}
where $\Lambda$ is a large-momentum cut-off and the non-analytic dependence on the coupling constant $\alpha$ is clear. We loosely refer to it as a solution of the Cooper problem~\cite{Cooper1956}, in which the presence of the finite Fermi-momentum leads to a {\it dimensional reduction} from 3D to 2D, the effect discussed below in more detail, resulting in the 2D-like density of states and the pairing gap that closely follows  Eq.~(\ref{eq:SE_2D}); see also Ref.~\cite{Galitskii}.

In 3D, the integral in~\eqref{eq:SCeq_Cooper} is finite for $\Delta\!\rightarrow \! 0$, so that the eigenvalue equation has a solution only for a coupling strength that exceeds a threshold value  $\alpha_c\!\sim\!\mathcal{O}(1/m)$.  

The power of these statements is that they rely only on the density of states in D dimensions and are expected to apply to any short-range attractive potential for non-relativistic (massive) particles.

As all the models in this study are 2D, the result in Eq.~(\ref{eq:SE_2D}) seems most relevant. However,  considering the other dimensions will also be important, because of the dimensional reduction and enhancement discussed below.

\subsubsection{Partial waves and dimensional enhancement}
\label{Sec:partial_waves}

Another systematic simplification of the SE (\ref{eq:2pSE}) for a more general interaction than in the simplest case discussed above uses the symmetry classification of the BSs. While a strict classification is available only at the high-symmetry ${\bf K}$-points, the symmetry labeling can be extended away from such points, following the continuity of the BS branches. Moreover, some of the symmetries of the states are retained along the high-symmetry ${\bf K}$-directions for a given lattice. 

In the continuum, the ${\bf K}\!=\!0$ point allows for an approach that is known as the partial-wave decomposition~\cite{Callaway1964}. As in the case of single-particle scattering in an external spherically-symmetric potential, the pair wave-function can be divided into orthogonal orbital waves $\psi_{\gamma}$, with $\gamma\!=\!\{s,p,d,\dots\}$. Accordingly, the pair's interaction potential $V({\bf q},{\bf p})$ can be written as a sum of the partial-wave terms, 
\begin{align}
V({\bf q},{\bf p})= \sum_\gamma V_\gamma({\bf q},{\bf p}),
\label{eq:Vkq_orth}
\end{align}
responsible for pairing in different orbital channels $\gamma$, which break up the SE in Eq.~(\ref{eq:2pSE}) into a system of independent integral equations for a BS in each symmetry channel. 

To expose additional insights provided by this decomposition, make further progress, and demonstrate generic properties of the solutions of the BS problems, we make a mild assumption that in each partial-wave channel the interaction potential $V_\gamma({\bf q},{\bf p})$ is separable, 
\begin{align}
V({\bf q},{\bf p})= -\sum_\gamma \alpha_\gamma R^*_\gamma({\bf q}) R_\gamma({\bf p}).
\label{eq:Vkq_orth_harmonics}
\end{align}
where $\alpha_\gamma$ are constants, and the orthogonality of the $V_\gamma({\bf q},{\bf p})$ terms follows from that of the harmonics
\begin{align}
\frac{1}{N}\sum_{\bf p}   R^*_\gamma({\bf p}) R_{\gamma'}({\bf p})=\delta_{\gamma,\gamma'}.
\label{eq:orthogonal_harmonics}
\end{align}
Note that for the magnon pairing problem on the lattice, the pair potential naturally splits into the lattice harmonics, which lead to a similar construction; see Sec.~\ref{sec:magSE}. Using Eq.~(\ref{eq:Vkq_orth_harmonics}), the SE in Eq.~(\ref{eq:2pSE}) reduces to a set of algebraic equations for the wave-functions in each channel
\begin{equation}
\psi_{\gamma}({\bf q})\!=\!\frac{\alpha_\gamma C_{\gamma}R_\gamma^*({\bf q})}{2\Delta_\gamma+{\bf q}^2/m}, \ \
C_\gamma \!=\! \frac{1}{2N}\sum_{{\bf p}} R_{\gamma} ({\bf p})\psi_{\gamma}({\bf p}),
\label{eq:psiCk_partialwave}
\end{equation}
with the selfconsistency conditions leading to the implicit eigenvalue equations for their binding energies $\Delta_\gamma$ 
\begin{align}
1=\frac{\alpha_\gamma}{2N}\sum_{\bf p}\frac{|R_{\gamma}({\bf p})|^2}{2\Delta_\gamma+{\bf p}^2/m},
\label{eq:SE_Egamma}
\end{align}
which can be seen as a generalization of Eq.~(\ref{eq:SCeq_Cooper}).

Crucially, the symmetry of the pairing potential $V_\gamma$ must follow that of the corresponding partial wave, prescribing a nodal structure to the pairing potential in momentum space for all partial waves except the $s$-wave. In the continuum, $R_s\!=\! const$ for the $s$-wave, $R_{p_{x(y)}}({\bf q}) \!\propto\! q_{x(y)}$ for the $p_{x(y)}$-wave, etc. Therefore, with the exception of the $s$-wave channel, which is clearly identical to the contact-interaction problem discussed in Sec.~\ref{simplest}, all other channels will have nodes in the pairing potentials that pass through the continuum minimum, ${\bf p}\!=\!0$. 

For the partial waves higher than the $s$-wave, this leads to additional powers of ${\bf p}$ in the numerator of the integral in (\ref{eq:SE_Egamma}), rendering it equivalent to a higher-dimensional problem---an effect we refer to as the {\it dimensional enhancement}. Given the generality of this consideration, it is believed that the weak attraction in a 2D pairing problem can create only an $s$-wave BS, while for a higher partial-wave the corresponding coupling strength $\alpha_\gamma$ must exceed a threshold value for a BS to exist.

\subsection{SE for two-magnon BS}
\label{sec:magSE}

\subsubsection{Differences of the magnon pairing problem}

With the insights of Sec.~\ref{sec:continuum}, it is useful to outline where the differences between the magnon pairing and a generic BS problem in the continuum considered above can come from. A direct comparison of the SE for the potential-like pairing in the continuum in Eq.~(\ref{eq:2pSE}) and the two-magnon SE on a lattice in Eq.~(\ref{eq:2MagSE}) suggests two sources: the structure of the two-magnon energy continuum on the left-hand side of the SE and the structure of the magnon-magnon interaction on its right-hand side.

For the first, the following effects will be important. The two-particle kinetic energy $\varepsilon_{\frac{\bf K}{2}+{\bf q}}+\varepsilon_{\frac{\bf K}{2}-{\bf q}}$ on a lattice is not trivially split into that of the center-of-mass and that of the relative motion of the particles, so the binding energy will generally acquire a dependence on the pair-momentum ${\bf K}$. For the models with only nearest-neighbor exchanges and for high-symmetry momenta ${\bf K}$, the two-magnon continuum can produce highly degenerate ${\bf q}$-bands, effectively reducing the dimensions of the SE, the effect we refer to as the {\it dimensional reduction}; see Sec.~\ref{Sec:BS_FM}. Lastly, the more general models with competing exchanges can render the magnon-band topography different from that in the continuum limit by shifting band minima away from the nodes of the higher partial waves, undermining the generality of the dimensional enhancement argument and violating the partial-wave pairing hierarchy discussed above~\cite{Zhitomirsky2010,NematicShengtao2023}.

The structure of the magnon-magnon interaction in Eq.~(\ref{eq:2MagSE}) does not conform to the potential-like form, containing terms that explicitly depend on the pair momentum (or momenta of individual particles), not just on the momentum transfer. Not only does it lead to another source of the ${\bf K}$-dependence for the binding energy in the SE, but it also renders some of the seemingly iron-clad logic of the mandatory $s$-wave pairing in lower dimensions moot. The presence of the AFM couplings in the models also introduces repulsive terms, resulting in a magnon interaction that can be purely repulsive throughout entire regions of ${\bf K}$-space.

\subsubsection{Lattice partial waves and magnon-BS SE}
\label{Sec:magSE_general}

For spin models on a lattice, the couplings between spins usually extend over a finite range. Therefore, for the fully polarized state, the momentum-dependence of the magnon-magnon interaction $V_{\bf K}({\bf q},{\bf p})$ in Eq.~\eqref{eq:2MagVKqp} on the Bravais lattice is simple, because it is composed of the exchanges ${\cal J}_{\bf q}\!=\!\sum_{\bm \delta} J_{\bm \delta}e^{-i{\bf q}{\bm \delta}}$, which consist of a finite sum of cosine functions of various momenta---the so-called lattice harmonics~\cite{Zhitomirsky2010}. 

This property suggests a generic procedure, which is similar, but not analogous, to the partial-wave decomposition discussed above.  Each lattice harmonic in $V_{\bf K}({\bf q},{\bf p})$ can {\it always} be cast in a separable form using trigonometric identities, leading to an expansion of the magnon interaction in a finite set of terms
\begin{align}
V_{\bf K}({\bf q},{\bf p}) = -\sum_\gamma \alpha_\gamma R_{\gamma}({\bf q}) \widetilde{R}_{\gamma,{\bf K}}({\bf p}),
\label{eq:Vkqp_expansion}
\end{align}
where $\gamma$ enumerates harmonics, $\alpha_\gamma$ are constants, $\{R_{\gamma}({\bf q})\}$ is the chosen basis of lattice harmonics, and $\widetilde{R}_{\gamma,{\bf K}}( {\bf p}) \!=\!R_{\gamma}({\bf p})-R_{\gamma}({\bf K}/2)$ renders the separable kernels of the integral SE asymmetric due to the non-potential-like terms in Eq.~\eqref{eq:2MagVKqp}. Although different terms of this expansion are not necessarily orthogonal, they often are.  

Using this expansion transforms SE in Eq.~\eqref{eq:2MagSE} into
\begin{align}
\label{eq:2magSE_algebraic1}
\psi_{\bf K}({\bf q})=-\frac{1}{E-E_{\bf K}({\bf q})}
\sum_{\gamma}\alpha_\gamma R_{\gamma}({\bf q})\,C_{\gamma,{\bf K}}, 
\end{align}
where $E_{\bf K}({\bf q})\!=\!\varepsilon_{\frac{\bf K}{2}+{\bf q}}+\varepsilon_{\frac{\bf K}{2}-{\bf q}}$ as before, and the  auxiliary constants for the fixed momentum of the pair ${\bf K}$ are
\begin{align}
\label{eq:2magSE_algebraic2}
C_{\gamma,{\bf K}} &= \frac{1}{2N}\sum_{{\bf p}}\widetilde R_{\gamma,{\bf K}}({\bf p})\psi_{\bf K}({\bf p}),
\end{align}
with the algebraic structure above similar to the one in Eq.~\eqref{eq:psiCk_partialwave}. The selfconsistency in~\eqref{eq:2magSE_algebraic2} leads to a system of coupled linear equations for the auxiliary constants  $C_{\gamma,{\bf K}}$, 
\begin{align}
{\bf C}_{\bf K}={\bf\hat{\rm\bf M}}_{\bf K}{\bf C}_{\bf K}, 
\label{eq:C_algebraic}
\end{align}
with ${\bf C}^T_{\bf K}\!=\!\{C_{\gamma,{\bf K}}\}$ and matrix elements of ${\bf\hat{\rm\bf M}}_{\bf K}$ 
\begin{align}
{M}_{\gamma \gamma',{\bf K}} &= -\frac{\alpha_{\gamma'}}{2N}\sum_{\bf p}  \frac{R_{\gamma'}(\mathbf{p}) \, \widetilde{R}_{\gamma,{\bf K}}(\mathbf{p})}{E- E_{\bf K}(\mathbf{p})}.
\label{eq:M_gamma}
\end{align}
From Eq.~\eqref{eq:C_algebraic}, the implicit eigenvalue equation for the magnon BS energies is given by 
\begin{equation}
\mbox{det}\big({\bf\hat{\rm\bf M}}_{\bf K}-{\bf\hat{\rm\bf I}}\big)=0,
\label{eq:detM=0}
\end{equation}
yielding the energies $E$ of such BSs that occur below the minimum of the two-magnon continuum $E_{\bf K}({\bf q})$ at a given ${\bf K}$. While appearing somewhat more complex than the eigenvalue problem in Eq.~(\ref{eq:SE_Egamma}),  Eq.~(\ref{eq:detM=0}) reduces to a set of similar ones if one assumes that all lattice harmonics are orthogonal, making the matrix ${\bf\hat{\rm\bf M}}_{\bf K}$ diagonal and producing eigenvalue equations ${M}_{\gamma \gamma,{\bf K}}\!=\!1$.

One can see the method outlined here as a more general and yet a particular version of the simplified partial-wave decomposition procedure discussed in Sec.~\ref{Sec:partial_waves}. For an arbitrary ${\bf K}$ of the pair, the SE is still reduced to a set of algebraic equations for the wavefunction~\eqref{eq:2magSE_algebraic1}, but now they are generally coupled. 

Formally, the corresponding eigenvalue problem in Eq.~\eqref{eq:detM=0} requires evaluation of the finite number $\sim\!N_\gamma^2$ of the integrals in Eq.~\eqref{eq:M_gamma}, where $N_\gamma$ is the number of lattice harmonics in the expansion in Eq.~(\ref{eq:Vkqp_expansion}). However, as we will demonstrate below, for the high-symmetry ${\bf K}$-points and along the high-symmetry ${\bf K}$-directions, the matrix ${\bf\hat{\rm\bf M}}_{\bf K}$ can be systematically reduced to a block-diagonal form, of which only certain blocks are relevant to the BS problem. Moreover, for the relatively simple models considered here, such blocks often appear as singlets or are degenerate, and are also amenable to exact or asymptotically exact analytic solutions.

Most frequently, the minimal approach for the simple models has used the basis of cosines for the lattice harmonics $R_{\gamma}({\bf q})$~\cite{Zhitomirsky2010, Rastelli2013, Fukuda1963, Wortis63}. However, as we advocate in the present study, the most natural, physical, and technically advantageous basis for the lattice problems is provided by the basis of lattice partial waves, whose symmetry offers a systematic understanding of their orthogonalities and mixings; see also Refs.~\cite{Callaway1965, Paulson1975, Loly1976}.

For the purposes of this study, we refer to the linear combinations of the ``elemental'' lattice harmonics that  transform according to the symmetry of the lattice as the {\it lattice partial waves}, with the simplest examples on the square lattice being the generalized $s$-wave $\propto\!(\cos q_x+\cos q_y)$ and $d$-wave $\propto\!(\cos q_x-\cos q_y)$. Their basis allows one to remove redundancy in the expansion of the magnon potential in Eq.~(\ref{eq:Vkqp_expansion}) and lend a more informative structure to the matrix ${\bf\hat{\rm\bf M}}_{\bf K}$. It also provides a convenient symmetry classification of the resultant BSs either at the high-symmetry ${\bf K}$ or away from them.

\section{magnon bound states in the square-lattice ferromagnet}
\label{Sec:BS_FM}

In this Section, we revisit one of the most studied 2D models that realize magnon BSs, the square-lattice Heisenberg ferromagnet~\cite{Loly1986, Mattis2006book, Rastelli2013, Wortis63}. The BSs within this model exhibit several puzzling features mentioned in Sec.~\ref{Sec:Intro}: the super-exponentially-shallow BS near the BZ center and two anomalously deep BSs along the high-symmetry momentum directions, with the second BS existing only in part of the BZ and the first spanning its entirety. Because these features arise in such a simple and symmetric model, we can expose the roles of dimensional reduction and enhancement and illustrate the usefulness of lattice partial waves in the magnon pairing problem.

\subsection{Lattice partial waves}
\label{Sec:BS_FM_partialwaves}

The model~(\ref{eq:HH}) for the square-lattice Heisenberg ferromagnet contains only the nearest-neighbor term $J_1\!=\!-1$, which we set as the energy unit for the rest of this study. The field term in~(\ref{eq:HH}) can be dropped as the ground state is fully polarized. The sum in the spin-interaction part of the model involves four nearest-neighbor vectors ${\bm \delta}$,  $(\pm1,0)$ and $(0,\pm1)$, in units of the lattice spacing.

The Fourier transform of the exchange and the tight-binding-like single-magnon energy in Eq.~\eqref{eq:ek_Jq} become 
\begin{equation}
\label{eq:J1_1mag}
{\cal J}_{\bf k}=-4\gamma_{\bf k} \ \ \mbox{and}\ \ \varepsilon_{\bf k}=4S(1-\gamma_{\bf k}),
\end{equation}
respectively, where we introduced a symmetric combination of the nearest-neighbor lattice harmonics 
\begin{align}
\gamma_{\bf q}=\frac12(\cos q_x+\cos q_y).\label{eq:gamma+}
\end{align}
For the magnon-magnon interaction in Eq.~\eqref{eq:2MagVKqp}, straightforward algebra using ${\cal J}_{\bf k}$ from Eq.~(\ref{eq:J1_1mag})  leads to 
\begin{equation}
\label{eq:VKq_J1expansion}
V_{\bf K}({\bf q},{\bf p})=-\alpha_s R_{s}({\bf q})\widetilde{R}_{s,{\bf K}}({\bf p})-\alpha_d R_{d}({\bf q})\widetilde{R}_{d,{\bf K}}({\bf p}),
\end{equation}
which explicitly conforms to the lattice partial-wave expansion suggested in Eq.~\eqref{eq:Vkqp_expansion}, with the two partial waves, $\gamma\!=\!\{s,d\}$, and their coupling constants and harmonics
\begin{align}
&\alpha_s\!=\!8, \quad R_{s}({\bf q})=\gamma_{\bf q}, \quad \widetilde{R}_{s,{\bf K}}({\bf p})=\gamma_{\bf p}-\gamma_{\frac{\bf K}2},\nonumber\\
&\alpha_d\!=\!8, \quad R_{d}({\bf q})=\gamma^{-}_{\bf q}, \quad \widetilde{R}_{d,{\bf K}}({\bf p})=\gamma^-_{\bf p}-\gamma^-_{\frac{\bf K}2},
\label{eq:V_J1only}
\end{align}
where we have also introduced an antisymmetric combinations of the nearest-neighbor lattice harmonics,
\begin{align}
\gamma^-_{\bf q}&=\frac12(\cos q_x-\cos q_y).\label{eq:gamma-}
\end{align}

The two lattice partial waves, $R_{s}({\bf q})$ and $R_{d}({\bf q})$ in Eq.~\eqref{eq:VKq_J1expansion}, correspond to the generalized $s$-wave and the $d_{x^2-y^2}$-wave (referred to as the $d$-wave), respectively. The generalized $s$-wave is finite at the ${\bf q}\!=\!0$, but has nodal lines elsewhere in the BZ.  The $d$-wave has the nodal structure that crosses the ${\bf q}\!=\!0$ point, as is expected; see Sec.~\ref{Sec:s_and_d_wave} for a  discussion of their symmetries.

\begin{figure*}[t]
\includegraphics[width=.8\linewidth]{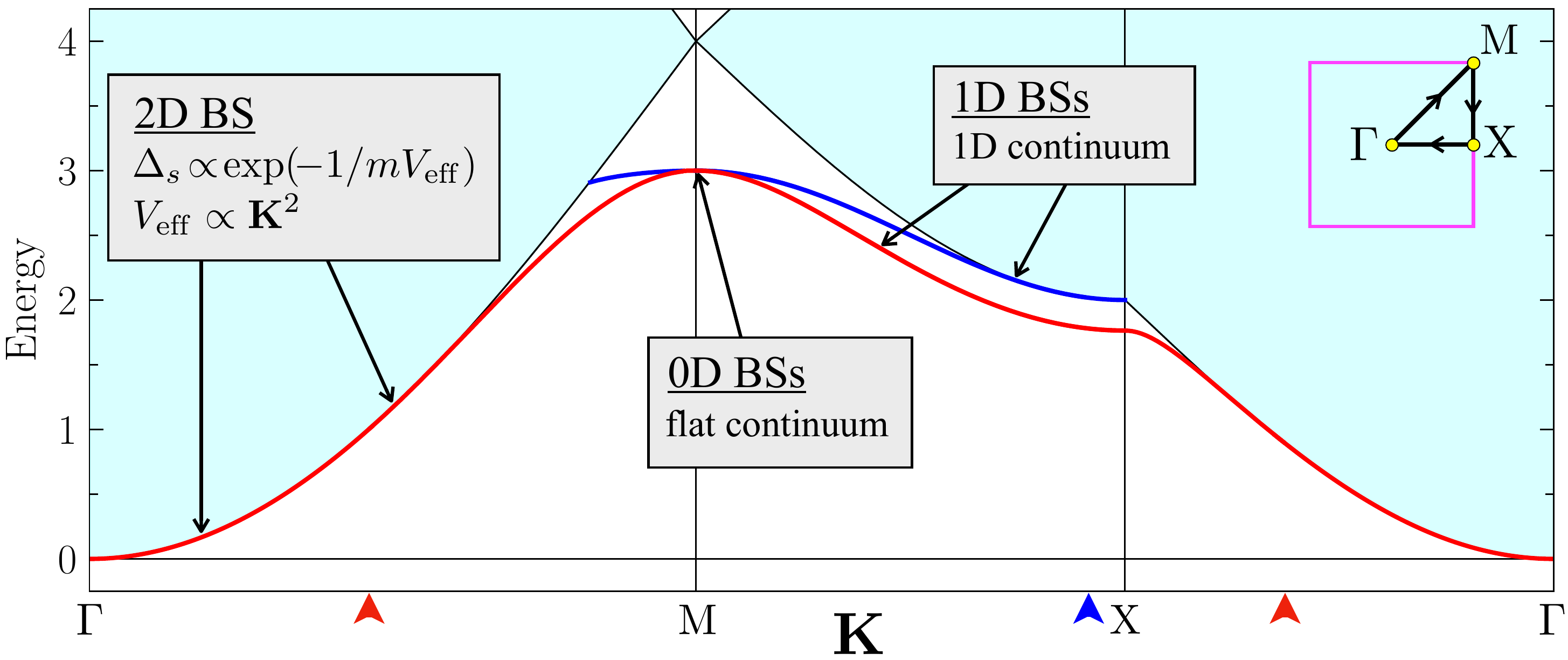}
\vskip -0.3cm
\caption{Red and blue solid lines: the energies of the BSs in the $S\!=\!1/2$ square-lattice Heisenberg ferromagnet vs pair momentum ${\bf K}$ along the high-symmetry $\Gamma$MX$\Gamma$ path in the BZ; see the inset, $\Gamma\!=\!(0,0)$, M$=\!(\pi,\pi)$, and X$=\!(\pi,0)$. The shaded region is the two-magnon continuum energy relative to its minimum; the black solid line is its boundary. Arrows point to the regions of the super-exponentially shallow $s$-wave BS near the $\Gamma$ point, the zero-dimensional (0D) degenerate BSs at the M point, and the 1D BSs along the MX line discussed below. Some of the binding energies are below numerical detectability for a significant portion of the $\Gamma$MX$\Gamma$ path, with the arrows along the horizontal axis that indicate the ${\bf K}$ values where the BS energies become numerically indistinguishable from the bottom of the two-magnon continuum; see the text.}
\label{fig:EkJ1}
\vskip -0.3cm
\end{figure*}

Using $\varepsilon_{\bf k}$ from~(\ref{eq:J1_1mag}), the two-magnon energy also decouples in a compact form
\begin{align}
E_{\bf K}({\bf p})=8S\big(1-\gamma^{\phantom -}_{\frac{\bf K}2}\gamma^{\phantom -}_{\bf p}-\gamma^-_{\frac{\bf K}2}\gamma^-_{\bf p}\big).
\label{eq:2magJ1}
\end{align}
Simple algebra in~(\ref{eq:2magJ1}) shows that the bottom of the two-magnon energy continuum for any total momentum of the pair ${\bf K}$ is given by its value at ${\bf p}^*\!=\!0$, $E_{\bf K}({\bf p}^*)\!=\!8S\big(1-\gamma_{\frac{\bf K}2}\big)$. For the matrix elements of the eigenvalue problem in Eq.~\eqref{eq:M_gamma}, it is convenient to introduce a binding energy $\Delta\!>\!0$ relative to the bottom of the two-magnon energy continuum, so that the energy denominator in the SE~(\ref{eq:2magSE_algebraic1}) becomes
\begin{equation}
E-E_{\bf K}({\bf p})=-2\Delta-8S\big( \gamma^{\phantom -}_{\frac{\bf K}2}-\gamma^{\phantom -}_{\frac{\bf K}2}\gamma_{\bf p}^{\phantom{-}}-\gamma^{-}_{\frac{\bf K}2}\gamma_{\bf p}^{{-}} \big).
\label{eq:E-2magJ1}
\end{equation}

\subsection{Solving the eigenvalue problem}
\label{Sec:BS_FM_evalueFig2}

With the magnon-magnon interaction in Eq.~\eqref{eq:VKq_J1expansion} adhering to the lattice partial-wave expansion, the SE can be rewritten in the algebraic form of Eq.~(\ref{eq:2magSE_algebraic1}), from which it is apparent that the pair wavefunction $\psi_{\bf K}({\bf q})$ in the square-lattice Heisenberg ferromagnet is necessarily a combination of the  $s$- and $d$-wave lattice partial waves. 

For a generic pair-momentum ${\bf K}$ that is away from high-symmetry points or directions, these partial-wave channels are generally mixed, so that the BS eigenvalue problem in Eq.~\eqref{eq:detM=0} is that for the $2\times 2$ matrix,
\begin{align}
\big|{\bf\hat{\rm \bf M}}_{\bf K}-{\bf\hat{\rm\bf I}}\big| =
\begin{vmatrix}
M_{{ss},{\bf K}}-1 & M_{{sd},{\bf K}} \\
M_{{ds},{\bf K}} & M_{{dd},{\bf K}} -1
\end{vmatrix}=0,
\label{eq:M2x2J1}
\end{align}
where the elements are  given by Eq.~\eqref{eq:M_gamma}, with the harmonics $R_{\gamma}({\bf q})$ and $\widetilde{R}_{\gamma,{\bf K}}({\bf p})$ and the coupling constants $\alpha_\gamma$ from Eq.~(\ref{eq:V_J1only}), and the energy denominator from Eq.~(\ref{eq:E-2magJ1}). We also note that since the separable form in the expansion of the interaction in Eq.~(\ref{eq:VKq_J1expansion}) is not symmetric, all matrix elements $M_{{\gamma\gamma'},{\bf K}}$ are different regardless of the expansion basis, so that $M_{{sd},{\bf K}}\!\neq\!M_{{ds},{\bf K}}$.

Therefore, finding the BSs energy in this model for a generic ${\bf K}$ amounts to numerically evaluating four integrals associated with the matrix elements $M_{{\gamma\gamma'},{\bf K}}$ and solving Eq.~\eqref{eq:M2x2J1}  within an iterative procedure in $E$.

With the analytical results for the BSs discussed below, in Fig.~\ref{fig:EkJ1} we present their energies obtained numerically as a function of the total momentum ${\bf K}$ along the high-symmetry $\Gamma$MX$\Gamma$ path for the $S\!=\!1/2$ FM square-lattice Heisenberg model; $\Gamma\!=\!(0,0)$, M$=\!(\pi,\pi)$, and X$=\!(\pi,0)$.  Their behavior in different regimes, which are explored in the following Sections, is highlighted throughout the BZ. The shaded area is the two-magnon continuum---the allowed energy range of the two-particle energies $E_{\bf K}({\bf q})\!=\!\varepsilon_{\frac{\bf K}{2}+{\bf q}}+\varepsilon_{\frac{\bf K}{2}-{\bf q}}$---and the black line indicates its boundaries. The solid blue and red lines denote the two BS branches identified by solving Eq.~\eqref{eq:M2x2J1}. Near the $\Gamma$ point and near the X point (for the upper branch), their binding energies are numerically indistinguishable from the minimum of the two-magnon continuum. 

To obtain the BS energies, our standard procedure was the following. At each ${\bf K}$ and for the energy $E$ below the two-magnon continuum, the 2D integrals in the $M_{{\gamma\gamma'},{\bf K}}$ matrix elements in Eq.~\eqref{eq:M_gamma} were evaluated using Gaussian quadratures on a $1000\times1000$ grid, and it was checked whether Eq.~\eqref{eq:M2x2J1} had any solutions to the precision of $10^{-5}$. To ensure that no BS solutions were missed, we performed a direct scan of the determinant in Eq.~\eqref{eq:M2x2J1} over a broad range of energies, from the bottom of the two-magnon continuum down. With the accuracy of our standard procedure, the smallest detectable binding energy $\Delta$ was $\sim\!10^{-5}$ in units of $|J_1|$.  In the case of shallow BSs, higher-resolution calculations and adaptive integration methods were employed, which enabled a decrease of this threshold to $\sim\!10^{-7}$. Still, for large ${\bf K}$-regions in Fig.~\ref{fig:EkJ1}, the BS energies were beyond numerical detectability; see  the discussion in Sec.~\ref{Sec:s_wave} below.

Needless to say, the BS spectrum shown in Fig.~\ref{fig:EkJ1} is in complete agreement with the earlier calculations in Refs.~\cite{Mattis2006book, Wortis63, Loly1986, Rastelli2013}, which used a different basis for the SE, provided by the ``elemental'' lattice harmonics, $\cos q_x$ and $\cos q_y$. Obviously, our use of the lattice partial waves does not alter the numerical results, but provides a better symmetry-based framework for their interpretation. 

Specifically, the resultant BS spectrum consists of two branches, with the lower branch spanning the entire BZ, becoming exceedingly shallow near the $\Gamma$ point, while being deeply bound between M and X points. The upper branch exists only within a limited region of momentum space, and the two branches are degenerate at the M point, where both have their largest binding energies. These results constitute the main puzzling features of the square-lattice ferromagnet BS problem, whose origin we analyze and disentangle next.

\begin{figure}[t]
\includegraphics[width=\linewidth]{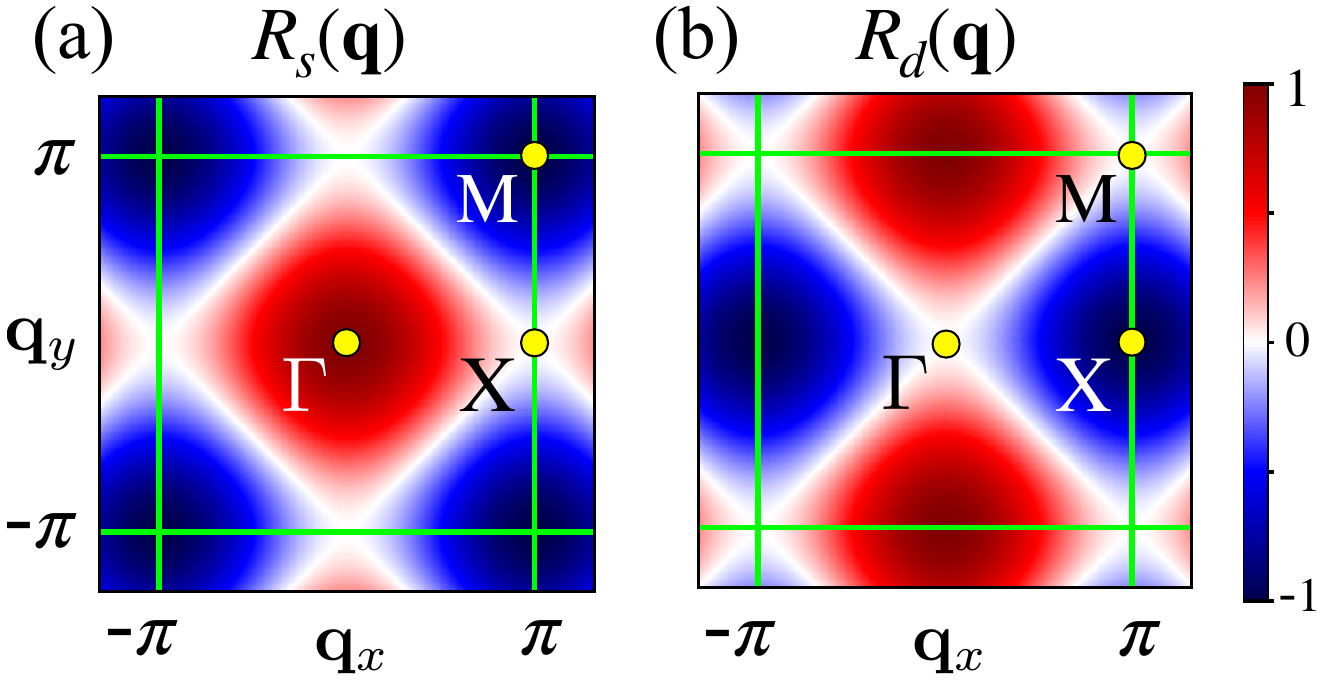}
\vskip -0.2cm
\caption{Intensity plots of the (a) generalized $s$-wave and (b) $d$-wave lattice partial waves $R_{s}({\bf q})$ and $R_{d}({\bf q})$ in momentum space; see Eqs.~\eqref{eq:V_J1only}. Nodal lines and the high-symmetry points in the BZ are indicated.}
\label{fig:HarmonicsJ1}
\vskip -0.4cm
\end{figure}

\subsection{$\Gamma$M direction: $s$- and $d$-wave bound states}
\label{Sec:s_and_d_wave}

The high-symmetry $\Gamma$M-direction of ${\bf K}$ offers significant simplifications and insights. 

In Fig.~\ref{fig:HarmonicsJ1}, we show the intensity plots of the generalized lattice partial waves in Eqs.~\eqref{eq:V_J1only}, $R_{s}({\bf q})$ and $R_{d}({\bf q})$, in momentum space. For the current consideration, it is important to note that the two partial waves have opposite parity under the mirror reflection about the $q_x\!=\!q_y$ line, the same direction as $\Gamma$M, with the $s$-wave harmonic being even, and the $d$-wave harmonic being odd. 

\begin{figure}[t!]
\includegraphics[width=\linewidth]{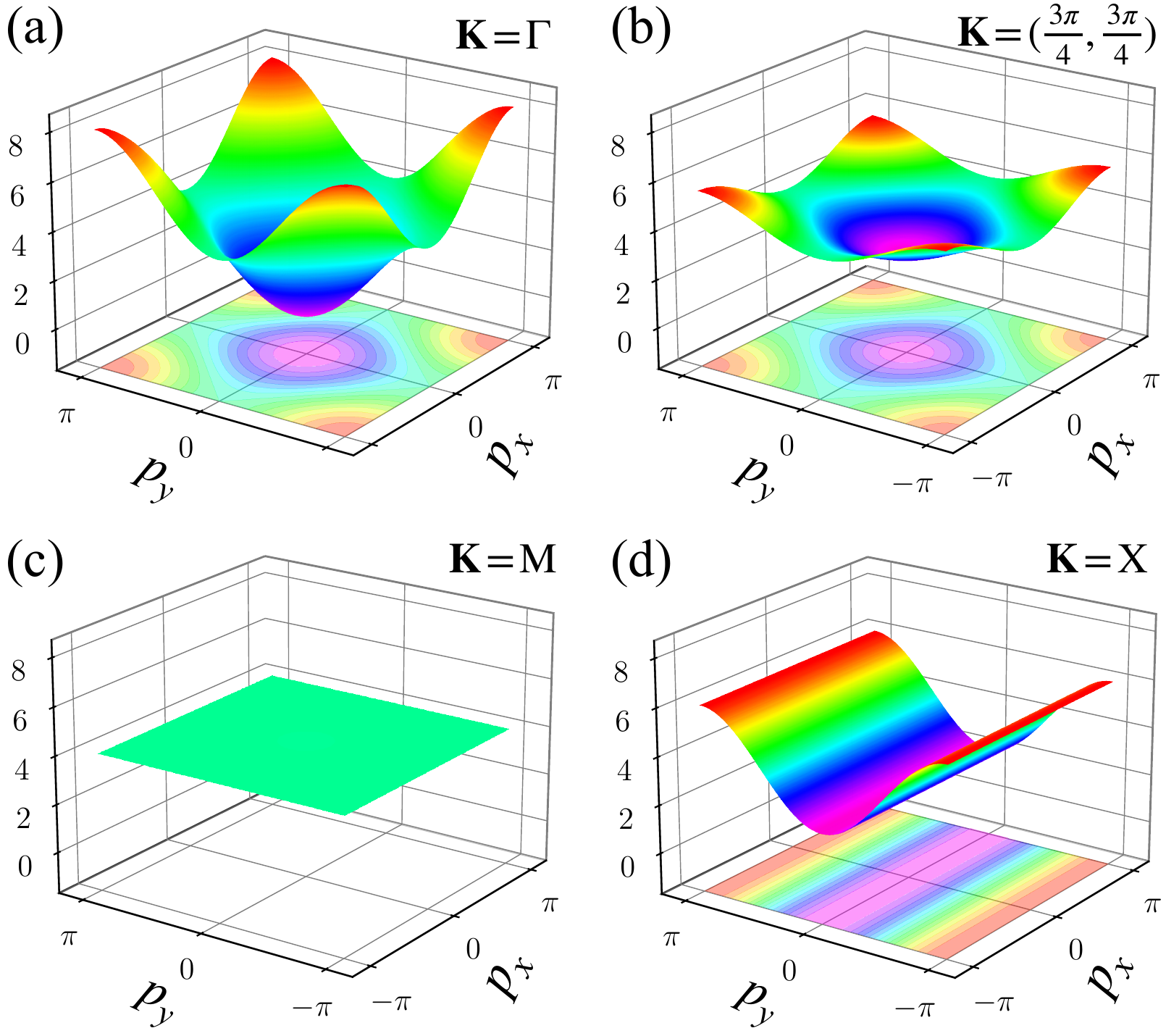}
\vskip -0.2cm
\caption{The two-magnon continuum energy, $E_{\bf K}({\bf p})\!=\!\varepsilon_{\frac{\bf K}{2}+{\bf p}}+\varepsilon_{\frac{\bf K}{2}-{\bf p}}$, Eq.~\eqref{eq:2magJ1} for $S\!=\!1/2$, vs internal momentum ${\bf p}$ for (a) ${\bf K}\!=\!\Gamma$, (b) ${\bf K}\!=\!(3\pi/4,3\pi/4)$, (c) ${\bf K}\!=$M, and (d) ${\bf K}\!=$X. Projection maps in (a), (b), and (d) are included for clarity.}
\label{fig:EkqJ1}
\vskip -0.4cm
\end{figure}

Because $\gamma^-_{{\bf K}/2}$ vanishes along this high-symmetry $\Gamma$M line, the two-magnon energy in Eq.~\eqref{eq:E-2magJ1} is symmetric under the same operation; see also Figs.~\ref{fig:EkqJ1}(a) and \ref{fig:EkqJ1}(b), in which we plot the two-magnon continuum energy, $E_{\bf K}({\bf p})\!=\!\varepsilon_{\frac{\bf K}{2}+{\bf p}}\!+\!\varepsilon_{\frac{\bf K}{2}-{\bf p}}$, vs internal momentum ${\bf p}$ for select {\bf K} values. Along $\Gamma$M, the separable parts of the interaction for the $d$-wave channel, $R_{d}({\bf q})$ and $\widetilde{R}_{d,{\bf K}}({\bf p})$ in Eqs.~\eqref{eq:V_J1only}, become equal, ${\bf K}$-independent, and orthogonal to the $s$-wave harmonics, $R_{d}({\bf q})\!=\!\widetilde{R}_{d,{\bf K}}({\bf p})\!=\!\gamma^-_{\bf p}$.

As a result, the mixed-channel matrix elements in Eq.~(\ref{eq:M2x2J1}), $M_{sd,{\bf K}}$ and $M_{ds,{\bf K}}$, vanish identically along the entire $\Gamma$M line, so that the matrix ${\bf \hat{\rm \bf M}}_{\bf K}$ is diagonal,
\begin{align}
\hat{\rm \bf M}_{\bf K} =
\begin{pmatrix}
M_{{ss},{\bf K}} & 0 \\
0 & M_{{dd},{\bf K}}
\end{pmatrix},
\label{eq:MGammatoMJ1}
\end{align}
and the eigenvalue problem for the two BS branches (\ref{eq:M2x2J1}) decouples into the ones for the ``pure'' $s$- and $d$-waves.

The resultant diagonal form of ${\bf \hat{\rm \bf M}}_{\bf K}$ also makes explicit the practical advantage of the lattice partial waves basis over the ``elemental'' cosine lattice harmonics, used in the earlier works. The two formulations are equivalent and yield identical spectra, but in the cosine basis the ${\bf \hat{\rm \bf M}}_{\bf K}$ matrix is not diagonal, with the higher symmetry showing only indirectly, through the identity relations among the matrix elements, rather than via an explicit orthogonality of the partial-wave symmetry channels. 

As is shown in Fig.~\ref{fig:EkJ1}, the two branches display qualitatively different behavior along the $\Gamma$M line. The $s$-wave BS solution exists for the entire line, except at the $\Gamma$ point, where its binding energy vanishes. By contrast, the $d$-wave BS solution appears only above a finite threshold momentum. Finally, the two branches become degenerate at the M point, where their binding energy is also maximal. 
In the remainder of this Section, we analyze the origin of these behaviors.

\subsubsection{$s$-wave}
\label{Sec:s_wave}

For the $s$-wave BS, the following  consideration is illuminating. For small total and relative momenta of the pair, the two-magnon energy can be split into that of the center of mass and the relative kinetic energy, making the energy denominator for the SE in Eq.~(\ref{eq:E-2magJ1}) the same as in the continuum pairing problem in Sec.~\ref{simplest},
\begin{align}
E- E_{\bf K}({\bf p})\approx -2\Delta-\frac{{\bf p}^2}{m}, \label{eq:2mag_approx_GammatoM} 
\end{align}
where $m\!=\!(2S)^{-1}$ is the magnon mass that can be obtained from $\varepsilon_{\bf k}$ in Eq.~(\ref{eq:J1_1mag}).

If,  for a moment, one assumes that the interaction in Eq.~\eqref{eq:V_J1only} has no hard-core, non-potential-like contributions of Eq.~(\ref{eq:2MagVKqp}), so that $\widetilde{R}_{s,{\bf K}}({\bf p})\!=\!R_{s}({\bf p})\!=\!\gamma_{\bf p}\!\approx\!1$, this would render the pairing problem in the $s$-wave channel identical to the one in Sec.~\ref{simplest}, with the resultant binding energy given by Eq.~\eqref{eq:SE_2D},
\begin{align}
\tilde{\Delta}_s\approx \frac{\Lambda}{2m}\, \exp\Big(-\frac{8\pi}{\alpha_s m}\Big)\propto S\, \exp\Big(-2\pi S\Big),
\label{eq:Delta_s_naiive}
\end{align}
which is finite at ${\bf K}\!=\!0$ as one na\"ively expects. 

The actual form of the magnon-magnon interaction, with all terms properly included as in Eq.~\eqref{eq:V_J1only}, yields $\widetilde{R}_{s,{\bf K}}({\bf p})\!=\!\gamma_{\bf p}-\gamma_{\frac{\bf K}2}\!\approx\!{\bf K}^2/16$, which modifies the na\"ive result in Eq.~(\ref{eq:Delta_s_naiive}) in a fairly straightforward way
\begin{align}
\Delta_s=\tilde{\Delta}_s\left(\alpha_s\rightarrow\alpha_s\,\frac{{\bf K}^2}{16}\right)\propto S\, \exp\left(-\frac{32\pi S}{{\bf K}^2}\right).
\label{eq:Delta_s_mag}
\end{align}
While Eq.~(\ref{eq:Delta_s_mag}) reproduces the result for the 2D Heisenberg FM from Refs.~\cite{Mattis2006book, Wortis63, Rastelli2013} and others (see also App.~\ref{A:swave} for the technical details), the discussion provided here yields important new insights.

As in the textbook $s$-wave pairing problem, exposed in Sec.~\ref{simplest}, the 2D magnon BS is guaranteed to occur and is exponentially shallow. However, unlike in the standard case, the effective magnon-magnon attractive potential vanishes with the pair momentum ${\bf K}$, not only amputating the BS from the $\Gamma$ point, but also making it super-shallow for a large portion of the momentum space. These distinct traits of the magnon BS can be traced directly to the non-potential terms in Eq.~\eqref{eq:2MagVKqp} associated with the hard-core repulsion between spin flips.

As was noted in the earlier works~\cite{Wortis63}, in order for a BS to be physically meaningful, its radius $\ell_s\!\propto\!1/\sqrt{m\Delta_s}$ should not exceed the linear dimensions of a realistic crystal. Using  Eq.~(\ref{eq:Delta_s_mag}), an easy estimate for $\ell_s\!\sim\!10^8$ lattice spacings suggests that the $s$-wave BS is not physically meaningful in about half of the BZ ($|{\bf K}|\!\alt\!0.4\pi$).

This makes it even more puzzling how such a super-shallow state can transition to the anomalously deep BS for large ${\bf K}$ near the M-point and along the MX direction. To pave the way for the answer, we note that for the $\Gamma$M line, the $s$-wave binding energy can be obtained without the small-${\bf K}$ approximation as
\begin{align}
\Delta_{s}\propto \exp\left(-\frac{8\pi}{m_{\bf K}^*V^{\rm eff}_{\bf K}}\right),\label{eq:Delta_s_Veff}
\end{align}
where $V^{\rm eff}_{\bf K}\!=\!\alpha_s (1-\gamma_{\frac{\bf K}2})$ is the effective attractive potential and $m_{\bf K}^*\!=\!(2S\gamma_{\frac{\bf K}2})^{-1}$ is the effective mass. 

One may see that this result must break down at the M point, ${\bf K}\!=\!(\pi,\pi)$, where the effective mass diverges and the two-magnon continuum becomes flat; see Fig.~\ref{fig:EkqJ1}(c). We defer the discussion of this effect to Sec.~\ref{Sec:0DBS} and continue with the $d$-wave BS for the $\Gamma$M line.

\subsubsection{$d$-wave}\label{Sec:GammaM_dwaveJ1}

Using the same approximations as for the $s$-wave, the $d$-wave BS also conforms to the pairing problem in the continuum, but for the higher partial wave; see Sec.~\ref{Sec:partial_waves}. The magnon-magnon potential in the $d$-wave channel in Eq.~(\ref{eq:V_J1only}) has nodes that pass through the minimum of the two-magnon energy; see Fig.~\ref{fig:HarmonicsJ1}(b). As a result, the $d$-wave eigenvalue problem, which takes the form of Eq.~(\ref{eq:SE_Egamma}), is subject to the dimensional enhancement, with the integrand in~(\ref{eq:SE_Egamma}) acquiring higher powers of momentum that render it non-divergent as in the  3D case. This puts a threshold ${\cal O}(1/m^*_{\bf K})$ on the coupling strength for the existence of the $d$-wave BS.

However, since the effective mass of the pair's relative motion $m^*_{\bf K}$ increases away from the $\Gamma$ point and diverges at the M point, such a threshold decreases accordingly and must vanish at the M point. This suggests that the $d$-wave BS must occur in the ${\bf K}$ region in proximity to the M point, with the boundary at some ${\bf K}_c$ along the $\Gamma$M line. Since ${\bf K}_c$ corresponds to the vanishing binding energy, it can be straightforwardly found from $M_{dd,{\bf K}_c}(\Delta_d\!=\!0)\!=\!1$, yielding 
\begin{align}
|{\bf K}_c|=2\sqrt{2}\arccos\left( \frac{4-\pi}{2S\pi} \right),
\label{eq:kc}
\end{align}
in agreement with Refs.~\cite{Wortis63, Rastelli2013}; see App.~\ref{A:dwaveBSJ1} for details. 

\subsection{M point: 0D degenerate bound states}
\label{Sec:0DBS}

Due to its simple tight-binding form, the magnon dispersion in the nearest-neighbor Heisenberg FM models is highly symmetric, leading to some special degeneracies in their two-magnon continua~\cite{Kagome_FM}. 

For the square lattice, one such degeneracy occurs at the total pair-momentum ${\bf K}$ at the M point, where the energies of the two magnons in Eq.~(\ref{eq:E-2magJ1}) add up to a constant, $E_{\bf K}({\bf p})\!=\!\varepsilon_{\frac{\bf K}{2}+{\bf p}}+\varepsilon_{\frac{\bf K}{2}-{\bf p}}\!=\!8S$, for any relative momentum ${\bf p}$ in the entire BZ. This corresponds to  making the two-magnon continuum flat, as is shown in Fig.~\ref{fig:EkqJ1}(c), and the effective mass of the pair's relative motion $m^*_{\bf K}$, introduced above, infinite. This is easy to understand by noting that for ${\bf k}\!=\!{\bf K}/2$, the single-magnon energy $\varepsilon_{\bf k}$ in (\ref{eq:J1_1mag}) is exactly midway through its symmetric band. 

This continuum flatness for the total momentum of the pair ${\bf K}\!=\!{\rm M}$ means that in the mapping of the two-particle pairing problem onto a problem of a single particle in an attractive potential, such a particle has no kinetic energy, making the problem 0D. The resultant pairs are fully localized at the nearest-neighbor sites---a separation dictated by their attractive potential. 

One can see this from the SE in Eq.~(\ref{eq:2magSE_algebraic1}), where the pair's wavefunction structure in momentum space for the case of the constant energy denominator will restrict itself to that of the nearest-neighbor partial-wave harmonics, $\psi_{\gamma,{\bf K}}({\bf q})\!\propto\!R_{\gamma}({\bf q})$, or $\gamma_{\bf q}$ and $\gamma^-_{\bf q}$ for the $s$- and $d$-waves, respectively~(\ref{eq:V_J1only}).

Although the degeneracy of the $s$- and $d$-wave BSs at ${\bf K}\!=\!{\rm M}$ may already be obvious from the real-space perspective of the immobile nearest-neighbor pairs, the explicit evaluation of their matrix elements for the eigenvalue problem in Eq.~(\ref{eq:M_gamma})  yields
\begin{align}
{M}_{\gamma \gamma,{\bf K}} = \frac{\alpha_{\gamma}}{4N\Delta_\gamma}\sum_{\bf p}  R_{\gamma}(\mathbf{p})^2,
\label{eq:M_gamma_atM}
\end{align}
where we have used that all non-potential-like terms in the magnon interaction (\ref{eq:V_J1only}) vanish at the M point, yielding $\widetilde{R}_{\gamma,{\bf K}}({\bf p})\!=\!R_{\gamma}({\bf p})$. Since the $s$- and $d$-harmonics are related by a mirror reflection about the $(\pi/2,p_y)$ line [see Fig.~\ref{fig:HarmonicsJ1} and Eqs.~(\ref{eq:gamma+}) and~(\ref{eq:gamma-})], one finds that the integrals in Eq.~(\ref{eq:M_gamma_atM}) are identical, making $M_{ss,{\bf K}}\!=\!M_{dd,{\bf K}}$ and yielding a doubly-degenerate BS with 
\begin{align}
\Delta_{s(d)} = \frac12,
\label{eq:Delta_atM}
\end{align}
in units of $|J_1|$, in agreement with Refs.~\cite{Wortis63, Rastelli2013}; see also Fig.~\ref{fig:EkJ1}. The independence of their energies from the spin value also follows directly from the fully local character of the BSs, with the nearest-neighbor attraction being the binding energy, $|J_1|\!=\!2\Delta$. 

The number of such degenerate BSs and that of the affiliated branches indeed appear to be equal to the dimension of the lattice, in agreement with the conventional wisdom~\cite{Mattis2006book, Akhiezer1968SW, Wortis63}. In actuality, this number corresponds to the number of independent bond directions of the lattice, which is equal to D for the much-studied FM linear chain, square, and simple cubic lattices. However, the true reason for such a dimensional count to work is the ultimate dimensional reduction of the pairing problem to 0D, making the pairs fully localized, a reason neither previously elucidated nor expected to be realized in a more generic model.

This discussion of the 0D BSs explains why the largest binding energy occurs at the M point,  why the two purely local BSs are degenerate, and how the super-shallow $s$-wave solution develops into the deepest one. It also connects naturally to the problem of the BSs along the MX high-symmetry line, which is considered next.

\subsection{MX direction: 1D BSs}
\label{Sec:MX}

One can see in Fig.~\ref{fig:EkJ1} that away from the highly degenerate M point and along the MX line, the lower branch of the BSs remains anomalously deep and the upper branch does not terminate the same way as it does in the M$\Gamma$ direction, persisting all the way to the X point. 

Following an analysis similar to the one above, this behavior can again be attributed to an effective dimensional reduction of the pairing problem. In this case, the high symmetry of the two-magnon continuum in the nearest-neighbor FM model for the MX line leads to a 1D form of the two-magnon energy, with the effective mass being infinite for one direction of the relative momentum for all ${\bf K}\!=\!(\pi, K_y)$ values; see Fig.~\ref{fig:EkqJ1}(d). 

Explicitly, the energy denominator for the two-particle SE in Eq.~(\ref{eq:E-2magJ1}) for the MX line becomes
\begin{equation}
E-E_{\bf K}({\bf p})=-2\Delta-4S\cos\frac{K_y}{2} \, \big(1-\cos p_y\big),
\label{eq:2mag_MToX}
\end{equation}
which is independent of $p_x$. For small $p_y$, it assumes the continuum-like form as in Sec.~\ref{sec:continuum}
\begin{equation}
E-E_{\bf K}({\bf p})\approx -2\Delta-\frac{p_y^2}{m^*_y},
\label{eq:2mag_MToXapp}
\end{equation}
with the 1D effective mass $m^*_y\!=\!\big(2S\cos (K_y/2)\big)^{-1}$.

With Eq.~(\ref{eq:2mag_MToXapp}) and using the partial-wave harmonics in Eq.~(\ref{eq:V_J1only}), one can find that all elements of the eigenvalue matrix are divergent as $M_{{\gamma\gamma'},{\bf K}}\!\propto\!1/\sqrt{\Delta}$, conforming to the 1D $s$-wave universality, which should guarantee that the BSs' binding energies are $\Delta_\gamma\!\sim\!\alpha_\gamma^2m^*_y$ as in the 1D continuum case of Sec.~\ref{simplest}, Eq.~(\ref{eq:SE_1D}).

The situation is a little more involved because the $s$- and $d$-wave pairing channels are coupled away from the M point, with all matrix elements in the eigenvalue problem in Eq.~(\ref{eq:M2x2J1}) nonzero. However, the mirror symmetry relation between partial-wave harmonics, $R_d({\bf p})\!\Leftrightarrow\!-R_s({\bf p})$ for the reflection about the $(\pi/2,p_y)$ line (see Fig.~\ref{fig:HarmonicsJ1}), leaves the two-particle energy in Eq.~(\ref{eq:2mag_MToX}) invariant and makes the diagonal and off-diagonal matrix elements pairwise equal, $M_{ss,{\bf K}}\!=\!M_{dd,{\bf K}}$ and $M_{sd,{\bf K}}\!=\!M_{ds,{\bf K}}$, for all ${\bf K}$ along the MX line. This reduces the eigenvalue problem in Eq.~(\ref{eq:M2x2J1}) to the pair of equations for the two orthogonal BS branches, referred to as the symmetric and antisymmetric ones, 
\begin{equation}
\label{eq:M_eq_forMX}
M_{ss,{\bf K}}\pm M_{sd,{\bf K}}=1.
\end{equation}
An approximate calculation for both branches can be pursued using a small $p_y$ expansion in Eq.~(\ref{eq:2mag_MToXapp}), and, after straightforward algebra using Eq.~(\ref{eq:M_gamma}) and partial-wave harmonics from Eq.~(\ref{eq:V_J1only}), one obtains, 
\begin{equation}
\label{eq:Delta_+-}
\Delta_\pm\approx\frac{\alpha_\pm^2 m^*_y}{32}, 
\end{equation}
in a complete formal identity with the 1D continuum result for the $s$-wave solution in Eq.~(\ref{eq:SE_1D}), with
\begin{equation}
\label{eq:alpha_+-} 
\alpha_+=2, \ \ \alpha_-=4\bigg(1-\cos \frac{K_y}{2} \bigg).
\end{equation}
Near the ${\rm X}\!=\!(\pi,0)$ point, Eq.~(\ref{eq:Delta_+-}) yields 
\begin{align}
\Delta_{+} \approx \frac{1}{16 S},\quad
\Delta_{-} \approx \frac{K_y ^4}{256 S}.
\label{eq:DeltaX_antisymmetric}
\end{align}
Intriguingly, {\it both} solutions are $s$-wave-like, with no sign of the dimensional enhancement that is characteristic of the higher partial waves. This is because neither of the BSs has nodes that go through the entire 1D line of minima of the two-magnon continuum. 

Another interesting observation is that the symmetric solution is unaffected by the hard-core, non-potential-like terms in the magnon interaction, while the antisymmetric one is, showing a projection out of the small-$K_y$ region similar to that of the $s$-wave state along the $\Gamma$M line in Sec.~\ref{Sec:s_wave}; see Eq.~(\ref{eq:Delta_s_mag}). As in that case, this suppression of the pairing is due to an effective magnon-magnon attraction $\alpha_-$ in Eq. (\ref{eq:alpha_+-}), which vanishes for ${\bf K}\!\rightarrow$X,
\begin{align}
\Delta_-=\Delta_+\bigg(\alpha_+\rightarrow\alpha_+\,\frac{K_y^2}{4}\bigg)\propto K_y^4,
\label{eq:Delta_-}
\end{align}
leading to the 1D version of the super-shallow $s$-wave energy in the vicinity of the X point and removing this BS from the X point entirely; see Fig.~\ref{fig:EkJ1}.

We also note that the integration in the eigenvalue equations (\ref{eq:M_eq_forMX}) can be performed without any approximations, yielding fully analytic, if somewhat cumbersome, results for the binding energies; see App.~\ref{A:1DBSJ1}. For the symmetric case, the solution for any spin value is
\begin{align}
\label{eq:Delta_+_exact}
\Delta_+=\frac12 \sqrt{16S^2\cos^2\frac{K_y}{2}+1}- 2S\cos \frac{K_y}{2} ,
\end{align}
which, at the X point, is 
\begin{align}
\label{eq:Delta_+_exactX}
\Delta_+=\frac12 \sqrt{16S^2+1}- 2S\approx \frac{1}{16S}+{\cal O}(S^{-3}),
\end{align}
in agreement with the approximate result of Eq.~(\ref{eq:Delta_+-}). For the antisymmetric case, the solution assumes a compact form only for $S\!=\!1/2$; see App.~\ref{A:1DBSJ1},
\begin{align}
\label{eq:Delta_-_exact}
\Delta_-=\frac12\bigg(1-\cos \frac{K_y}{2} \bigg)^2,
\end{align}
also in agreement with Eq.~(\ref{eq:Delta_+-}). Both results also successfully interpolate to the degenerate, $S$-independent BS energies at the M point for $K_y\!=\!\pi$; see Eq.~(\ref{eq:Delta_atM}).

The analytical results in Eqs.~(\ref{eq:Delta_+_exact}) and (\ref{eq:Delta_-_exact}) have been previously obtained in   Refs.~\cite{Wortis63, Rastelli2013}, but their physical discussion has been lacking until now.

Altogether, as can be seen in Fig.~\ref{fig:EkJ1}, both BSs are present for the entire MX line, with the lower one being anomalously deep because of the 1D-like form of the two-magnon continuum, and the upper one adhering to the same 1D $s$-wave universality, but with the effective attractive potential that vanishes at the X point, leading to another instance of a super-shallow state.

\subsection{X$\Gamma$ direction}
\label{Sec:XG_J1}

We conclude the discussion of the BSs in the square-lattice Heisenberg FM with the X$\Gamma$ high-symmetry line. 

Along this line, all four matrix elements of $\hat{\bf M}_{\bf K}$ are distinct and the eigenvalue problem (\ref{eq:M2x2J1}) does not admit simple analytical solutions. However, for $\bf K$ near $\Gamma$, one can show that only the $s$-wave component is divergent in the small-$\Delta$ limit, yielding the same exponential asymptotes for the binding energy as those of the $s$-wave BS along the $\Gamma$M line in Eq.~\eqref{eq:Delta_s_mag}; see also Ref.~\cite{Rastelli2013}. 

Since for $\bf K$ away from the X point the two-magnon continuum exhibits no degeneracy that would result in a dimensional reduction, and since the second BS branch has already terminated at the X point, only one bound state solution exists along the X$\Gamma$ line. 

\subsection{Summary of the BSs in the $J_1$-only model}

In this section, we have provided a thorough dissection of the pairing problem in the nearest-neighbor 2D square-lattice Heisenberg FM model, with a focus on exposing the origin of the puzzling, if not contradictory, behaviors of the magnon BSs, especially when contrasted with the expectations for this problem in the 2D continuum.  

We have shown that the complicated behavior of the two-magnon BSs stems from a convolution of several distinct mechanisms rather than a single universal pairing scenario. Near the $\Gamma$ point, the $s$-wave BS is made super-exponentially shallow by the non-potential terms in the magnon-magnon interaction, which suppress the effective attraction as the pair momentum vanishes. By contrast, the $d$-wave channel follows a conventional expectation for higher partial waves: dimensional enhancement imposes a threshold for binding, so the $d$-wave appears only close to the M point, where a different mechanism is at play.

The anomalously deep BSs at large pair momenta near the BZ boundary avoid the non-potential terms in the magnon-magnon interaction by projecting them out with their orbital structure, and originate instead from accidental degeneracies of the $J_1$-only two-magnon continuum. At the M point, the continuum becomes completely flat, reducing the pairing problem to a 0D one and producing degenerate, fully local BSs. Along the MX line, the two-magnon continuum is effectively 1D, which provides deeper BSs according to the 1D universality. The vanishing of the energy for one of the 1D BSs at the X point is a 1D analogue of the super-shallow BSs in the $\Gamma$-point proximity---with the non-potential terms leading to a vanishing effective attraction potential.  

In what follows, we explore how these BSs evolve upon including the AFM $J_2$ term in the model. The anticipated changes stem from two effects: $J_2$ makes the magnon band, and hence the two-magnon continuum, more generic and less prone to degeneracies, while its AFM character also introduces repulsive interaction channels. These trends are expected to lead to a more conventional 2D universality for the BSs. However, one important qualitative change in the $J_1$--$J_2$ model follows from the transformation of the magnon dispersion, with the minima at the AFM ordering vectors---a feature that reshuffles the favorable kinematics to the $d$-wave pairing channel. These effects are examined next.

\section{Bound states in the square-lattice $J_1$--$J_2$ FM-AFM model}
\label{Sec:BS_FM_AFM}

In this section, we extend the previous analysis to the square-lattice Heisenberg ferromagnet with an antiferromagnetic next-nearest-neighbor exchange $J_2$. The mixed $J_1$--$J_2$ ferro-antiferromagnetic exchange produces a more general magnon band structure that lifts the degeneracies of the two-magnon continuum present in the $J_1$-only model. For sufficiently large $J_2$, the BS spectrum in the polarized phase of the model transforms from the two-branch structure of the $J_1$-only model to a single $d$-wave branch, which also becomes the ground state, prompting significant interest in the potential nematic phases associated with this state~\cite{NematicShengtao2023, Shannon2006}. Here, we examine this transformation and the evolution of the deep BSs of the $J_1$-only model at the high-symmetry momenta in the more general setting of the $J_1$--$J_2$ model.

\subsection{Lattice partial waves}
\label{Sec:BS_FMAFM_partialwaves}

To investigate the evolution of the magnon BSs in the square-lattice $J_1$--$J_2$ FM-AFM model, one needs to remain in the polarized FM-like phase, which is favorable for pairing. For that, the field $H\!\geq\!4S(2J_2-1)$ must be retained in Eq.~\eqref{eq:HH} for $J_2\!\geq\!0.5$ to stabilize this phase as a ground state; see Ref.~\cite{NematicShengtao2023} for the phase diagram.

\begin{figure}[t]
\includegraphics[width=\linewidth]{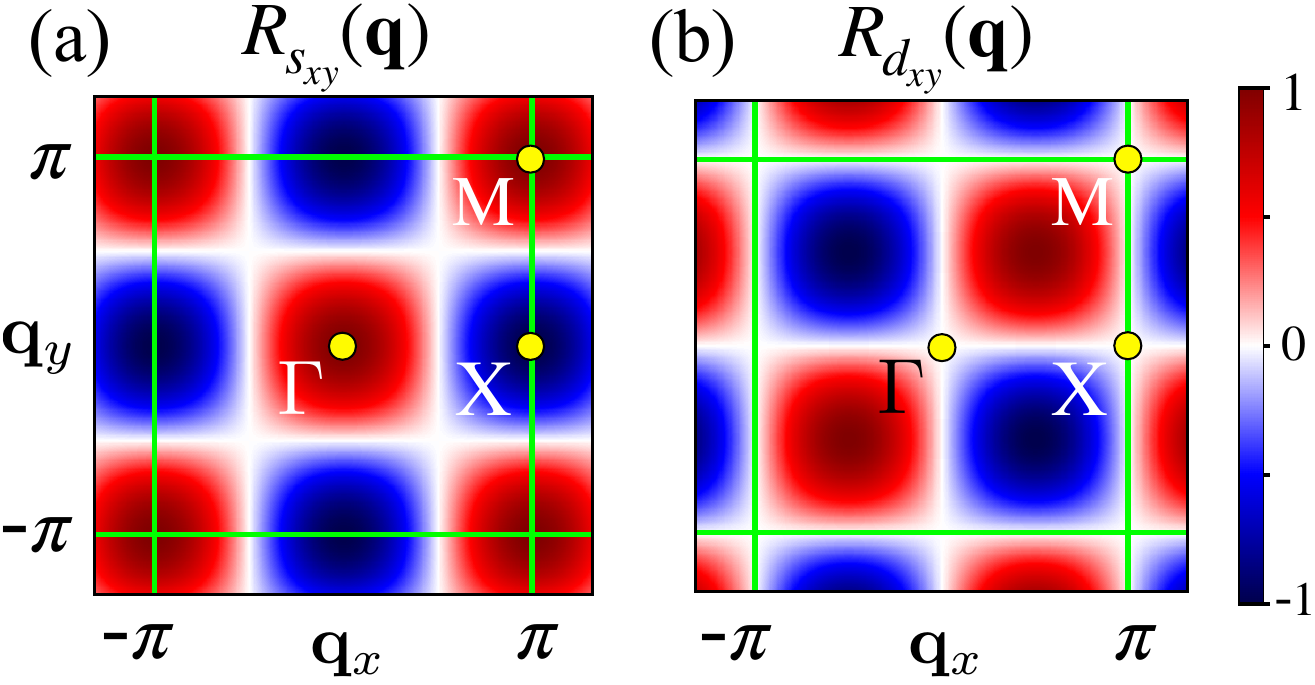}
\vskip -0.2cm
\caption{Intensity plots of the (a) generalized $s_{xy}$ and (b) $d_{xy}$ lattice partial waves, $R_{s_{xy}}({\bf q})$ and $R_{d_{xy}}({\bf q})$, respectively, in momentum space; see Eq.~\eqref{eq:V_J2}. Nodal lines and the high-symmetry points in the BZ are indicated.}
\label{fig:HarmonicsJ2}
\vskip -0.4cm
\end{figure}

In addition to the nearest neighbors in the $J_1$ term in~\eqref{eq:HH}, the AFM exchange $J_2$ involves the next-nearest-neighbor vectors ${\bm \delta}^{(2)}\!=\!\pm(1,\pm 1)$ (in units of the lattice spacing), so the Fourier transform of the exchanges and the tight-binding magnon energy~\eqref{eq:ek_Jq} are given by
\begin{align}
{\cal J}_{\bf k}&=-4\big(\gamma_{\bf k}-J_2\gamma^{(2)}_{\bf k}\big),
\label{eq:J2_1mag} \\
\varepsilon_{\bf k}&=H+4S\big(1-\gamma_{\bf k}\big)-4SJ_2\big(1-\gamma^{(2)}_{\bf k}\big),
\label{eq:ekJ1J2_1mag}
\end{align}
respectively, where the second-neighbor lattice harmonic
\begin{align}
\gamma_{\bf q}^{(2)}=\cos q_x \cos q_y,\label{eq:gamma2}
\end{align}
is introduced. Straightforward algebra in~\eqref{eq:2MagVKqp} using ${\cal J}_{\bf k}$ from Eq.~\eqref{eq:J2_1mag} gives the lattice partial-wave expansion for the magnon-magnon interaction, 
\begin{align}
V_{\bf K}({\bf q},{\bf p})=-\sum_\gamma\alpha_\gamma R_{\gamma}({\bf q})\widetilde{R}_{\gamma,{\bf K}}({\bf p})\, ,
\label{eq:VKq_J1J2expansion}
\end{align}
with $\gamma\!=\!\{s,d,s_{xy},d_{xy}\}$. The $s$- and $d$-waves are the same as in the $J_1$-only model, Eq.~\eqref{eq:VKq_J1expansion}, and two additional ones, $s_{xy}$ and $d_{xy}$, are brought in by the $J_2$ exchange, with their coupling constants $\alpha_{s_{xy}}\!=\!\alpha_{d_{xy}}\!=\!-8J_2$ and their partial-wave harmonics given by
\begin{align}
\label{eq:V_J2}
&R_{s_{xy}}\!({\bf q})\!=\!\gamma_{\bf q}^{(2)}, \, \  \widetilde{R}_{s_{xy},{\bf K}}({\bf p})\!=\!\gamma_{\bf p}^{(2)}\!-\!\gamma_{\frac{\bf K}2}^{(2)},\\
&R_{d_{xy}}\!({\bf q})\!=\!\gamma^{d_{xy}}_{\bf q}, \, \ \widetilde{R}_{d_{xy},{\bf K}}({\bf p})\!=\!\gamma^{d_{xy}}_{\bf p}\!-\!\gamma^{d_{xy}}_{\frac{\bf K}2},\nonumber
\end{align}
where the $d_{xy}$-symmetric lattice harmonic is 
\begin{align}
\gamma^{d_{xy}}_{\bf q}&=\sin q_x\sin q_y.\label{eq:gammadxy}
\end{align}
Similarly to the generalized $s$- and $d$-waves, the partial-wave harmonics $R_{s_{xy}}({\bf q})$ and $R_{d_{xy}}({\bf q})$ in Eq.~\eqref{eq:V_J2} correspond to the generalized $s_{xy}$- and $d_{xy}$-waves, respectively.
The generalized $s_{xy}$-wave is finite at ${\bf q}\!=\!0$ while having nodal lines elsewhere in the BZ. The $d_{xy}$-wave has nodal lines along $q_{x(y)}\!=\!0$ that cross the $\Gamma$ point; see Fig.~\ref{fig:HarmonicsJ2}.

Since $J_2$ is positive, the coupling constants $\alpha_{s_{xy}}$ and $\alpha_{d_{xy}}$ are negative, indicating that both interaction channels are repulsive and disfavor pairing. The $s$- and $d$-wave channels from the FM $J_1$-term in~(\ref{eq:VKq_J1J2expansion}) are still attractive.

The two-magnon energy can also be written in terms of the same lattice harmonics as in Eqs.~(\ref{eq:V_J1only}) and (\ref{eq:V_J2})
\begin{align}
E_{\bf K}({\bf p})&=2H+
8S\big(1-\gamma^{\phantom -}_{\frac{\bf K}2}\gamma^{\phantom -}_{\bf p}-\gamma^-_{\frac{\bf K}2}\gamma^-_{\bf p}\big) \nonumber \\
&-8J_2S(1-\gamma_{\frac{{\bf K}}{2}}^{(2)}\gamma_{\bf p}^{(2)}
-\gamma_{\frac{{\bf K}}{2}}^{d_{xy}}\gamma_{\bf p}^{d_{xy}}).
\label{eq:2magJ1J2}
\end{align}
Unlike in the $J_1$-only case,  the minimum of the two-particle energy is no longer fixed at ${\bf p}^*\!=\!0$, but depends on both $J_2$ and the pair momentum ${\bf K}$.

\subsection{Solving the eigenvalue problem}

Following the general formalism for the magnon pairing in the FM-polarized states described in Sec.~\ref{Sec:magSE_general} and using the partial-wave decomposition of the magnon-magnon interaction in Eq.~\eqref{eq:VKq_J1J2expansion}, the magnon BS SE for the $J_1$--$J_2$ FM-AFM square-lattice model can be written in algebraic form with the pair wavefunction $\psi_{\bf K}({\bf q})$ in Eq.~\eqref{eq:2magSE_algebraic1} splitting into the $s$-, $d$-, $s_{xy}$-, and $d_{xy}$-wave components. For a generic pair-momentum ${\bf K}$, the partial-wave channels mix and the BS eigenvalue problem in Eq.~\eqref{eq:detM=0} is that for the $4\times 4$ matrix $\hat{\rm \bf M}_{\bf K}$, with its elements given by Eq.~\eqref{eq:M_gamma}, the partial waves and coupling constants from Eqs.~(\ref{eq:V_J1only}) and (\ref{eq:V_J2}), and the two-magnon continuum energy from Eq.~(\ref{eq:2magJ1J2}).

\begin{figure}[t!]
\includegraphics[width=\linewidth]{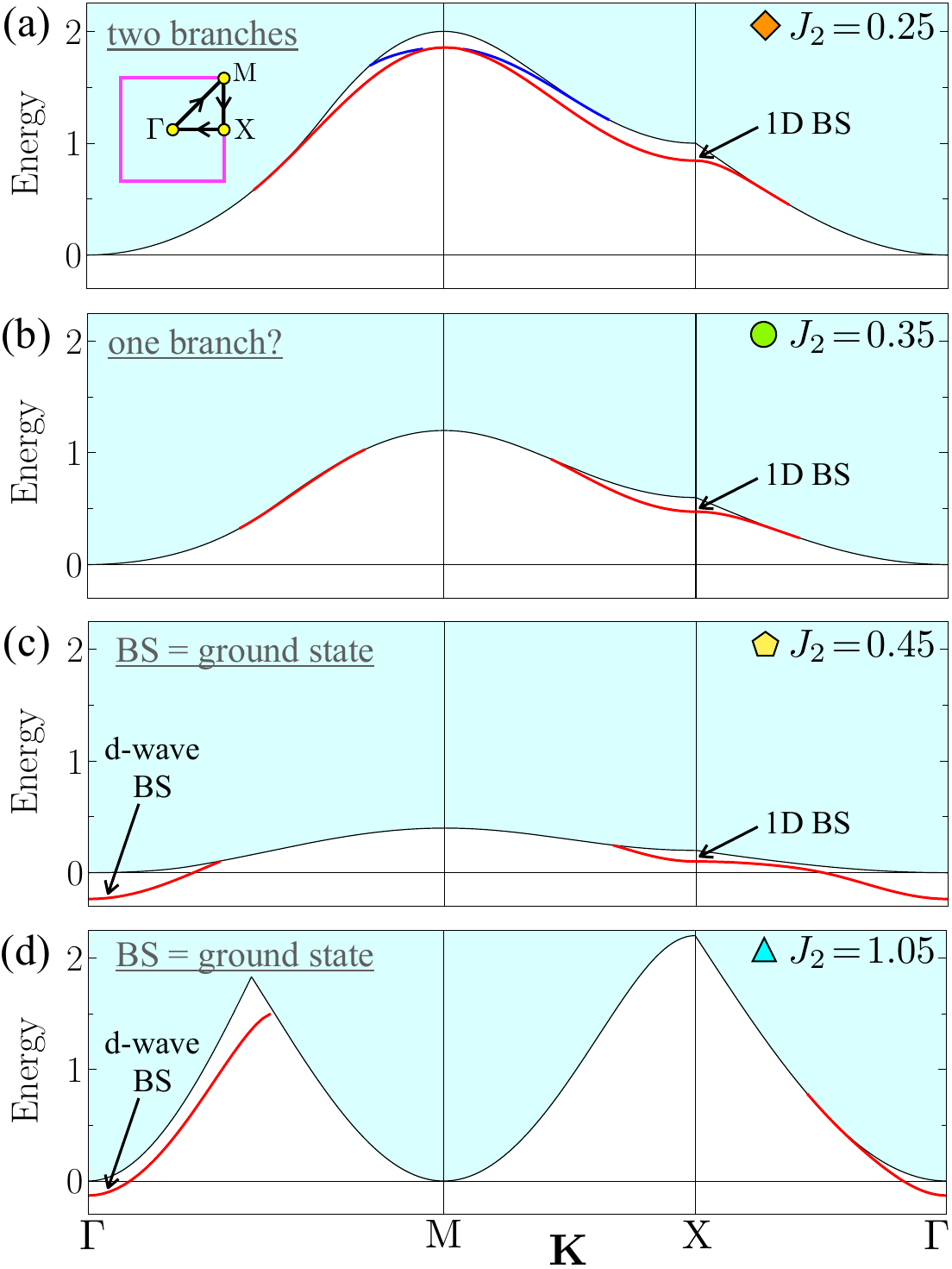}
\vskip -0.2cm
\caption{Same as in Fig.~\ref{fig:EkJ1} for the $S\!=\!1/2$ $J_1$--$J_2$ FM-AFM model for representative $J_2$ (in units of $|J_1|$): (a) $J_2\!=\!0.25$, (b) $J_2\!=\!0.35$, (c) $J_2\!=\!0.45$, and (d) $J_2\!=\!1.05$, also marked by the symbols used in Fig.~\ref{fig:DeltasJ1J2}. The $J_2$ values are chosen to highlight the changes in the BS spectrum; see the text.}
\label{fig:EkJ1J2}
\vskip -0.5cm
\end{figure}

To investigate the key transformations of the BS spectrum for the $S\!=\!1/2$ FM-AFM $J_1$--$J_2$ model, we employ the numerical procedure described in Sec.~\ref{Sec:BS_FM_evalueFig2}, which was used to obtain the results shown in Fig.~\ref{fig:EkJ1} for the $J_1$-only model. Figures~\ref{fig:EkJ1J2}(a)-(d) mirror Fig.~\ref{fig:EkJ1} and present the BS energy spectrum for the representative values of $J_2$, from smaller to larger, to highlight the changes. As in Fig.~\ref{fig:EkJ1}, the BS energies are shown as a function of the total pair momentum ${\bf K}$ along the $\Gamma$MX$\Gamma$ path, the shaded region is the two-magnon continuum, and the black line is its boundary. The solid blue and red lines denote the BS branches obtained by solving the eigenvalue problem for $\hat{\rm \bf M}_{\bf K}$ described above, with their behavior explored in the following sections. An animation of the evolution of the BSs vs $J_2$ from 0 to $1.2|J_1|$ along the same $\Gamma$MX$\Gamma$ contour but in finer steps of $0.05|J_1|$ is provided in the Supplemental Material (SM)~\cite{SM}.

\begin{figure}[t!]
\includegraphics[width=\linewidth]{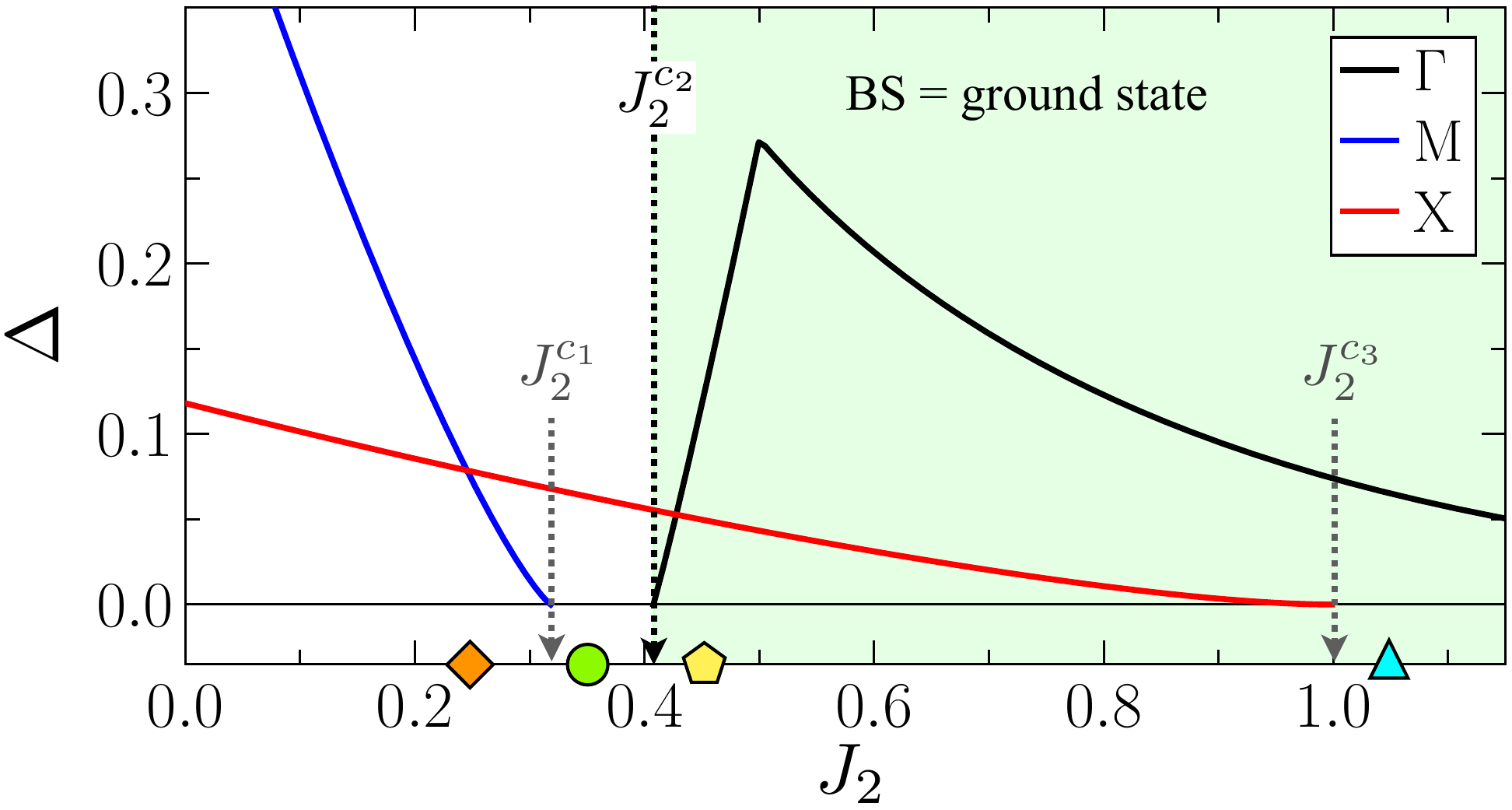}
\vskip -0.3cm
\caption{Binding energies $\Delta$ for the $S\!=\!1/2$ $J_1$--$J_2$ model at the $\Gamma\!=\!(0,0)$, M$=\!(\pi,\pi)$, and X$=\!(\pi,0)$ points as functions of $J_2$. The arrows show $J_2$ values at which the BSs at the M and X points vanish, and the onset of a paired $d$-wave ground state at the $\Gamma$ point; the filled symbols are $J_2$'s in Fig.~\ref{fig:EkJ1J2}.} 
\label{fig:DeltasJ1J2}
\vskip -0.4cm
\end{figure}

Figure~\ref{fig:DeltasJ1J2} complements our results for the BS spectrum in Fig.~\ref{fig:EkJ1J2} by showing the binding energies for the $S\!=\!1/2$ model at the high-symmetry points, $\Gamma=(0,0)$, M$\ =(\pi,\pi)$, and X$\ =(\pi,0)$, as functions of $J_2$. The filled symbols correspond to the values of $J_2$ in Figs.~\ref{fig:EkJ1J2}(a)-(d). As discussed in more detail in the following sections, the qualitative changes in the BS spectrum are closely associated with the emergence or disappearance of the BS solutions at these high-symmetry points.

Specifically, as we show below, the binding energy for both BS solutions at the M point vanish at $J_{2}^{c_1}\!=\!1/\pi$, beyond which the second BS branch lingers for a narrow range of $J_2$ as a super-shallow state and disappears from the entire BZ; see Figs.~\ref{fig:DeltasJ1J2}, \ref{fig:EkJ1J2}(a), and \ref{fig:EkJ1J2}(b), and SM~\cite{SM}. At $J_{2}^{c_2}\!\approx \!0.41$, the $d$-wave BS appears at the $\Gamma$ point, indicating the onset of the paired ground state as the {\it global} energy minimum; see Figs.~\ref{fig:DeltasJ1J2}, \ref{fig:EkJ1J2}(c), and \ref{fig:EkJ1J2}(d). The kink in $\Delta_\Gamma$ at $J_{2}\!=\!0.5$ corresponds to a switch of the magnon band minimum from the $\Gamma$ point to the X points. Lastly, at $J_{2}^{c_3}\!=\!1$, the most resilient among the anomalously deep BSs of the $J_1$-only model at the X point vanishes; see Figs.~\ref{fig:DeltasJ1J2} and \ref{fig:EkJ1J2}(d). Below, we focus on illuminating these qualitative and quantitative changes.

\subsection{BSs at the ${\rm M}$ and ${\rm X}$ points}
\label{Sec:M_and_X_J1J2}

The high-symmetry M and X points are home to the deepest BSs in the $J_1$-only FM model, but their evolution with the AFM $J_2$ exchange is very different; see Figs.~\ref{fig:EkJ1J2} and \ref{fig:DeltasJ1J2}. Here we elucidate these differences in detail.

\vspace{-0.2cm}
\subsubsection{BSs at the $\rm M$ point}
\label{Sec:M_point_J1J2}

In the $J_1$-only model, the two BSs at the M point are the deepest, owing to the complete flatness of the two-magnon band at that point, which implies a fully localized nature of these BSs and makes their degeneracy obvious, from both the partial-wave and the real-space perspectives; see Figs.~\ref{fig:EkJ1} and \ref{fig:EkqJ1}(c), and Sec.~\ref{Sec:0DBS}.

Introducing an AFM $J_2$-term in the model is expected to remove the high degeneracy of the two-magnon continuum, delocalize the BSs, and lift their degeneracy. A more generic form of the continuum is anticipated to transform the BSs' behavior into a 2D universality, in which one exponentially shallow BS should survive while the other would be pushed out into the continuum. It is also possible that both BSs could be destroyed by a finite $J_2$ due to the repulsive magnon-magnon interaction channels provided by the AFM exchange.

As one can see in Figs.~\ref{fig:EkJ1J2}(a) and \ref{fig:EkJ1J2}(b), these expectations are not straightforwardly fulfilled, but only partially so, and in an unexpected combination. Surprisingly, the two BSs are {\it not} split by a finite $J_2$, and {\it both} of them disappear upon crossing a threshold value of $J_{2}^{c_1}$. This is because the $J_2$-term, while introducing finite dispersion into the continuum and lifting its degeneracy,  still preserves the high symmetry of the lattice. 

As in the $J_1$-only model, the high symmetry of the M point renders the ``hard-core''  components of the three partial waves zero, $\gamma_{\frac{\bf K}2}^{\phantom -}\!=\!\gamma_{\frac{\bf K}2}^{-}\!=\!\gamma_{\frac{\bf K}2}^{(2)}\!=\!0$. This makes the separable interactions in the partial-wave decomposition in Eq.~\eqref{eq:VKq_J1J2expansion} fully symmetric for the $\gamma\!=\!\{s,d,s_{xy}\}$ channels, $\widetilde{R}_{\gamma,{\bf K}}({\bf p})\!\equiv\!R_{\gamma}({\bf p})$.  

Moreover, the two-magnon energy in Eq.~(\ref{eq:2magJ1J2}) simplifies considerably at the M point, reducing the denominator of the two-particle SE in Eq.~\eqref{eq:M_gamma} to
\begin{align}
E-E_{\rm M}({\bf p})=-2\Delta-8J_2S\big(1+\gamma_{\bf p}^{d_{xy}}\big),
\label{eq:Ek2_MJ2}
\end{align}
which depends only on the $d_{xy}$-harmonic, Eq.~(\ref{eq:gammadxy}), and has minima at the ${\bf Q}\!=\!\pm(\pi/2,-\pi/2)$ points. 

It is easy to show that all of the above lead to the fully diagonal form of the matrix $\hat{\rm \bf M}_{\bf K}$ for the pair momentum ${\bf K}\!=\!{\rm M}$, with the eigenvalue equation~\eqref{eq:detM=0} reducing to ${M}_{\gamma \gamma,{\bf K}}\!=\!1$ for each of its diagonal elements.  

For the four pairing channels, three symmetry operations are useful: two mirror reflections, about the $p_x\!=\!p_y$ and $(\pi/2,p_y)$ lines, and an inversion about the $(\pi/2,\pi/2)$ point; see Figs.~\ref{fig:HarmonicsJ1} and~\ref{fig:HarmonicsJ2}, which make these transformations transparent. The $d_{xy}$-harmonic, and hence the energy denominator in the SE, Eq.~(\ref{eq:Ek2_MJ2}), are even under all three transformations.  The $d$-wave decouples from the rest of the channels because it is the only one that is odd under the $p_x\!=\!p_y$ mirror reflection. The $s$-wave is odd under the $(\pi/2,\pi/2)$ inversion, while the $s_{xy}$- and $d_{xy}$-waves are even. The $s_{xy}$ and $d_{xy}$ channels decouple because the $s_{xy}$-wave is odd under the $(\pi/2,p_y)$ mirror reflection.

Since the $s_{xy}$ and $d_{xy}$ channels are repulsive, they do not yield BSs, leaving the two attractive $\gamma\!=\!\{s,d\}$-wave channels unmixed as in the $J_1$-only model in Sec.~\ref{Sec:0DBS}, with their eigenvalue equations now given by
\begin{align}
{M}_{\gamma \gamma,{\bf K}} = \frac{\alpha_{\gamma}}{4N}\sum_{\bf p}  \frac{R_{\gamma}(\mathbf{p})^2}{\Delta_\gamma+4J_2S\big(1+\gamma_{\bf p}^{d_{xy}}\big)}=1.
\label{eq:M_gamma_atMJ2}
\end{align}
The degeneracy of the resulting BSs follows from the same mirror operation about the $(\pi/2,p_y)$ line, which, as in Sec.~\ref{Sec:0DBS}, transforms $R_{s}(\mathbf{p})\!\Leftrightarrow\!-R_{d}(\mathbf{p})$ but leaves the energy denominator invariant.

With the reasons for the persistent degeneracy of the two BSs at the M point elucidated, the remaining puzzle is the origin of their destruction upon reaching the threshold value of $J_{2}^{c_1}$. Formally, the two-magnon energy from the $J_2$-term in Eq.~\eqref{eq:Ek2_MJ2} changes the effective dimensions of the pairing problem at the M point from the 0D case of the $J_1$-only model to a generic 2D problem. More specifically, the minima of the continuum are at the ${\bf Q}\!=\!\pm(\pi/2,-\pi/2)$ points, where the energy in Eq.~\eqref{eq:Ek2_MJ2} can be approximated as
\begin{align}
E-E_{\rm M}({\bf p})\approx-2\Delta-\frac{({\bf p}\!-\!{\bf Q})^2}{m_{\rm M}^*},
\label{eq:Ek2_MJ2approx}
\end{align}
where $m_{\rm M}^*\!=\!(4J_2S)^{-1}$ is the effective mass due to $J_2$.

The crucial point is that these continuum minima occur at finite momenta ${\bf Q}$ that lie precisely on the nodal lines of {\it both} $s$ and $d$ partial waves (see Fig.~\ref{fig:HarmonicsJ1}). This makes their pairing problem equivalent to that of higher partial waves, referred to as the dimensional enhancement mechanism, which requires the coupling strength to exceed the kinetic energy, ${\cal O}(1/m_{\rm M}^*)\!\sim\! {\cal O}(J_2S)$, to form a BS. Obviously, at small $J_2$, the effective mass is large and the pairing still yields a degenerate pair of BSs, but they disappear when mass $m_{\rm M}^*$ becomes light enough.

As we show in App.~\ref{A:DeltaMJ1J2}, the integral in the eigenvalue problem for the $s$- and $d$-wave BSs at the M point, Eq.~(\ref{eq:M_gamma_atMJ2}), can be evaluated in a compact form for any $\Delta$:
\begin{align}
M_{\gamma \gamma,{\rm M}}&=\frac{1}{2\pi S J_2\kappa} \Big(E\big(\kappa\big)-\big(1-\kappa^2\big) K\big(\kappa\big)\Big),
\label{eq:MM_J1J2}
\end{align} 
where $\delta\!=\!\Delta/(4SJ_2)$, $\kappa\!=\!1/(1+\delta)$, and $K(\kappa)$ and $E(\kappa)$ are the complete elliptic integrals of the first and second kind, respectively. 

The BSs disappear at $J_{2}^{c_1}$, which is obtained by solving $M_{\gamma \gamma,{\rm M}}\!=\!1$ at $\Delta\!=\!0$. Using $\lim_{\kappa \to 1} (1-\kappa) K(\kappa)\!=\! 0$ and $E(1)\!=\!1$, one straightforwardly obtains
\begin{equation}
J_{2}^{c_1}=\frac1{2\pi S},
\end{equation}
which agrees with $J_{2}^{c_1}\!=\!1/\pi$ value for $S\!=\!1/2$ in Fig.~\ref{fig:DeltasJ1J2} and with the discussion in Ref.~\cite{NematicShengtao2023}.

Further insights into the $J_2$-dependence of the binding energy at the M point, such as the small-$J_2$ expansion near the $J_1$-only value of $\Delta\!=\!\frac{1}{2}$, Eq.~(\ref{eq:Delta_atM}), and the nonlinear character of its approach to $J_{2}^{c_1}$ that can be seen in Fig.~\ref{fig:DeltasJ1J2}, are discussed in App.~\ref{A:DeltaMJ1J2}.

Altogether, the persistence of the degeneracy of the two BSs at the M point, despite the lifting of the high degeneracy of the two-magnon continuum by a finite $J_2$, is shown to come from the remaining high symmetry of both the $J_1$--$J_2$ model and the eigenvalue SE. The eventual disappearance of the degenerate BSs is {\it not} the result of the repulsive magnon-magnon interaction associated with $J_2$, but of the dimensional enhancement for both $s$ and $d$ partial waves due to their nodal lines crossing the two-magnon continuum minima. 

\subsubsection{X point: 1D BS}
\label{Sec:1DBSJ1J2}

While the evolution of the two BS solutions at the M point discussed above is due to a combination of the $J_2$-induced dispersion of the continuum and a dimensional enhancement, the behavior of the 1D-like BS at the X point must be different. 

This is because the two-magnon energy band for ${\bf K}\!=$X remains 1D-like as in Fig.~\ref{fig:EkqJ1}(d), with the energy denominator of the two-particle SE~\eqref{eq:M_gamma} unaffected by $J_2$,
\begin{align}
\label{eq:E2_J1J2_X_point}
E-E_{\rm X}({\bf p})= -2\Delta-4S\big(1-\cos p_y\big),
\end{align}
suggesting that a stronger, 1D-like pairing, referred to as the dimensional reduction, should persist for any $J_2$. 

As is observed in Figs.~\ref{fig:EkJ1J2} and \ref{fig:DeltasJ1J2}, the BS at the X point, referred to as the symmetric one for the $J_1$-only problem in Sec.~\ref{Sec:MX}, is, indeed, remarkably robust and is the longest-surviving branch inherited from the $J_1$-only model, lasting up to $J_{2}^{c_3}\!=\!1$.  

Since the energy denominator of the two-particle SE in Eq.~(\ref{eq:E2_J1J2_X_point}) is unaffected by $J_2$, the disappearance of the BS {\it must} come from the mixing with the repulsive $J_2$-channels, leading to the change of the magnon interaction from attraction to repulsion. For the discussion below, it is worth noting that such a change {\it does not} remove the divergences in the eigenvalue integrals that guarantee the BSs in 1D and 2D as discussed in Sec.~\ref{simplest}, but switches the sign of these divergences, which, in turn, deprive the eigenvalue problem of a solution.  

 The symmetry analysis of the partial waves at the X point shows that the BS-yielding symmetric combination of $s$ and $d$ waves mixes only with the $s_{xy}$-wave harmonic because both are odd under the mirror reflection about the $(\pi/2,p_y)$ line. The antisymmetric combination of $s$ and $d$ waves, the $d_{xy}$-wave harmonic, and the energy denominator in Eq.~(\ref{eq:E2_J1J2_X_point}) are all even under it. This also remains true for ${\bf K}$ along the entire MX line. At the X point, reflection about the $p_x$ or $p_y$ axis further decouples the antisymmetric and $d_{xy}$ channels. However, neither yields a BS: the BS in the antisymmetric channel is suppressed by the hard-core repulsion as in the $J_1$-only limit, and the $d_{xy}$ channel is purely repulsive.

 It is worth noting that, for the entire MX line, the symmetric and antisymmetric combinations of the $s$ and $d$ partial waves are more optimally represented by the ``elemental'' lattice partial waves 
\begin{align}
\label{eq:alt_rep_MX}
R_{S(A)}({\bf p})=\gamma_{\bf p}\pm\gamma^{-}_{\bf p}=\cos p_x\,(\cos  p_y), 
\end{align}
which, together with their pairwise mixing with the $s_{xy}$ and $d_{xy}$ harmonics, reduce the eigenvalue matrix $\hat{\rm \bf M}_{\bf K}$ to block-diagonal form with two $2\times 2$ blocks; see App.~\ref{A:MmatrixMXJ1J2}. The ``hard-core'' components of the symmetric and $s_{xy}$ partial waves are also zero along the MX line,  $\cos K_x/2\!=\!\gamma_{\frac{\bf K}2}^{(2)}\!=\!0$, providing additional simplifications. 

For the X point, the eigenvalue problem for the mixture of the symmetric $s$-$d$ and $s_{xy}$ harmonics can be advanced analytically since the 1D profile of the continuum in Eq.~\eqref{eq:E2_J1J2_X_point} renders the momentum integrals in the matrix elements of $\hat{\rm \bf M}_{\bf K}$, Eq.~\eqref{eq:M_gamma}, effectively one-dimensional. Straightforward algebra yields a compact, if cumbersome, relation between the binding energy $\Delta$ and $J_2$; see App.~\ref{A:DeltaXJ1J2}. It simplifies at a threshold value of $J_2$ for $\Delta\!\rightarrow\!0$, making explicit the sign change of the 1D-like, $1/\sqrt{\Delta}$-divergence in the eigenvalue equation (\ref{eq:SE_1D}),
\begin{align}
\label{eq:X_point_app_EV}
1-\frac{1}{4S}\, \frac{J_2}{J_2^{c_3}}\approx \frac{1}{2\sqrt{4S\Delta}}\left(1-\frac{J_2}{J_2^{c_3}}\right),
\end{align}
which renders the BS nonexistent for $J_2\!>\!J_2^{c_3}$. Here
\begin{align}
\label{eq:X_J2c}
J_2^{c_3}=\frac{1}{2(1-1/4S)},
\end{align}
which yields $J_2^{c_3}\!=\!1$ for $S\!=\!1/2$ in agreement with Fig.~\ref{fig:DeltasJ1J2}.

The solution of Eq.~(\ref{eq:X_point_app_EV}) for $\Delta$ can be cast in the standard 1D-like form of Eq.~(\ref{eq:SE_1D})
\begin{align}
\Delta_{\rm X} &\approx \frac{\alpha_{\rm eff}^2m_y^*}{32}, \label{eq:DeltaXapproxJ1J2}
\end{align}
with the effective coupling $\alpha_{\rm eff}\!=\!4\big( J_2^{c_3}- J_2 \big)$ and effective mass $m_y^*\!=\!(2S)^{-1}$ of the continuum in Eq.~(\ref{eq:E2_J1J2_X_point}). 

The evolution of both symmetric and antisymmetric BSs for $S\!=\!1/2$ along the MX line for $J_2$ between its threshold values at the M and X points, $J_2^{c_1}\!=\!1/\pi$ and $J_2^{c_3}\!=\!1$, respectively, is considered next.

\subsection{MX line: Suppression of the 1D BSs}
\label{Sec:MXJ1J2}

As is shown above, the BSs at the M and X points are suppressed by two {\it different} mechanisms, the dimensional enhancement and the admixture of the repulsive channel, respectively. The next question is: what is the mechanism for the eventual demise of the 1D-like BSs of the $J_1$-only model for the pair momentum along the MX line, ${\bf K}\!=\!(\pi,K_y)$, as a function of the AFM $J_2$?

From the numerical results in Fig.~\ref{fig:EkJ1J2} and SM~\cite{SM}, it may seem that the upper BS branch gradually shrinks toward the M point as $J_2$ approaches the threshold value $J_2^{c_1}$, and disappears from the entire BZ {\it together} with the $d$-wave BS at the M point; see also Fig.~\ref{fig:DeltasJ1J2}. However, since this branch is very shallow for ${\bf K}$ close to the X point already in the $J_1$-only model, the numerical (un)detectability of the BS prevents one from concluding definitively about its disappearance.

The evolution of the lower BS branch seems more straightforward: starting with the disappearance of the deepest BS of the $J_1$-only model at the M point at $J_2^{c_1}$, the termination point of this branch steadily moves from the M point toward the X point until it vanishes completely at $J_2^{c_3}$; see Figs.~\ref{fig:EkJ1J2} and \ref{fig:DeltasJ1J2}, and SM~\cite{SM}. However, which of the two distinct mechanisms, if any, is responsible for these trends is an open question.

Below we demonstrate that despite the lower symmetry of the MX line, which allows mixing of different channels, the lower branch inherits a crucial symmetry from the M point and is suppressed by a combination of {\it both} the dimensional enhancement and repulsive channel mechanisms. The behavior of the upper BS branch turns out to be significantly richer than suggested by the numerical phenomenology of Fig.~\ref{fig:EkJ1J2}, and we demonstrate that it survives for $J_2$ {\it above} the $J_2^{c_1}$ threshold, albeit for a narrow range of $J_2$ and with exceptionally low binding energy. All results below are for $S\!=\!1/2$ unless stated otherwise.

\subsubsection{${\rm MX}$ line: two-magnon band}
\label{Sec:MX_continuum}

\begin{figure}[t!]
\includegraphics[width=\linewidth]{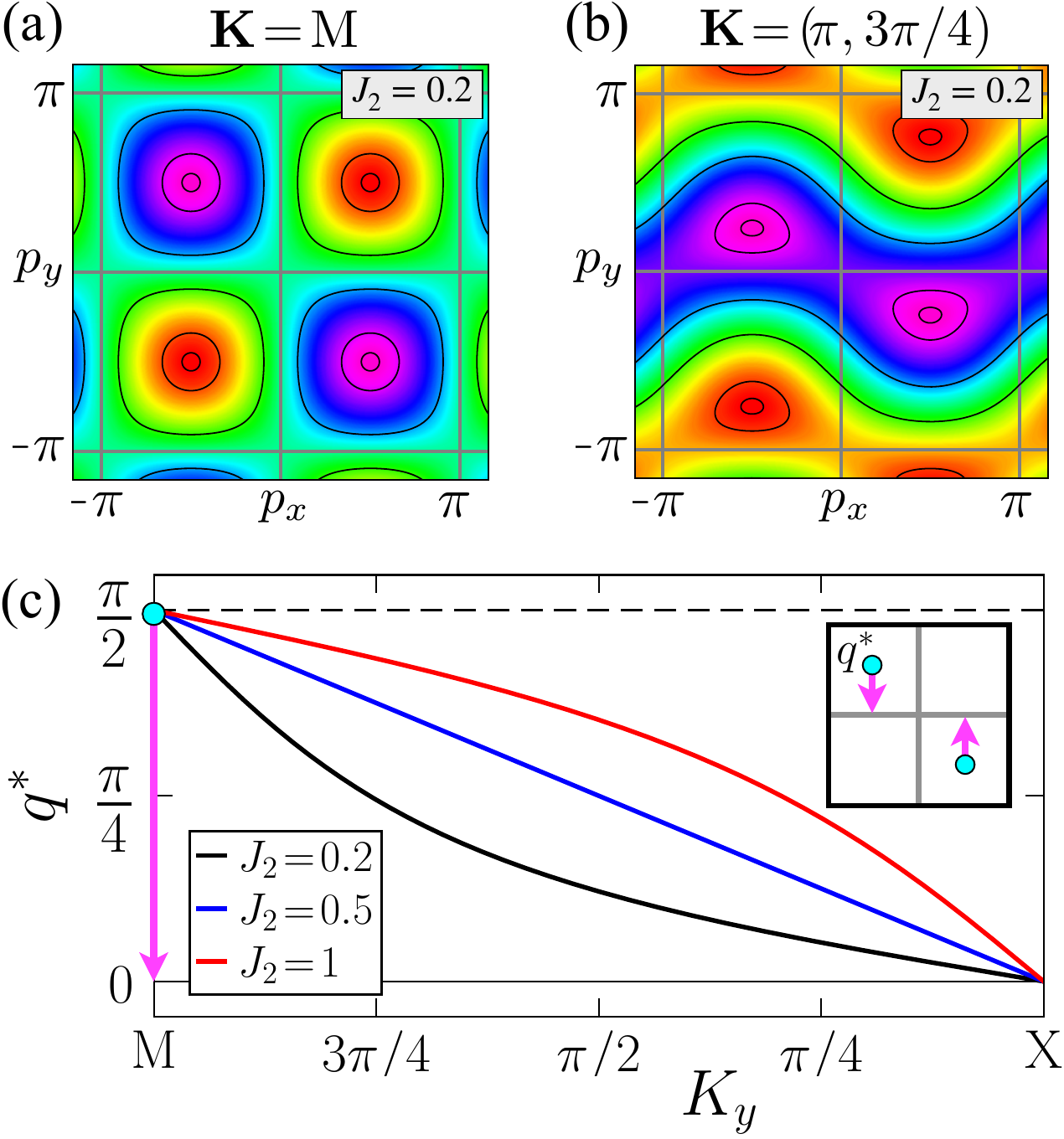}
\vskip -0.3cm
\caption{Intensity maps of the two-magnon energy, Eq.~(\ref{eq:MX_2mag_E}), for $J_2\!=\!0.2$. (a) ${\bf K}\!=\!(\pi,\pi)$ (M point) and (b) ${\bf K}\!=\!(\pi,3\pi/4)$. (c) $q^*$ vs $K_y$, Eq.~(\ref{eq:MX_2mag_q*}), along the MX line for representative values of $J_2\!=$0.2, 0.5, and 1.0. Inset: migration of the two-magnon band minima for $K_y$ from M to X.}
\label{fig:MX_E_q*}
\vskip -0.4cm
\end{figure}

Along the MX line, ${\bf K}\!=\!(\pi,K_y)$, and at a finite $J_2$, the two-magnon continuum must interpolate between the symmetric form at the M point, Eq.~(\ref{eq:Ek2_MJ2}), with the minima at ${\bf Q}\!=\!\pm(\pi/2,-\pi/2)$, and the degenerate 1D shape at the X point, Eq.~(\ref{eq:E2_J1J2_X_point}). Using Eq.~(\ref{eq:2magJ1J2}) and the fact that some of the ``hard-core'' partial waves are zero along the MX line,  $\cos K_x/2\!=\!\gamma_{\frac{\bf K}2}^{(2)}\!=\!0$, one finds that the two-magnon energy indeed directly mixes the two ingredients: the 1D-like dispersion of the X point and the 2D-like one of the M point, leading to an intricate form,
\begin{align}
\label{eq:MX_2mag_E}
E-E_{\bf K}({\bf p})=&-2\Delta-\Delta E_{\bf K}({\bf p}^*)   \\
&+4S\Big(\cos\frac{K_y}{2}\cos p_y - 2J_2 \sin\frac{K_y}{2} \, \gamma_{\bf p}^{d_{xy}}\Big),\nonumber
\end{align}
where $\Delta E_{\bf K}({\bf p}^*)\!=\!4S\sqrt{\cos^2(K_y/2)+4J_2^2\sin^2(K_y/2)}$ is a constant energy offset for the two-magnon energy minimum, which is reached at the incommensurate momenta, ${\bf p}^*\!=\!\pm \big( \frac\pi2, -q^* \big)$, with
\begin{align}
\label{eq:MX_2mag_q*}
q^*=\arctan\left(2J_2\tan \frac{K_y}{2}\right).
\end{align}
Figure~\ref{fig:MX_E_q*} showcases the resultant shape of the two-magnon energy band and the migration of its minima as a function of $K_y$ as the pair momentum moves from the M to the X point. Importantly, these minima continue to reside along the vertical $\pm(\pi/2, p_y)$ lines, with the band retaining an important symmetry.  

The same mirror reflection about the $(\pi/2,p_y)$ line that was used in the symmetry analysis for the M and X points in Secs.~\ref{Sec:M_point_J1J2} and \ref{Sec:1DBSJ1J2} is instrumental here. It leaves invariant not only the energy denominator in Eq.~(\ref{eq:MX_2mag_E}), but also the antisymmetric ($R_A({\bf p})\!=\!\cos p_y$) and $d_{xy}$ partial waves, while the symmetric ($R_S({\bf p})\!=\!\cos p_x$) and $s_{xy}$ partial waves are odd under this operation. By continuity with the $J_1$-only model, Sec.~\ref{Sec:MX}, the two BS solutions should come from the two pairwise mixtures of the partial waves, the upper one from the antisymmetric with the $d_{xy}$, and the lower one from the symmetric with the $s_{xy}$, each of which should be obtained from the eigenvalue problem for the corresponding $2\times 2$ block of the block-diagonal $4\times 4$ eigenvalue matrix $\hat{\rm \bf M}_{\bf K}$; see App.~\ref{A:MmatrixMXJ1J2}.

\subsubsection{${\rm MX}$ line: lower BS branch}
\label{Sec:MX_symmetric}

The important feature of the two-magnon continuum discussed above is that its minima stay on the $\pm(\pi/2, p_y)$ paths for ${\bf K}$ along the MX line. Since the lower BS branch comes from a mixture of the symmetric combination of the $s$ and $d$ partial waves, $R_S$ in Eq.~(\ref{eq:alt_rep_MX}), and the $s_{xy}$ partial wave, the nodal lines for both harmonics {\it necessarily} cross these two-magnon band minima for any ${\bf K}$; see Figs.~\ref{fig:HarmonicsJ1}, \ref{fig:HarmonicsJ2}(a), and \ref{fig:MX_E_q*}. 

As with the dimensional enhancement at the M point in Sec.~\ref{Sec:M_point_J1J2}, such crossings place the pairing into the higher-partial-wave class, which needs attraction to exceed the kinetic energy. For ${\bf K}$ near the M point, the mixing with the repulsive $s_{xy}$ channel is weak and the main mechanism for the suppression of the BS should remain the same as at the M point. In Fig.~\ref{fig:MX_E_q*}(b), one can see that for ${\bf K}$ away from the M point, the two-magnon band becomes flatter, suggesting heavier effective mass and, hence, an enhanced pairing. Naturally, this implies that larger values of $J_2$ are needed to suppress the BS compared to $J_2^{c_1}$ for the M point.

For ${\bf K}$ near the X point, the two-magnon band transforms into an lopsided shape with an anisotropic effective mass, approaching the degenerate 1D form in Fig.~\ref{fig:EkqJ1}(d) and making the dimensional enhancement less effective. On the other hand, given the analysis of the transition for the strongly mixed symmetric and $s_{xy}$ partial waves at the X point in Sec.~\ref{Sec:1DBSJ1J2}, the same strong admixture of the repulsive channel must suppress the BS in the vicinity of the X point by switching the effective coupling constant from attraction to repulsion. 

\begin{figure}[t]
\includegraphics[width=\linewidth]{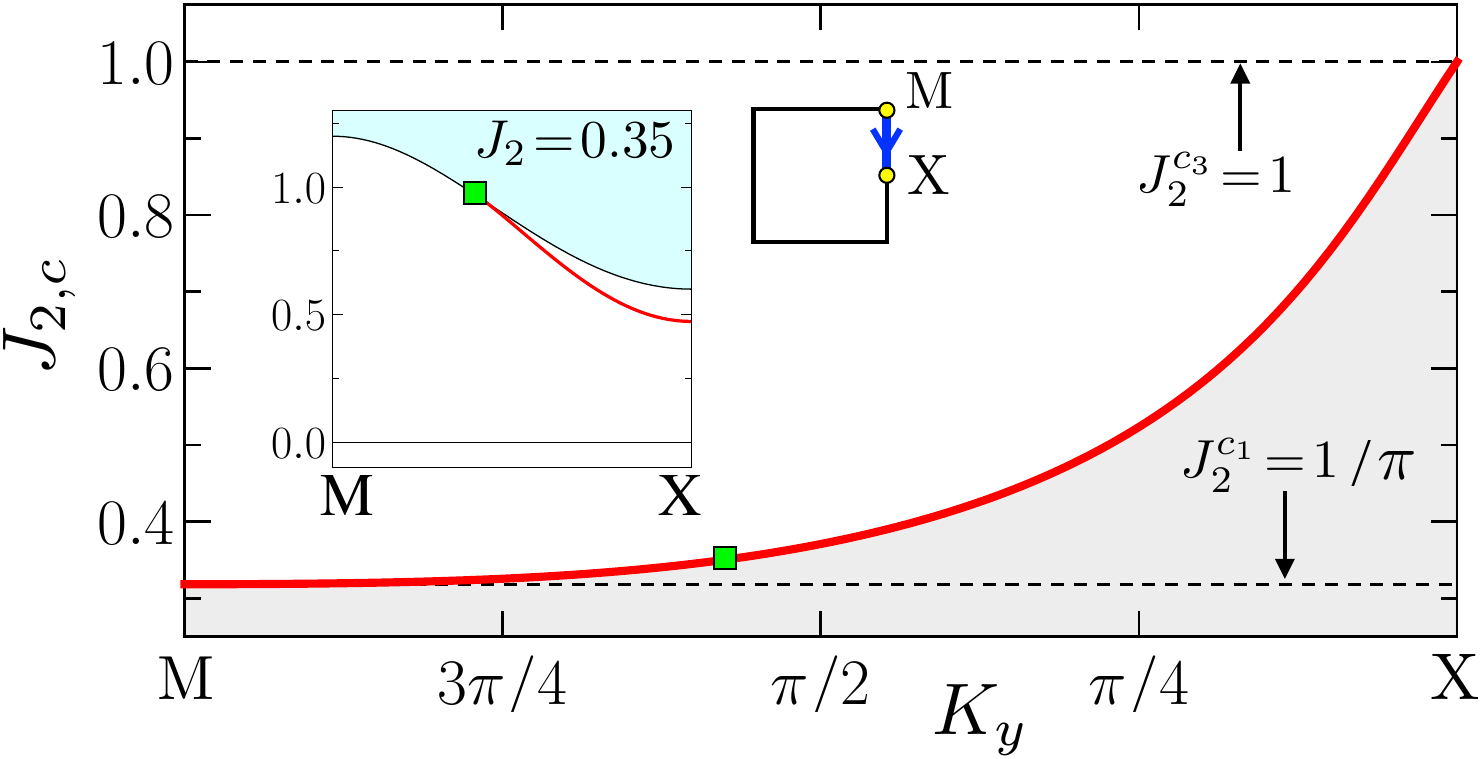}
\vskip -0.3cm
\caption{The BS threshold boundary $J_{2,c}$ as a function of $K_y$ along the MX line for $S\!=\!1/2$; the shaded area marks the phase space for BS existence; the symbol corresponds to the one in the inset. Insets: the MX panel from Fig.~\ref{fig:EkJ1J2}(b) for $J_2\!=\!0.35$; the symbol highlights the termination point of the BS branch for this $J_2$; the sketch of the BZ with the MX path.}
\label{fig:MX_Jc_vs_Ky}
\vskip -0.4cm
\end{figure}

Figure~\ref{fig:MX_Jc_vs_Ky} summarizes and quantifies these trends by showing the threshold boundary $J_{2,c}$ for the suppression of the lower BS as a function of $K_y$ from the M to the X point for $S\!=\!1/2$. The gray area marks the phase space for BS existence. Indeed, $J_{2,c}$ continuously interpolates between $J_2^{c_1}\!=\!1/\pi$ for the M point and $J_2^{c_3}\!=\!1$ for the X point, in agreement with the expectations outlined above. The inset shows the MX panel from Fig.~\ref{fig:EkJ1J2}(b) for $J_2\!=\!0.35$, with the highlighted termination-like point of the BS branch, at which it ceases to exist. The corresponding value of $J_{2,c}$ is also shown in the main panel. 

The values for $J_{2,c}$ vs $K_y$ in Fig.~\ref{fig:MX_Jc_vs_Ky} were obtained using numerical integration with Gaussian quadratures on the $N_p\!\times\! N_p$ grid of ${\bf p}$-points in the matrix elements $M_{\gamma\gamma',{\bf K}}$ of the eigenvalue matrix, Eq.~\eqref{eq:M_gamma}, for the $2\times 2$ sector of the symmetric and $s_{xy}$ partial waves; see App.~\ref{A:MmatrixMXJ1J2}. For each ${\bf K}$ along the MX line, the solution of the eigenvalue problem, $|\hat{\rm \bf M}_{\bf K}-\hat{\rm \bf I}|\!=\!0$, was sought for $\Delta\!=\!0$, using the linear bisection method to a precision of $10^{-8}$. Because of the dimensional enhancement, none of the integrals are divergent and one can directly use $\Delta\!=\!0$ in their energy denominators, Eq.~(\ref{eq:MX_2mag_E}). The resultant values of $J_{2,c}$ are the same to an accuracy of $10^{-6}$ for $N_p\!=\!2000$ and 12000.

\subsubsection{${\rm MX}$ line: upper BS branch}
\label{Sec:MX_asymmetric}

Understanding the evolution of the upper BS branch with $J_2$ constitutes a significantly more challenging task. The numerical phenomenology in Fig.~\ref{fig:EkJ1J2} and SM~\cite{SM} suggests that it shrinks toward the M point as $J_2\!\rightarrow\!J_2^{c_1}$ and disappears together with the BS  at the M point. 

The difficulty of proving or disproving this scenario lies in the smallness of the binding energies in this regime, which makes it challenging to extract the actual values of $\Delta$ from the eigenvalue problem numerically with any reasonable accuracy---the situation that is already familiar from the $J_1$-only model for ${\bf K}$ close to the X point, where the BS branch becomes super-shallow due to hard-core repulsion that suppresses the strength of the effective attraction. While there is analytical guidance in the $J_1$-only case (see Sec.~\ref{Sec:MX}), none exists for the $J_1$--$J_2$ model because of the more complex shape of the two-magnon band and mixing with other partial waves.

The evolution of the two-magnon band from the $J_1$-only case, where it is 1D-like for all ${\bf K}$ along the MX line, to the shapes in Fig.~\ref{fig:MX_E_q*}(b), is described in Sec.~\ref{Sec:MX_continuum}. 

For finite $J_2$, the upper BS branch comes from the combination of the antisymmetric mixture of the $s$ and $d$ partial waves, $R_A$ in Eq.~(\ref{eq:alt_rep_MX}), and the $d_{xy}$ partial wave.  In contrast to the symmetric and $s_{xy}$ partial waves of the lower branch, the ``hard-core'' components for the antisymmetric and  $d_{xy}$ harmonics are not zero along the MX line (see App.~\ref{A:MmatrixMXJ1J2}), and they continue to play the same role as in the $J_1$-only model, suppressing the pairing.  

However, unlike for the lower BS branch, the nodal lines for the antisymmetric and $d_{xy}$ partial waves {\it do not} cross the two-magnon band minima in Fig.~\ref{fig:MX_E_q*}, implying no dimensional enhancement for ${\bf K}$ along the MX line; see Figs.~\ref{fig:HarmonicsJ1} and \ref{fig:HarmonicsJ2}(b). Crucially, this feature guarantees that all integrals in the eigenvalue matrix elements, Eq.~\eqref{eq:M_gamma}, diverge logarithmically, implying an $s$-wave-like BS in 2D regardless of the strength of attraction; see Sec.~\ref{simplest}. Somewhat unexpectedly, this suggests an intriguing alternative scenario in which the lower BS branch discussed in Sec.~\ref{Sec:MX_symmetric} may be suppressed more effectively because of the dimensional enhancement mechanism, while the upper one, being free from it, survives for larger, or even arbitrary $J_2$, however shallow it may become.

Altogether, three different effects shape the destiny of the upper BS branch: the lifting of the 1D-like degeneracy of the two-magnon band, with incommensurate minima avoided by the nodal lines of the relevant partial waves; the mixing of the attractive partial wave with the repulsive channel; and the continuing presence of the hard-core terms in the interactions.

In order to shed light on this problem, it is important to note that while the BS binding energy can be too small to be detected, the {\it sign} of the divergence in the eigenvalue determinant, Eq.~\eqref{eq:detM=0}, should still be able to show whether the eigenvalue equation has a solution.

\begin{figure}[t]
\includegraphics[width=\linewidth]{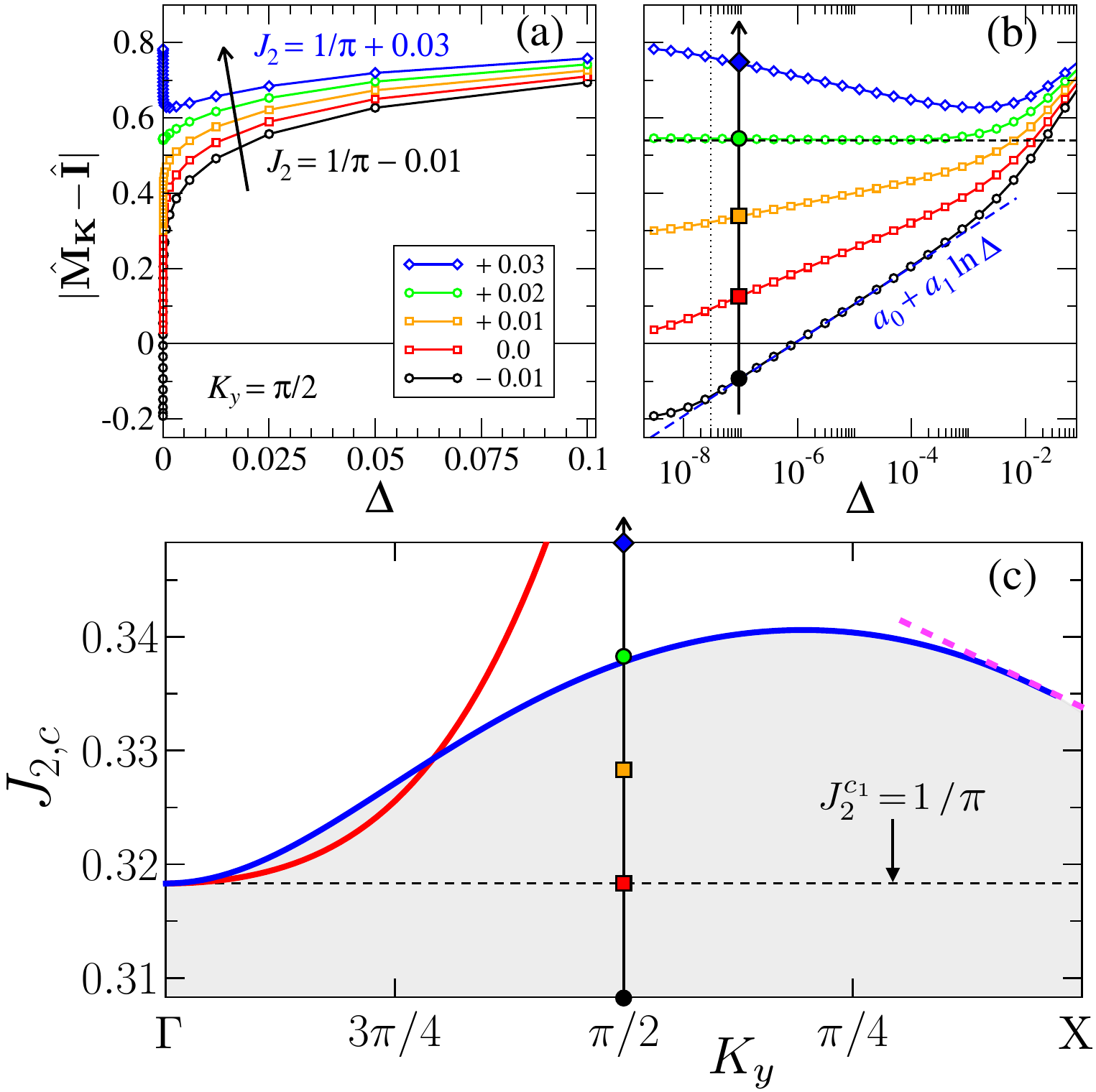}
\vskip -0.3cm
\caption{(a) $|\hat{\rm \bf M}_{\bf K}-\hat{\rm \bf I}|$ for the upper BS branch as a function of $\Delta$ for $K_y\!=\!\pi/2$, $J_2$ near $J_2^{c_1}\!=\!1/\pi$. (b) The same on a semi-log plot. The vertical dotted line is the finite-size-effect boundary, $3\!\times\! 10^{-8}$. The colored solid symbols and the black arrows in (a) and (b) serve to cross-correlate with the data in (c). (c) The BS threshold boundary $J_{2,c}$ for the upper BS branch (blue line) and the shaded phase space where this BS exists. The red solid line is the threshold boundary for the lower branch from Fig.~\ref{fig:MX_Jc_vs_Ky}. The dashed magenta line is an extrapolation, see the text.}
\label{fig:MX_AS_MvsDelta_log}
\vskip -0.5cm
\end{figure}

Figures~\ref{fig:MX_AS_MvsDelta_log}(a) and \ref{fig:MX_AS_MvsDelta_log}(b) illustrate such a study for a representative value of $K_y\!=\!\pi/2$, several values of $J_2$ between $1/\pi-0.01$ and $1/\pi+0.03$ near $J_2^{c_1}\!=\!1/\pi$, and $S\!=\!1/2$. They show the eigenvalue determinant, $|\hat{\rm \bf M}_{\bf K}-\hat{\rm \bf I}|$, for the upper BS branch, i.e., the $2\times 2$ sector of the antisymmetric and $d_{xy}$ partial waves (see App.~\ref{A:MmatrixMXJ1J2}) as a function of $\Delta$. Of interest is its behavior for small binding energies, which clearly indicates a change of the sign of the divergence for different choices of $J_2$. 

Figure~\ref{fig:MX_AS_MvsDelta_log}(b) shows the same data as Fig.~\ref{fig:MX_AS_MvsDelta_log}(a), but on a semi-log scale. It demonstrates the distinct logarithmic nature of the divergence of the determinant at small $\Delta$ and makes explicit the change of its sign with $J_2$. For the data in these figures, we have used the numerical integration with Gaussian quadratures on the $N_p\times N_p$ grid of ${\bf p}$-points with $N_p\!=\!12000$ in the elements $M_{\gamma\gamma',{\bf K}}$ of the eigenvalue matrix, Eq.~\eqref{eq:M_gamma}. The vertical dotted line in Fig.~\ref{fig:MX_AS_MvsDelta_log}(b) marks the boundary, $3\!\times\! 10^{-8}$, below which the results become affected by the finite-size effects of the integration. This boundary is inferred from a systematic increase of the integration grid density from $N_p\!=\!1000$ to 12000 and a comparison of the results. 

The linear fit in Fig.~\ref{fig:MX_AS_MvsDelta_log}(b), spanning four orders of magnitude in $\Delta$, provides  undeniable evidence of the logarithmic term in the determinant. There is also a clear change of the sign of this term with the increase of $J_2$, indicating that the effective attraction changes to repulsion, underpinning the admixture of the repulsive channel as the key mechanism for suppressing the upper BS branch for MX line. The dashed horizontal line in Fig.~\ref{fig:MX_AS_MvsDelta_log}(b) shows that the change occurs at $J_{2,c}\!\approx\!1/\pi+0.02$ for this $K_y$.

The colored solid symbols in Fig.~\ref{fig:MX_AS_MvsDelta_log}(b) and the black arrow in  Fig.~\ref{fig:MX_AS_MvsDelta_log}(a)  and \ref{fig:MX_AS_MvsDelta_log}(b) cross-correlate with the presentation in Fig.~\ref{fig:MX_AS_MvsDelta_log}(c), which shows the threshold boundary $J_{2,c}$ for the suppression of the upper BS as a function of $K_y$ along the MX line, as in Fig.~\ref{fig:MX_Jc_vs_Ky}. The $J_{2,c}$ values are found numerically by the linear bisection method as for Fig.~\ref{fig:MX_Jc_vs_Ky}, but applied to finding a zero of the logarithmic derivative of the determinant $|\hat{\rm \bf M}_{\bf K}-\hat{\rm \bf I}|$ at $\Delta\!=\!10^{-7}$.

This approach becomes less reliable very close to the X point, where the near-1D-like band structure and the near-decoupling from the repulsive $d_{xy}$ channel compete with the hard-core repulsion, which diminishes the kinematic range for the divergence in the eigenvalue determinant. While this complex competition  makes analytical insights difficult in this limit, an extrapolation using the magenta dashed line in Fig.~\ref{fig:MX_AS_MvsDelta_log}(c) points to $J_{2,c}({\rm X})\!=\!1/3$.

As one can see in Fig.~\ref{fig:MX_AS_MvsDelta_log}(c), the $J_{2,c}$ curve for the upper BS is not a monotonic function of $K_y$ and has a reentrant trend, indicating that there is a range of $J_2$ with {\it two} termination points, for which the BS branch retreats from {\it both} endpoints, M and X.

From Fig.~\ref{fig:MX_AS_MvsDelta_log}(c), it is clear that although the upper BS branch does not disappear together with the BS at the M point at $J_2^{c_1}\!=\!1/\pi$, the extent of its additional range of existence is limited to only about $1/\pi+0.02$. Another quantitative measure of the limited significance of the discussed effect can be obtained from the data in Fig.~\ref{fig:MX_AS_MvsDelta_log}(b), from which one can estimate the maximal depth of the BS for a ``typical'' $K_y\!=\!\pi/2$ at $J_2\!=\!J_2^{c_1}$, when the BS at the M point disappears. While the extrapolation of the red curve is outside the numerical confidence of the calculations, there are no physical reasons to doubt the validity of such an extrapolation, which yields $\Delta\!\approx\!10^{-9}$. 

With such a small binding energy of the surviving BS branch, which clearly puts it in the class of states that are not physically meaningful according to our own discussion in Sec.~\ref{Sec:s_wave}, and the narrow extra range of $J_2$ within which it manages to survive, one may wonder if these results are purely academic. They are not. 

The key insight from the approach discussed here is that while finding the BS binding energy for the super-shallow states might be out of numerical reach for any reasonable computational effort, the  {\it existence} or the {\it absence} of the BS and the phase boundary for a BS state can be determined with high confidence and precision.

Another interesting observation is presented in Fig.~\ref{fig:MX_AS_MvsDelta_log}(c) by the red curve, which is the $J_{2,c}$ boundary for the lower BS branch from Fig.~\ref{fig:MX_Jc_vs_Ky}. It supports the intuition behind the intriguing alternative scenario suggested earlier, in which the lower BS branch was hypothesized to be suppressed more effectively than the upper one, which indeed survives till larger $J_2$ for a range of ${\bf K}$. 

It is worth noting that one can perform a numerical analysis of the behavior of the eigenvalue determinant as a function of $\Delta$, with great success and accuracy, for different lower-dimensional types of divergences, such as the 1D-like $1/\sqrt{\Delta}$ behavior for the problem of the BS at the X point, discussed in Sec.~\ref{Sec:1DBSJ1J2}. In that particular case, we have a fully analytic solution, Eq.~(\ref{eq:X_point_app_EV}), with a $J_2$-dependent factor that explicitly switches the sign of the divergence from the guaranteed BS solution to the absence of such a solution at a threshold $J_2^{c_3}$.

Altogether, the upper BS branch for the MX line presented a significantly richer behavior than suggested by the numerical phenomenology of Fig.~\ref{fig:EkJ1J2}, surviving for larger $J_2$ than expected, outliving the lower branch for a range of ${\bf K}$, and exhibiting other complex trends, albeit in a narrow range of $J_2$ and with low binding energy. 

\subsection{$\Gamma$M line: kaleidoscope of the BSs}
\label{Sec:GammaM}

Building on the understanding of the evolution of the two BS branches along the MX line provided above, a similar exploration for the pair momentum along the $\Gamma$M line, ${\bf K}\!=\!(K_x,K_x)$, is in order. 

The numerical guidance from Fig.~\ref{fig:EkJ1J2} and SM~\cite{SM} suggests the gradual contraction of the upper BS branch, referred to as the $d$-wave in the $J_1$-only model, toward the M point, and its demise together with the BSs at that point. The lower BS branch, the $s$-wave in the $J_1$-only limit, although made super-shallow by the hard-core repulsion in the broad ${\bf K}$ range surrounding the $\Gamma$ point already in that limit, manages to remain numerically detectable up to larger values of $J_2$. This lower branch also seems to coexist over some $J_2$-range with the appearance of the new BS branch, which nucleates at the $\Gamma$ point and expands outward from it, signifying the stabilization of a BS as a global energy minimum of the model at $J_2^{c_2}\!\approx\!0.41$; see also Fig.~\ref{fig:DeltasJ1J2}. 

Given the lessons of false expectations, unexpected resolutions, and subtle reasons behind them learned from the evolution of the BS branches for the MX line in Secs.~\ref{Sec:M_and_X_J1J2} and \ref{Sec:MXJ1J2}, one may be cautious about drawing conclusions from the numerical observations alone, expecting more rigorous insights from the combination of the symmetry considerations and additional dedicated numerical investigations. 

The questions that need to be resolved are the following. Does the upper BS branch disappear at the same  $J_2^{c_1}$ as the M-point one, or does it extend to a larger $J_2$ as its MX counterpart in Sec.~\ref{Sec:MX_asymmetric}? With no resilient base support, like the one at the X point illuminated in Sec.~\ref{Sec:1DBSJ1J2}, how does the lower BS branch manage to linger in the middle of the $\Gamma$M line? Does it truly disappear when the new, global-minimum BS branch emerges, or does it persist with the binding energies too low to detect? Are the $s$-wave-like branch and the new, global-minimum one truly orthogonal, or are they actually the same? If this new state is a $d$-wave, as was indicated in the earlier studies~\cite{NematicShengtao2023}, is it related to the $d$-wave-like upper branch that disappeared for smaller values of $J_2$?

Below we address these questions. We demonstrate that because of the high symmetry of the $\Gamma$M direction, the $d$-wave channel is still orthogonal to the rest of the partial wave manifold. Unlike the upper branch along the MX line, it also preserves its nodal-crossing of the two-magnon band minima and keeps the dimensional enhancement character associated with it. As a result, the $d$-wave BS branch is indeed shown to disappear together with the BS at the M point. 

However, an intriguing aspect of its behavior can be observed. The termination momentum ${\bf K}_c$ shows a non-monotonic behavior as a function of $J_2$: it first extends farther away from the M point compared to its $J_1$-only value, only to collapse onto the M point as $J_2\!\rightarrow\!J_2^{c_1}$. This is due to {\it anisotropic} dimensional reduction, which partially compensates the effect of the dimensional enhancement for the $d$-wave channel. This form of the dimensional reduction originates from the pinching-off of the two minima of the two-magnon band from the $\Gamma$ point at some $J_2$-dependent ${\bf K}^*$, which results in the highly anisotropic, non-parabolic dispersion of the band in one of the principal momentum directions. This yields a higher density of states in the low-energy sector, similar to the more straightforward 1D-shapes discussed above.  

The same mechanism can be shown to be responsible for an additional stabilization of the lower BS branch, which is free from the dimensional enhancement, as the nodal structures of its partial waves have no crossings of the two-magnon band minima. This effect modifies the divergence of the eigenvalue determinant for ${\bf K}\!\rightarrow\!{\bf K}^*$ from a logarithmic to a fractional power law, $1/\Delta^{1/4}$.

Aside from this additional enhancement, the lower BS branch for the $\Gamma$M line appears close in character to the upper BS branch for the MX line of Sec.~\ref{Sec:MX_asymmetric}, as it combines several opposing trends: the anisotropic dimensional reduction and the absence of the dimensional enhancement that favor pairing, while the mixing with the repulsive channels and the hard-core terms that limit the 2D Cooper-like divergence work against it. It is shown to disappear completely at $J_2\!=\!0.5$, with the ${\bf K}$ range of its existence contracting from the M point to the $\Gamma$ point.

The rest of the questions about mixing and relationships of different channels are addressed below in the course of our analysis and further in Sec.~\ref{Sec:dwave_GammaJ1J2}. All results below are for $S\!=\!1/2$ unless stated otherwise.

\subsubsection{$\Gamma$M line: two-magnon band transformation}
\label{Sec:GammaM_E_J1J2}

As before, the structure of the two-magnon continuum for the pair momentum ${\bf K}$ along the $\Gamma$M direction has to be analyzed first in order to see if it allows mixing of different interaction channels. 

 At the $\Gamma$ point, for small $J_2$, the minimum of the two-magnon band cannot change drastically from its position in the $J_1$-only limit and should remain at ${\bf p}^*\!=\!0$, as in Fig.~\ref{fig:EkqJ1}(a). At the M point, on the other hand, the dispersion of the two-magnon band is solely due to the $J_2$-term and has minima at ${\bf p}^*\!=\!\pm(\pi/2,-\pi/2)$; see Fig.~\ref{fig:MX_E_q*}(a). Clearly, these two limits should be interpolated as ${\bf K}\!=\!(K_x,K_x)$ moves from $\Gamma$ to M.  

The minimization of the two-magnon continuum energy in Eq.~\eqref{eq:2magJ1J2} for ${\bf K}$ along  the $\Gamma$M line yields the minimum located along the diagonal normal to the $\Gamma$M direction, ${\bf p}^*\!=\!\pm(q^*,-q^*)$, with a piecewise solution,
\begin{align}
\label{eq:GM_2mag_q*}
q^*&=
\left\{\begin{array}{ll} 
0, &    K_x \leq  K_x^*, \\
{\arccos\left(\frac{1}{2J_2}\,\cos \frac{K_x}{2}\right)}, &  K_x >  K_x^*,
\end{array} \right.
\end{align}
where $K_x^*\!=\!2\arccos (2J_2)$. The denominator of the two-particle SE in Eq.~(\ref{eq:M_gamma}) is
\begin{align}
\label{eq:MG_2mag_E}
&E-E_{\bf K}({\bf p})=-2\Delta-\Delta E_{\bf K}({\bf p}^*)   \\
&+8S\Big(\cos\frac{K_x}{2}\,\gamma_{\bf p}- J_2 \Big(\cos^2\frac{K_x}{2}\,\gamma^{(2)}_{\bf p}+\sin^2\frac{K_x}{2}\, \gamma_{\bf p}^{d_{xy}}\Big)\Big),\nonumber
\end{align}
where $\Delta E_{\bf K}({\bf p}^*)$ is a constant energy offset for the two-magnon energy minimum, given by
\begin{align}
\label{eq:MG_2mag_dE*1}
\Delta E_{\bf K}({\bf p}^*)=8S\left(\cos \frac{K_x}{2}-J_2\cos^2 \frac{K_x}{2}\right), 
\end{align}
for $K_x\! \leq\!  K_x^*$, and 
\begin{align}
\label{eq:MG_2mag_dE*2}
\Delta E_{\bf K}({\bf p}^*)=8SJ_2\left(1+\tan^2 \frac{K^*_x}{2}\,\cos^2 \frac{K_x}{2}\right),
\end{align}
for  $K_x\! >\!  K_x^*$.

\begin{figure}[t]
\includegraphics[width=\linewidth]{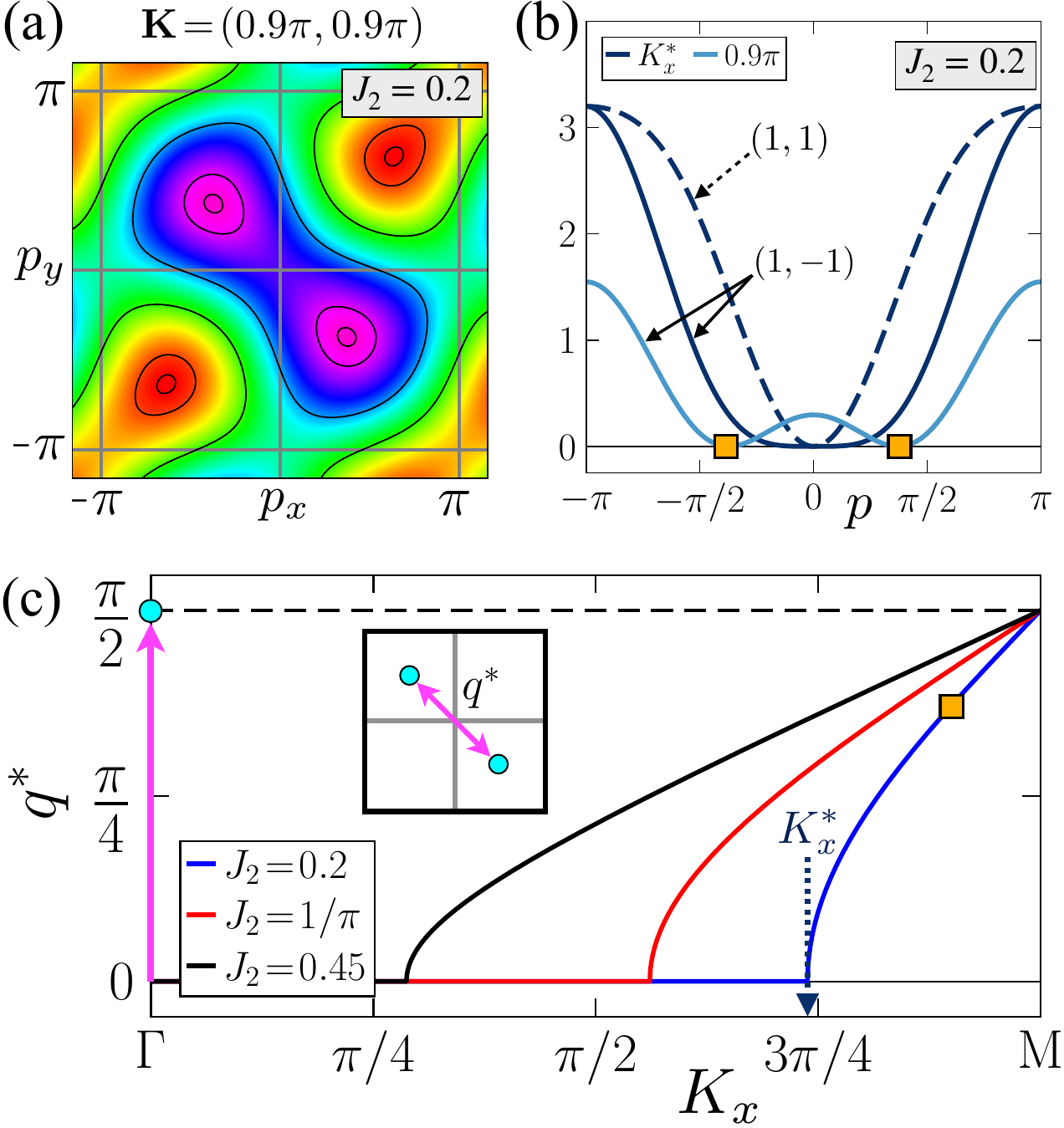}
\caption{(a) An intensity map of the two-magnon energy, Eq.~(\ref{eq:MG_2mag_E}), for $J_2\!=\!0.2$ and $K_x\!=\!0.9\pi$. (b) $E_{\bf K}({\bf p})-\Delta E_{\bf K}({\bf p}^*)$ for $J_2\!=\!0.2$ and for $K_x\!=\!0.9\pi$ and $K_x\!=\!K_x^*$ in the $(1,1)$ and $(1,-1)$ directions. (c) $q^*$ from Eq.~(\ref{eq:GM_2mag_q*}) as a function of $K_x$ for $J_2\!=$0.2, $1/\pi$, and 0.45. The square symbol in (b) and (c) marks $q^*$ for $J_2\!=\!0.2$ and $K_x\!=\!0.9\pi$. Inset: the direction of $q^*$ migration.}
\label{fig:MG_E_q*}
\end{figure}

Figure~\ref{fig:MG_E_q*} makes explicit the resultant shapes of the two-magnon energy band and its dependence on the pair momentum ${\bf K}$ and $J_2$. Figure~\ref{fig:MG_E_q*}(a) shows an intensity map of this band for a representative $J_2\!=\!0.2$ and the pair momentum close to the M point, ${\bf K}\!=\!(0.9\pi,0.9\pi)$. The band minima are at the incommensurate ${\bf p}^*$ along the $(1,-1)$ diagonal, Eq.~(\ref{eq:GM_2mag_q*}), and the asymmetric form of the band is evident. 

Figure~\ref{fig:MG_E_q*}(b) shows the two-magnon band dispersion $E_{\bf K}({\bf p})$ relative to the constant energy offset for the energy minimum $\Delta E_{\bf K}({\bf p}^*)$ from Eqs.~(\ref{eq:MG_2mag_dE*1}) and (\ref{eq:MG_2mag_dE*2}) for the same $J_2\!=\!0.2$ and for the two values of the pair momentum along the $\Gamma$M line, ${\bf K}\!=\!(0.9\pi,0.9\pi)$ and ${\bf K}\!=\!(K_x^*,K_x^*)$. The momentum $K_x^*$ corresponds to the pinch-point of the two-magnon band minima off the $\Gamma$ point for that $J_2$; see Fig.~\ref{fig:MG_E_q*}(c). 

In Fig.~\ref{fig:MG_E_q*}(b), the two-magnon energy for $K_x\!=\!0.9\pi$ has two incommensurate minima at ${\bf p}^*\!=\!\pm(q^*,-q^*)$ according to Eq.~(\ref{eq:GM_2mag_q*}), marked by the two square symbols. For $K_x\!=\!K_x^*$, the energy is plotted in two principal directions, along the $(1,1)$ and  $(1,-1)$ diagonals, demonstrating a non-parabolic $\propto\!p^4$ dispersion in the $(1,-1)$ direction, associated with the pinching-off of the band minima at this momentum. 

Figure~\ref{fig:MG_E_q*}(c) shows the evolution of the band-minimum momentum $q^*$ from Eq.~(\ref{eq:GM_2mag_q*}) as a function of the pair momentum $K_x$ along the $\Gamma$M line for three representative values of $J_2\!=$0.2, $1/\pi$, and 0.45. The square symbol marks the value of $K_x\!=\!0.9\pi$ on the $J_2\!=$0.2 curve corresponding to the energy band with two minima in Fig.~\ref{fig:MG_E_q*}(b). The inset in Fig.~\ref{fig:MG_E_q*}(c) shows the direction of the migration of the two-magnon band minima as the pair momentum moves from the $\Gamma$ point to the M point. For $J_2\!>$0.5, the two-magnon continuum experiences a transition to a new regime, as discussed in Sec.~\ref{Sec:dwave_GammaJ1J2}.

Importantly, for the entire $\Gamma$M line, the two-magnon band minima reside along the $(1,-1)$ diagonal, making the SE energy denominator in Eq.~(\ref{eq:MG_2mag_E}) invariant under a mirror reflection about the $p_x\!=\!p_y$ line. For the four lattice partial waves in Eq.~\eqref{eq:VKq_J1J2expansion}, only the $d$-wave harmonic is odd under this transformation; see Figs.~\ref{fig:HarmonicsJ1} and \ref{fig:HarmonicsJ2}. Because of that, the $d$-wave channel is decoupled from the rest of the harmonics and the upper BS branch in Fig.~\ref{fig:EkJ1J2} remains pure $d$-wave as in the $J_1$-only case. The same symmetry eliminates the hard-core term for the $d$-wave channel, $\gamma^-_{\frac{\bf K}{2}}=0$.

The three remaining harmonics are all coupled, forming a $3\times 3$ block of the eigenvalue matrix $\hat{\rm \bf M}_{\bf K}$; see App.~\ref{A:MmatrixGMJ1J2}. As discussed above, the $s_{xy}$ and $d_{xy}$ channels in this mix are repulsive, but the numerical phenomenology of Fig.~\ref{fig:EkJ1J2} suggests a survival of the lower, $s$-wave-like branch for values of $J_2$ exceeding $J_2^{c_1}\!=\!1/\pi$, the threshold for the disappearance of the BS at the M point

These two BS branches are discussed in the remainder of this section. 

\subsubsection{$\Gamma$M line: $d$-wave BS branch}
\label{Sec:GammaM_dwaveJ1J2}

The $d$-wave branch in the $J_1$-only model is present only in a finite range of the pair-momenta in the proximity of the M point, with the termination momentum ${\bf K}_c$ given in Eq.~(\ref{eq:kc}). It is free from the effects of the hard-core suppression of pairing, but is subject to the dimensional enhancement because of the nodes of its partial wave crossing the two-magnon band minimum, and it exists because the effective pair-mass diverges at the M point, as discussed in Sec.~\ref{Sec:GammaM_dwaveJ1}; see also Fig.~\ref{fig:EkJ1}. 

For a finite $J_2$, and unlike the upper branch along the MX line, the $d$-wave partial wave along the $\Gamma$M line preserves its nodal-crossing of the two-magnon band minima, because these minima are either at the $\Gamma$ point or along the $(1,-1)$ diagonal; see Fig.~\ref{fig:MG_E_q*} and discussion above. Therefore, the BS in this channel remains ``dimensionally enhanced'' for all $J_2\!<\!0.5$; see also Sec.~\ref{Sec:dwave_GammaJ1J2}. Because of that, it is natural to expect that the $d$-wave BS branch disappears at $J_2^{c_1}\!=\!1/\pi$, when the $J_2$-generated mass for the BS at the M point reaches a critical value for pairing in the higher partial-wave channel; see Sec.~\ref{Sec:M_point_J1J2}. This expectation is in agreement with the numerical evidence in Fig.~\ref{fig:EkJ1J2} and SM~\cite{SM}, which suggest a continuous contraction of the upper BS branch toward the M point as $J_2\!\rightarrow\!J_2^{c_1}$ and its complete disappearance together with the BSs at the M point, at $J_2\!=\!J_2^{c_1}$.

\begin{figure}[t]
\includegraphics[width=\linewidth]{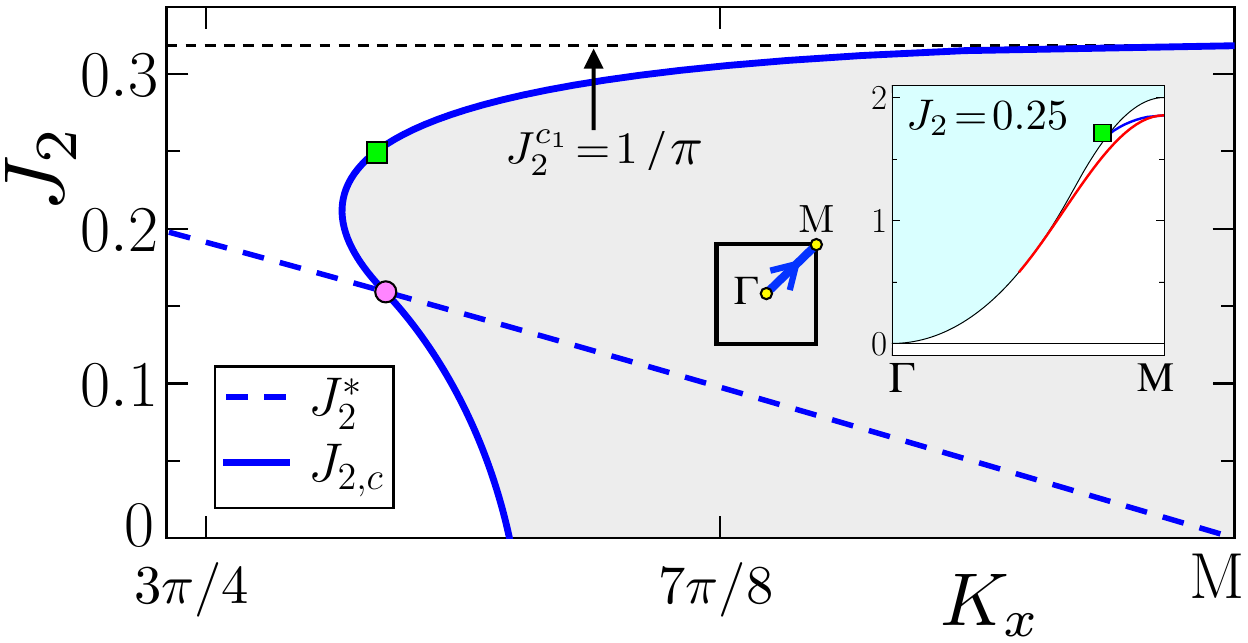}
\caption{The threshold boundary $J_{2,c}$ as a function of $K_x$ along the $\Gamma$M line for $S\!=\!1/2$ (solid line). The gray area marks the phase space for BS existence; the square corresponds to the one in the inset and the circle to the inflection point of the curve. The dashed line is for the $J_2^*$  as a function of $K_x$, see the text. Insets: the $\Gamma$M panel from Fig.~\ref{fig:EkJ1J2}(a) for $J_2\!=\!0.25$; the square highlights the termination point of the BS branch; the sketch of the BZ with the $\Gamma$M path.}
\label{fig:MG_Jc_vs_Kx}
\end{figure}

Although this expectation is correct, an intriguing non-monotonic behavior of the termination momentum ${\bf K}_c$ as a function of $J_2$ can be observed in Fig.~\ref{fig:MG_Jc_vs_Kx}, in which the $J_{2,c}$ BS boundary as a function of $K_x$ (solid curve) has a re-entrant shape. While it indeed collapses onto the M point for $J_2$ approaching $J_2^{c_1}$, its initial departure from its $J_1$-only value is {\it away} from the M point.  This curve is obtained by solving the eigenvalue problem for the $d$-wave channel with $\Delta\!=\!0$; see the detailed method description in Sec.~\ref{Sec:MX_symmetric} for the results related to Fig.~\ref{fig:MX_Jc_vs_Ky}. 

The non-monotonic behavior observed in Fig.~\ref{fig:MG_Jc_vs_Kx} can be traced to the transformation of the two-magnon band discussed above. Finite $J_2$ introduces anisotropy in the two-magnon band, providing a higher density of states in the low-energy sector and culminating in the pinching-off of the two minima from the $\Gamma$ point with a non-parabolic dispersion of the band in one of the principal momentum directions, as is shown in Fig.~\ref{fig:MG_E_q*}.  In Fig.~\ref{fig:MG_Jc_vs_Kx}, we also show the values of $J_2^*\!=\!\frac12\cos\frac{K_x}{2}$ at which such a  pinching-off occurs for a given $K_x$ (dashed curve). Its crossing with the threshold curve of $J_{2,c}$ for the BS boundary (solid curve) corresponds to the inflection point in the latter, where the effect of the described {\it anisotropic} dimensional reduction, partially compensating the effect of the dimensional enhancement for the $d$-wave channel, is maximal. 

\subsubsection{$\Gamma$M line: $s$-wave BS branch}
\label{Sec:GammaM_swaveJ1J2}

As discussed in Secs.~\ref{Sec:s_wave} and~\ref{Sec:0DBS}, the $s$-wave BS in the $J_1$-only model is super-exponentially shallow in the broad region near the $\Gamma$ point, but becomes deeper toward the M point where the hard-core effects diminish,  the effective pair-mass diverges, and the BS becomes effectively zero-dimensional; see Fig.~\ref{fig:EkJ1}.

At finite $J_2\!=\!J_2^{c_1}$, the BS at the M point disappears because the effective mass crosses the pairing threshold for the higher partial wave, which the $s$-wave is for that pair momentum ${\bf K}$; see Sec.~\ref{Sec:M_point_J1J2}. Away from the M point along the $\Gamma$M line, two repulsive channels, $s_{xy}$ and $d_{xy}$, mix with the attractive $s$-wave channel, forming the lower BS branch, whose evolution with $J_2$ can be observed in Fig.~\ref{fig:EkJ1J2} and in the SM~\cite{SM}. 

Formally, this lower BS branch for the $\Gamma$M line appears close in character to the upper BS branch for the MX line discussed in Sec.~\ref{Sec:MX_asymmetric}, because it is subject to similarly opposing trends.  Away from the M point, none of the contributing partial waves for this branch have nodal lines crossing the two-magnon band minima, suggesting 2D-like logarithmic divergences that favor pairing. However, hard-core terms limit the kinematic range for such a divergence in the attractive channel and mixing with the repulsive channels also works against the pair-formation.

The evidence from Fig.~\ref{fig:EkJ1J2} and SM~\cite{SM} suggests that this BS branch survives for $J_2$ above the critical $J_2^{c_1}$ for the M point, at least in a portion of the momentum space, where it manages to remain numerically detectable.

\begin{figure}[t]
\includegraphics[width=\linewidth]{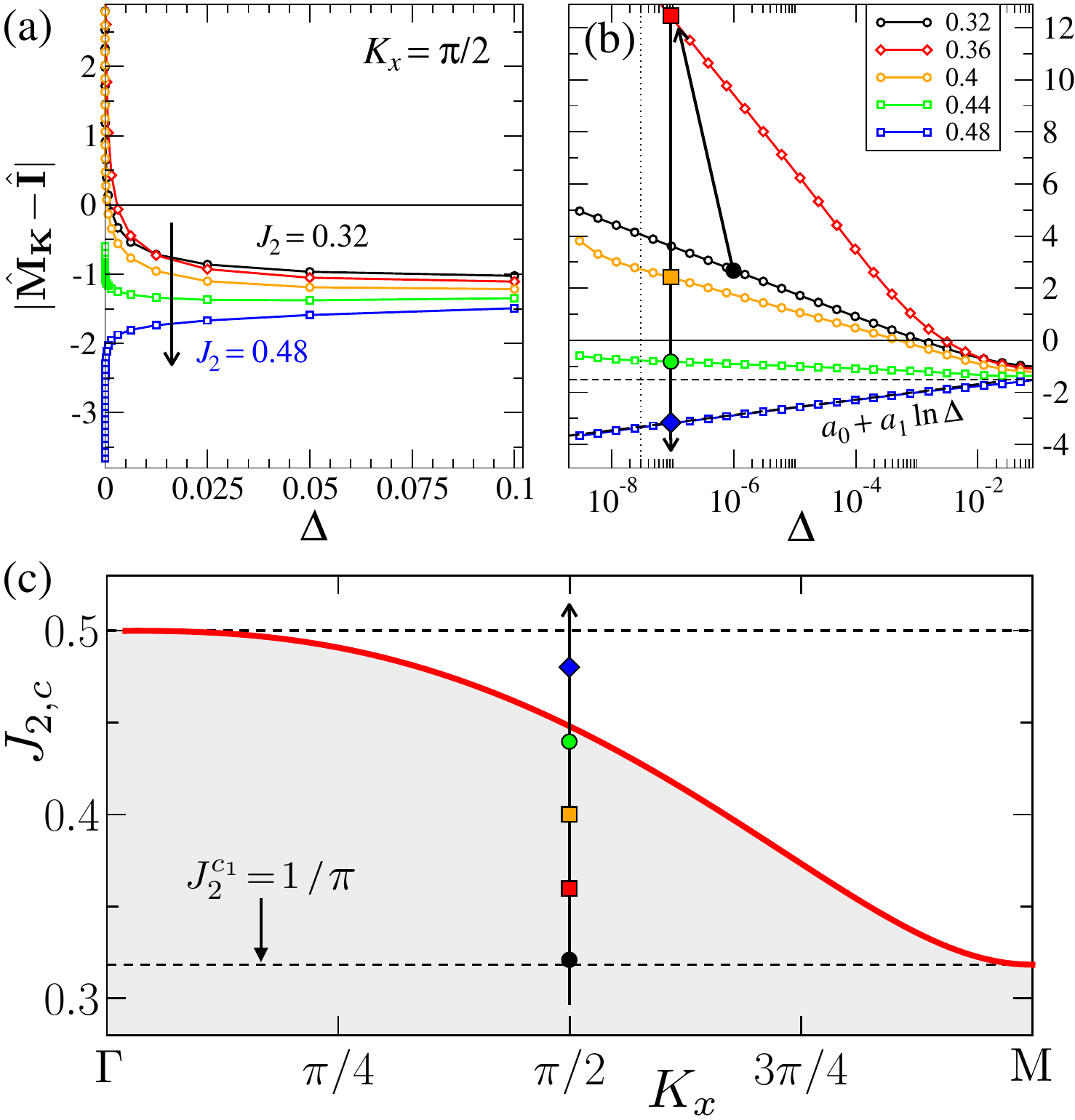}
\vskip -0.3cm
\caption{Same as in Fig.~\ref{fig:MX_AS_MvsDelta_log}, for the lower BS branch along the $\Gamma$M line. (a) and (b) $|\hat{\rm \bf M}_{\bf K}-\hat{\rm \bf I}|$ as a function of $\Delta$ for $K_x\!=\!\pi/2$, for several $J_2$ on the linear and semi-log plot, respectively. The solid symbols and arrows serve to cross-correlate with the data in (c). (c) The BS region (shaded) and the threshold boundary $J_{2,c}$ of this BS branch as a function of $K_x$.}
\label{fig:GM_ssd_MvsDelta_log}
\vskip -0.5cm
\end{figure}

Given that the lower BS branch can be shallow and may be numerically hard to detect, it requires the use of the approach described in Sec.~\ref{Sec:MX_asymmetric}, which allows one to study the sign of the divergence in the eigenvalue determinant to pinpoint whether the BS eigenvalue equation has a solution even if such a solution is beyond numerical reach. With the technical details of this approach discussed in the context of the results presented in Fig.~\ref{fig:MX_AS_MvsDelta_log}, our Fig.~\ref{fig:GM_ssd_MvsDelta_log} summarizes our findings for the lower BS branch along the $\Gamma$M line. 

Figures~\ref{fig:GM_ssd_MvsDelta_log}(a) and \ref{fig:GM_ssd_MvsDelta_log}(b) show the eigenvalue determinant, $|\hat{\rm \bf M}_{\bf K}-\hat{\rm \bf I}|$, for the lower BS branch, i.e., for its $3\times 3$ sector of the $s$, $s_{xy}$, and $d_{xy}$ partial waves (see App.~\ref{A:MmatrixGMJ1J2}), as a function of $\Delta$. The results are for a representative choice of pair-momentum $K_x\!=\!\pi/2$, several $J_2\!>\!J_2^{c_1}$, and on the linear and semi-log scale of $\Delta$, respectively. Similarly to the results for the upper BS branch along the MX line in Fig.~\ref{fig:MX_AS_MvsDelta_log}(b), the distinct logarithmic nature of the divergences of the determinant at small $\Delta$ is obvious in Fig.~\ref{fig:GM_ssd_MvsDelta_log}(b), and so is the change of its sign with $J_2$, indicating that the effective attraction changes to repulsion, underpinning the admixture of the repulsive channel as the key mechanism for suppressing this BS branch. 

The colored solid symbols and arrows in Figs.~\ref{fig:GM_ssd_MvsDelta_log}(a) and \ref{fig:GM_ssd_MvsDelta_log}(b) are used to cross-correlate with the data in Fig.~\ref{fig:GM_ssd_MvsDelta_log}(c), which shows the BS region of existence and the threshold boundary $J_{2,c}$ for the suppression of this BS branch as a function of $K_x$; see Sec.~\ref{Sec:MX_asymmetric} for the technical details of the method to obtain $J_{2,c}$. 

Overall, the ${\bf K}$-range of existence of the lower BS branch gradually contracts from the M point to the $\Gamma$ point as a function of $J_2$, as shown in Fig.~\ref{fig:GM_ssd_MvsDelta_log}(c), and disappears completely at $J_2\!=\!0.5$. However, this is not the full story, as is indicated by a non-monotonic behavior of the eigenvalue determinant in Figs.~\ref{fig:GM_ssd_MvsDelta_log}(a) and \ref{fig:GM_ssd_MvsDelta_log}(b) as a function of $J_2$. 

Before the pairing diminishes and the lower BS branch disappears because of the effective repulsion prevailing over the attraction, the binding energy at any given ${\bf K}$ along the $\Gamma$M line experiences a pairing {\it enhancement}. One can see this effect in Fig.~\ref{fig:GM_ssd_MvsDelta_log}(b), which shows that both the divergent slope of the eigenvalue determinant and the binding energy are {\it larger} for the $J_2$ that is farther away from $J_2^{c_1}$ than the minimal one (compare curves marked by the red square and black circle), indicating enhanced pairing in the former. This is followed by the monotonic decrease and eventual change of the sign of the slope of the eigenvalue determinant upon further increase of $J_2$, corresponding to a suppression of the BS. 

This effect is due to the {\it anisotropic} dimensional reduction discussed above, in which the pinching-off of the minima in the two-magnon band creates a non-parabolic dispersion in one of the principal momentum directions and is responsible for a higher density of states in the low-energy sector; see Fig.~\ref{fig:MG_E_q*}. 

While this effect is similar to the one described above for the $d$-wave channel, for the $s$-$s_{xy}$-$d_{xy}$ BS branch it modifies the divergence of the eigenvalue determinant for ${\bf K}\!\rightarrow\!{\bf K}^*$ from a logarithmic to a fractional power law. This law can be inferred from the asymptotic form of the two-magnon band that has a non-trivial $p^4$ dispersion in one of the two principal directions and a ``normal'' parabolic one in the other, as shown in Fig.~\ref{fig:MG_E_q*}(b), modifying the integral in the eigenvalue SE (\ref{eq:SCeq_Cooper}) to
\begin{equation}
\frac{1}{2N}\sum_{\bf p} \frac{1}{2\Delta+{\bf p}_1^2/m_2+{\bf p}_2^4/m_4}\propto\frac{m_2^{1/2}m_4^{1/4}}{\Delta^{1/4}}.
\label{eq:fractional}
\end{equation}
The resulting $1/\Delta^{1/4}$ divergence can be seen as intermediate between the standard 2D-like logarithmic and 1D-like $1/\sqrt{\Delta}$ laws discussed in Sec.~\ref{simplest}. 

This change of the type of the divergence in the eigenvalue determinant can be observed  directly for ${\bf K}\!=\!{\bf K}^*$ and we provide the corresponding plots in App.~\ref{A:MmatrixGMJ1J2}. We also note that the logarithmic behavior is restored fairly quickly for ${\bf K}$ away from ${\bf K}^*$. Thus, for ${\bf K}\!=\!\pi/2$ in Fig.~\ref{fig:GM_ssd_MvsDelta_log}(b), the value of $J_2^*$ at which the pinching-off of the minima occurs is $1/2\sqrt{2}\!\approx\!0.35355$, but already for the $J_2\!=\!0.36$ data set (marked by the red square), the divergence of the eigenvalue determinant, although enhanced, is clearly logarithmic. 

\subsection{The $d$-wave BS as a ground state}
\label{Sec:dwave_GammaJ1J2}

In this section, we discuss the appearance of the paired state as a true ground state of the $J_1$--$J_2$ model. As one can see in Figs.~\ref{fig:EkJ1J2}(c) and (d), as well as in the SM~\cite{SM}, the $d$-wave BS becomes the global energy minimum beyond a threshold value of $J_2^{c_2}$; see also Fig.~\ref{fig:DeltasJ1J2}. A transition to this regime is especially important for studies of the nature of spin-nematic phases, because it is the condensate of magnon BSs that corresponds to such a phase; see Ref.~\cite{NematicShengtao2023, Chubukov1991Nematic}.

We first discuss the appearance of such a paired ground state at the $\Gamma$ point, its transition through different regimes, and the reason for its stability for all $J_2\!>\!0.5$. Then we map out the region of existence of this $d$-wave branch in the pair-momentum space, as done for the other branches in Secs.~\ref{Sec:MXJ1J2} and \ref{Sec:GammaM}. 

\subsubsection{$\Gamma$ point}
\label{Sec:dwave_Gamma_pointJ1J2}

At $J_2\!=\!0.5$, the two-magnon continuum at the $\Gamma$ point undergoes a transition because the {\it single-magnon} bands, Eq.~(\ref{eq:ekJ1J2_1mag}), change from the FM-like, with the minimum at $(0,0)$, to the AFM-like, with the minima at $(\pi,0)$ and $(0,\pi)$ points. Precisely at $J_2\!=\!0.5$, these bands have cross-like lines of the degenerate minima along the $k_x$ and $k_y$ directions~\cite{Shannon_2004}, also leading to a completely flat minimum of the two-magnon continuum along the entire $\Gamma$MX$\Gamma$ contour in Fig.~\ref{fig:EkJ1J2}; see SM~\cite{SM}.

\begin{figure}[t]
\includegraphics[width=\linewidth]{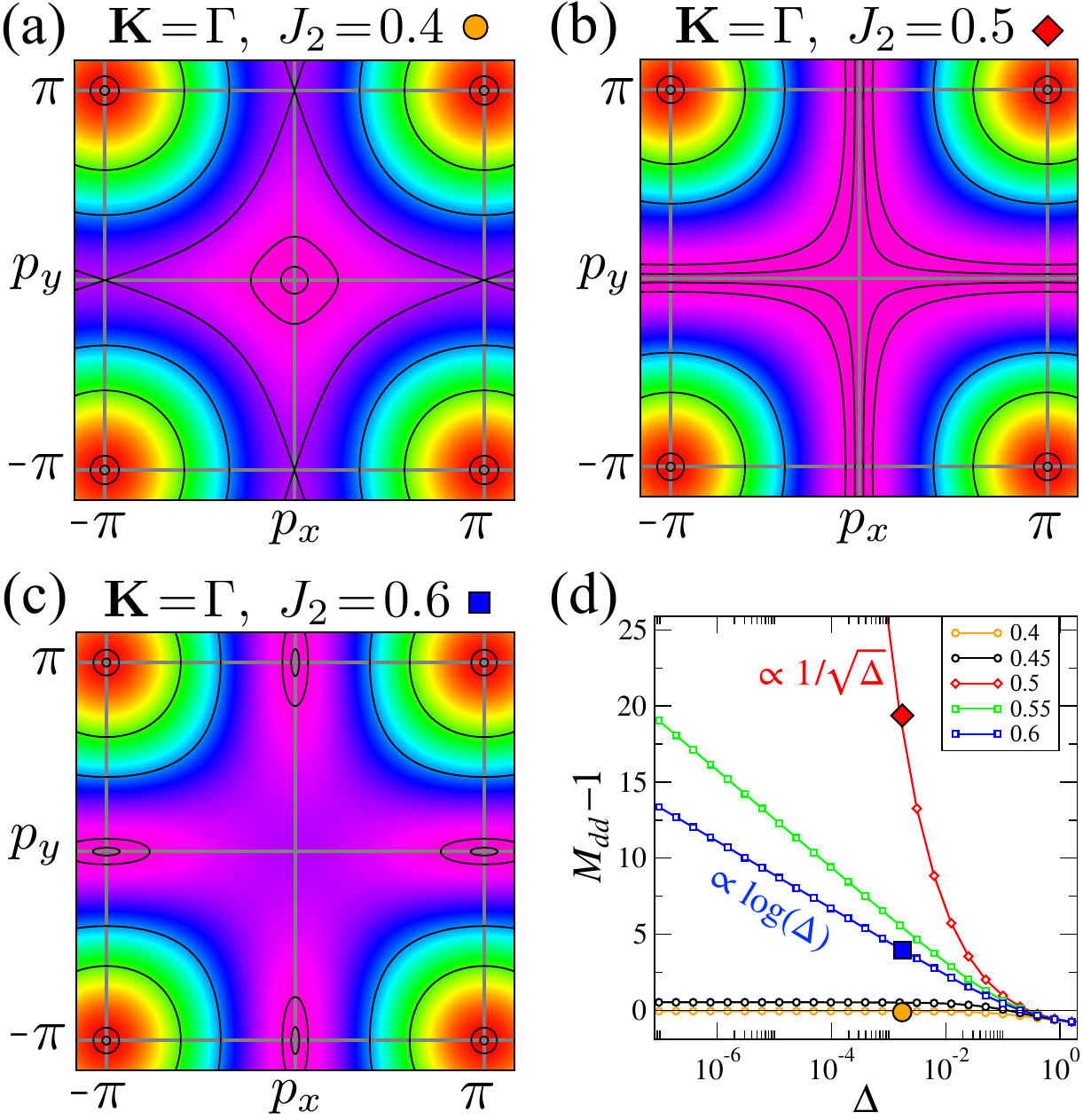}
\vskip -0.3cm
\caption{The intensity maps of the two-magnon energy bands at the $\Gamma$ point, $E_{\Gamma}({\bf p})$, for $J_2\!=\!0.4$,  0.5, and 0.6 in (a), (b), and (c), respectively. (d) The eigenvalue determinant, $M_{dd,\Gamma}-1$, as a function of $\Delta$ for $J_2\!=\!0.4$, 0.45, 0.5, 0.55, and 0.6. Characteristic behavior in three different regimes, $J_2\!<\! 0.5$, $J_2\!=\! 0.5$, and $J_2\!>\! 0.5$, is indicated; the solid symbols identify the data sets.}
\label{fig:GM_d_J2_0_5}
\vskip -0.4cm
\end{figure}

The corresponding transformation of the two-particle band for the pair-momentum ${\bf K}\!=\!\Gamma$ mirrors that of the single-particle bands, because $E_{\Gamma}({\bf p})\!=\!2\varepsilon_{{\bf p}}$. This transformation is exhibited in Figs.~\ref{fig:GM_d_J2_0_5}(a)-\ref{fig:GM_d_J2_0_5}(c), which show the intensity plots of $E_{\Gamma}({\bf p})$ relative to its minimum for $J_2\!=\!0.4$, 0.5, and 0.6, respectively. The developing 1D-like flatnesses in the $p_x$ and $p_y$ directions in Fig.~\ref{fig:GM_d_J2_0_5}(a) with $(0,\pi)[(\pi,0)]$ being the saddle points in the two-magnon band, the complete flatness in the same directions in Fig.~\ref{fig:GM_d_J2_0_5}(b), and the transition of the band minima to the $(\pi,0)[(0,\pi)]$ points in Fig.~\ref{fig:GM_d_J2_0_5}(c) are all made clear. 

The denominator of the two-particle SE in Eq.~(\ref{eq:M_gamma}) for ${\bf K}\!=\!\Gamma$ can be inferred from Eqs.~(\ref{eq:MG_2mag_E}), (\ref{eq:MG_2mag_dE*1}), and (\ref{eq:MG_2mag_dE*2})
\begin{equation}
\label{eq:G_2mag_E}
E-E_{\Gamma}({\bf p})\!=\!-2\Delta-\Delta E_{\Gamma}({\bf p}^*) +8S\big(\gamma_{\bf p}- J_2 \gamma^{(2)}_{\bf p}\big),
\end{equation}
with $\Delta E_{\Gamma}({\bf p}^*)$ being a constant energy offset 
\begin{align}
\label{eq:G_2mag_dE}
\Delta E_{\Gamma}({\bf p}^*)&=
\left\{\begin{array}{ll} 
8S\left(1-J_2\right),  &    J_2 \leq  0.5, \\
8SJ_2, &   J_2 >  0.5,
\end{array} \right.
\end{align}
where ${\bf p}^*\!=\!(0,0)$ for  $J_2\!\leq\! 0.5$ and ${\bf p}^*\!=\!(0,\pi)[(\pi,0)]$ for  $J_2\!>\! 0.5$, respectively.

As shown in Figs.~\ref{fig:EkJ1J2}(c) and~\ref{fig:EkJ1J2}(d), the $d$-wave BS branch reappears in the $J_1$--$J_2$ model spectrum for ${\bf K}$ at the $\Gamma$ point and in its vicinity already for $J_2\!<\!0.5$, that is, below the described band transformation. 

As discussed in Sec.~\ref{Sec:GammaM}, the $d$-wave partial wave is distinct from the rest of the harmonics in its symmetry, is free from admixtures of the repulsive channels, and is also not affected by the hard-core terms in the interaction for the entire $\Gamma$M line. It does not generally form a BS at the $\Gamma$ point because of the dimensional enhancement, as its nodes cross the band minimum at ${\bf p}^*\!=\!(0,0)$. This latter property remains true for $J_2\!\leq\! 0.5$, but the effective masses of the two-magnon band in the $p_x$ and $p_y$ directions diverge as $J_2\!\rightarrow\! 0.5$ [see Figs.~\ref{fig:GM_d_J2_0_5}(a) and \ref{fig:GM_d_J2_0_5}(b)], which suggests that a BS will be formed when the mass is sufficiently heavy, precipitating the appearance of the $d$-wave BS as a global energy minimum.

For $J_2\!=\!0.5$, Fig.~\ref{fig:GM_d_J2_0_5}(b) implies a different universality of the pairing problem, because the two-magnon band is 1D-like and the nodal crossing at the $\Gamma$ point is not effective at mitigating the associated divergence in the SE. In the $J_2\!>\! 0.5$ regime, exemplified in Fig.~\ref{fig:GM_d_J2_0_5}(c), the two-magnon band minima are at the ${\bf p}^*\!=\!(0,\pi)[(\pi,0)]$ points, so the $d$-wave nodal structure ceases to lead to a dimensional enhancement, making this partial wave equivalent to the one for the textbook 2D Cooper-like ``$s$-wave'' solution; see  Sec.~\ref{simplest} and Ref.~\cite{NematicShengtao2023}.

Below we study these pairing regimes for the $d$-wave BS at the $\Gamma$ point. 
Because the $d$-wave pairing channel is decoupled from the rest, its eigenvalue equation is just $M_{dd,\Gamma}\!=\!1$, with the matrix element from Eq.~(\ref{eq:M_gamma}), the coupling constant and partial wave harmonics from Eq.~(\ref{eq:V_J1only}), and the energy denominator from Eq.~(\ref{eq:G_2mag_E}).

Since in the $J_2\!\leq\! 0.5$ regime the $d$-wave channel is still dimensionally enhanced, its eigenvalue matrix element is independent of $\Delta$ for small enough $\Delta$---the behavior clearly exhibited in Fig.~\ref{fig:GM_d_J2_0_5}(d) by the $J_2\!=\!0.4$ and $0.45$ data sets. Therefore, the eigenvalue equation solved for $\Delta\!=\!0$ should yield the threshold value of $J_2^{c_2}$ at which this BS first occurs at the $\Gamma$ point. The corresponding 2D integral in the matrix element $M_{dd,\Gamma}$ can be evaluated analytically for $\Delta\!=\!0$ and $J_2\!\leq\! 0.5$ to yield a somewhat cumbersome combination of elementary functions of $J_2$ presented in App.~\ref{A:dwaveGammaMJ1J2}. The resulting equation for $J_2$ is solved numerically, giving a transition point into the paired ground-state regime for $S\!=\!1/2$ at 
\begin{align}
J_2^{c_2}\approx 0.4077593304754\dots,
\label{eq:J2c2}
\end{align}
which is consistent with the results in Fig.~\ref{fig:DeltasJ1J2} and Ref.~\cite{NematicShengtao2023}.

For $J_2\!=\!0.5$, the $d$-wave eigenvalue matrix element at the $\Gamma$ point can be evaluated in a compact form
\begin{equation}
M_{dd,\Gamma}=-1+\frac{\sqrt{\Delta+4}}{3\pi\sqrt{\Delta}}\Big(
\Delta K\left(\kappa\right)+(4-\Delta)E\left(\kappa\right)\Big),
\label{eq:Gamma_J2_0_5_exact_Mdd}
\end{equation}
where $\kappa\!=\!4/(\Delta+4)$ and $K$ and $E$ are complete elliptic integrals of the first and second kind, respectively. For small $\Delta$, the leading asymptotic behavior 
\begin{equation}
M_{dd,\Gamma}\sim\frac{8}{3\pi\sqrt{\Delta}},
\end{equation}
clearly points to the expected 1D-like divergence, which is also put on display by the $J_2\!=\!0.5$ data set in Fig.~\ref{fig:GM_d_J2_0_5}(d). The numerical solution of the eigenvalue equation with $M_{dd,\Gamma}$ from Eq.~\eqref{eq:Gamma_J2_0_5_exact_Mdd} gives
\begin{equation}
\Delta = 0.2727735681011576\dots,
\end{equation}
in agreement with the results in Fig.~\ref{fig:DeltasJ1J2} and Ref.~\cite{NematicShengtao2023}.

For the ``$s$-wave-like'' regime of $J_2\!>\!0.5$, the dimensional enhancement no longer suppresses pairing in the $d$-wave channel, with the eigenvalue matrix element expected to diverge logarithmically with $\Delta$, the behavior exhibited by the $J_2\!=\!0.55$ and 0.6 data sets in Fig.~\ref{fig:GM_d_J2_0_5}(d). 

For $J_2\!\gg\!1$, the asymptotic expression for the BS binding energy can be obtained analytically~\cite{NematicShengtao2023}. The effective mass near the two-magnon band minima, $(0,\pi)[(\pi,0)]$, becomes isotropic, $m^*\!\approx\!(4SJ_2)^{-1}$, and the pairing problem reduces to the 2D case in Eq.~\eqref{eq:SE_2D}
\begin{align}
\Delta_{\Gamma}\approx \frac{\Lambda}{2m^*}\, \exp\Big(\!\!-\!\frac{4\pi}{\alpha_d m^*}\Big)\propto J_2S\, \exp\Big(\!-\!2\pi S J_2\Big),
\label{eq:d_wave_asympt}
\end{align}
where the extra factor $1/2$ in the exponent compared to Eq.~\eqref{eq:SE_2D} originates from the two band minima at $(0,\pi)[(\pi,0)]$~\cite{NematicShengtao2023}. For smaller $J_2$, the effective masses at these minima become strongly anisotropic, signifying a transition to the 1D-like pairing regime at $J_2\!=\!0.5$, and the approximation in Eq.~\eqref{eq:d_wave_asympt} breaks down.

Thus, the appearance of the $d$-wave BS near the $\Gamma$ point for $J_2\!\alt\!0.5$ is not unlike the presence of the BS branch of the same symmetry for smaller $0\!<\!J_2\!<\!1/\pi$ near the M point, discussed in detail in Sec.~\ref{Sec:GammaM_dwaveJ1J2}. For the $d$-wave branch near the M point, the divergence of the effective pair-mass in the $J_1$-only model is removed by the $J_2$-induced dispersion, which eventually suppresses the pairing in this channel altogether, while for the $d$-wave branch near the $\Gamma$ point, the mass increase is due to the band transformation upon the crossing of $J_2\!=\!0.5$ described above. However, the difference between the evolution of the two branches is that the $d$-wave BS branch near the $\Gamma$ point survives this band transformation and emerges from it free from the dimensional enhancement, ensuring its stability and existence for {\it all} $J_2\!>\!0.5$.

Below we investigate the extent of this BS branch in the pair-momentum space. 

\subsubsection{$d$-wave BS branch}
\label{Sec:dwave_Kc_J1J2}

Below we mostly concentrate on the $\Gamma$M direction for the pair momentum ${\bf K}$, capitalizing on our detailed analysis of the BS evolution along this high-symmetry line from the $J_1$-only model for $0\!<\!J_2\!\leq\!0.5$ in Sec.~\ref{Sec:GammaM}. The $d$-wave harmonic remains orthogonal to the other partial waves for the entire $\Gamma$M line and for any $J_2$. Therefore, the $d$-wave BS branch that can be seen in Fig.~\ref{fig:EkJ1J2}(c) and in the SM~\cite{SM} should coexist (without mixing) with the super-shallow $s$-$s_{xy}$-$d_{xy}$ BS branch, discussed in Sec.~\ref{Sec:GammaM_swaveJ1J2}, for $J_2^{c_2}\!<\!J_2\!\leq\!0.5$ and even in a similar range of  ${\bf K}$.

As one can see in Figs.~\ref{fig:EkJ1J2}(c) and \ref{fig:EkJ1J2}(d), and in more detail in the SM~\cite{SM}, the ${\bf K}$-range for the $d$-wave BS branch along the $\Gamma$M direction expands rapidly as a function of $J_2$ from the nucleation value $J_2^{c_2}$ for the $\Gamma$ point, Eq.~(\ref{eq:J2c2}), but it always terminates at some finite $K_{x,c}$ and in a rather definitive manner, leaving no doubt that it does not extend beyond that ${\bf K}$ value. 

Such a behavior is natural for the $J_2\!\leq\!0.5$ range where the pairing in the $d$-wave channel remains dimensionally enhanced, so that a termination at a finite ${\bf K}$ can be related to the effective pair-mass crossing a pairing threshold.  However, given the discussion in Sec.~\ref{Sec:dwave_Gamma_pointJ1J2} for the $d$-wave pairing at the $\Gamma$ point, the persistence of this behavior for $J_2\!>\!0.5$ is puzzling.  If the $d$-wave harmonic does not mix with the repulsive channels and does not have the hard-core terms in its attractive interaction---both are true for the entire $\Gamma$M line---why would it cease to exist for $J_2\!>\!0.5$ where it should be a textbook $s$-wave-like 2D case for the pairing, which must lead to a BS as long as there is an attraction? 

\begin{figure}[t]
\includegraphics[width=\linewidth]{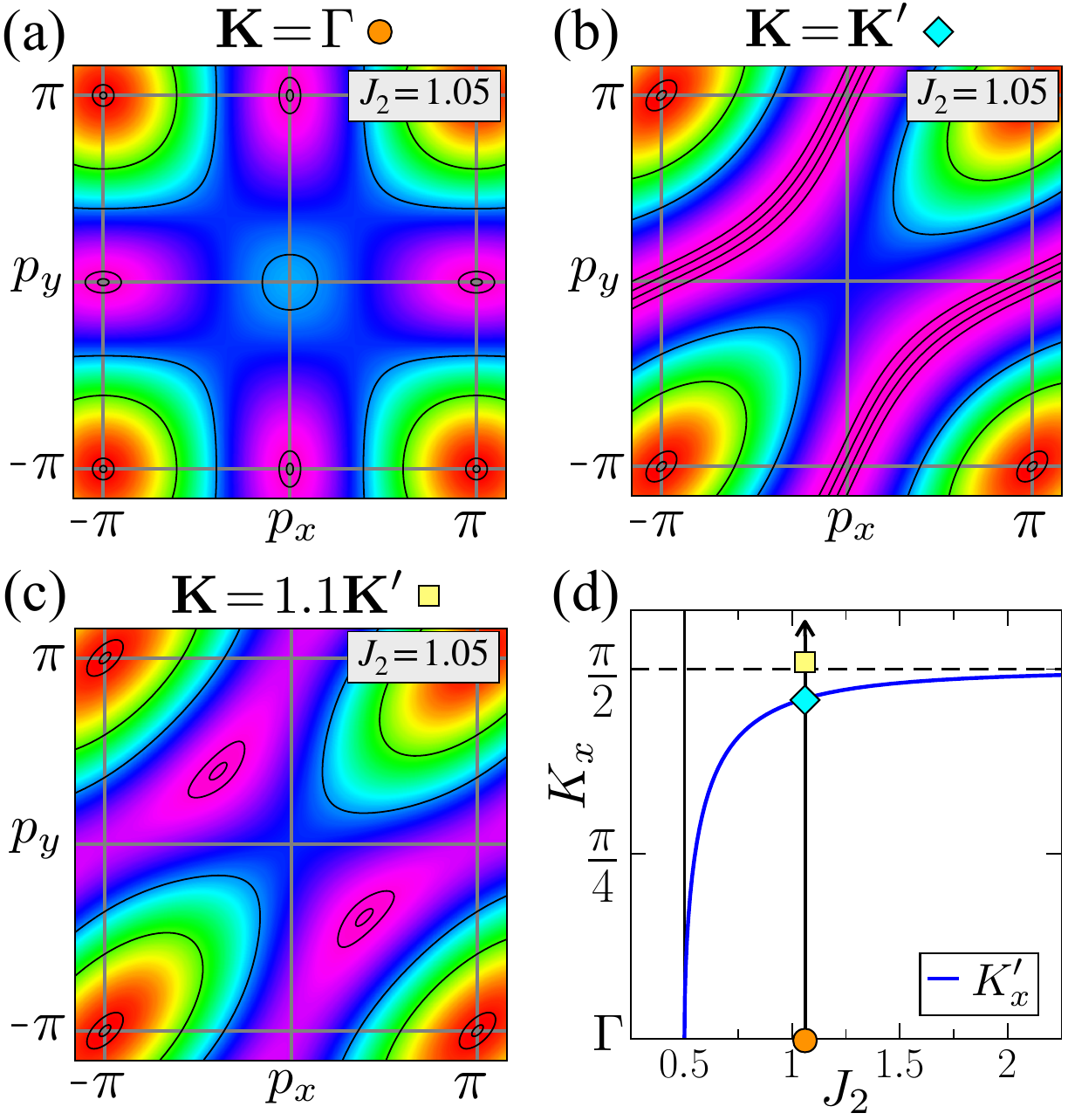}
\vskip -0.3cm
\caption{The intensity maps of the two-magnon energy band for $J_2\!=\!1.05$ at the (a) ${\bf K}\!=\!\Gamma$, (b) ${\bf K}\!=\!{\bf K}^{\prime}$, and (c) ${\bf K}\!=\!1.1{\bf K}^{\prime}$. (d) $K_x^{\prime}$ as a function of $J_2$. The arrow and solid symbols mark ${\bf K}$ values in (a)-(c) for $J_2\!=\!1.05$.}
\label{fig:GM_Ekq_largeJ2}
\vskip -0.4cm
\end{figure}

This question is addressed in our Fig.~\ref{fig:GM_Ekq_largeJ2}, which exposes the transformation of the two-magnon band as a function of the pair momentum ${\bf K}$ along the $\Gamma$M line for a representative $J_2$ from the  $J_2\!>\!0.5$ range ($J_2\!=\!1.05$).  

For $J_2\!>\!0.5$, the two-magnon band minima for ${\bf K}\!=\!\Gamma$ are at the $(0,\pi)[(\pi,0)]$ points; see Fig.~\ref{fig:GM_Ekq_largeJ2}(a) and Fig.~\ref{fig:GM_d_J2_0_5}(c) in Sec.~\ref{Sec:dwave_Gamma_pointJ1J2}. As a function of the pair momentum ${\bf K}$ along the $\Gamma$M line, the minima shift to ${\bf p}^*\!=\!\pm(q^*,-q^*)$ along the $(-1,1)$ diagonal with $q^*$ given in Eq.~(\ref{eq:GM_2mag_q*}); see Fig.~\ref{fig:GM_Ekq_largeJ2}(c) and Sec.~\ref{Sec:GammaM_E_J1J2}. 

The transition is at $K_x^{\prime}\!=\!\arccos\left(1/(8J_2^2-1)\right)$ via a two-magnon band transformation which is different from the one discussed in Sec.~\ref{Sec:GammaM_E_J1J2}. Instead, at ${\bf K}\!=\!{\bf K}^{\prime}$, the minima at the $(0,\pi)[(\pi,0)]$ points merge into the 1D-like channels shown in Fig.~\ref{fig:GM_Ekq_largeJ2}(b). Figure~\ref{fig:GM_Ekq_largeJ2}(d) shows $K_x^{\prime}$ as a function of $J_2$. The arrow and solid symbols mark ${\bf K}$ values in Figs.~\ref{fig:GM_Ekq_largeJ2}(a)-(c) for $J_2\!=\!1.05$.

Altogether, the resolution of the $d$-wave branch termination paradox is simple: it takes place only at ${\bf K}\!>\!{\bf K}^{\prime}$, where the two-magnon band topography is back to the regime in which the $d$-wave nodal lines cross its minima, making it subject to a dimensional enhancement. 

\begin{figure}[t]
\includegraphics[width=\linewidth]{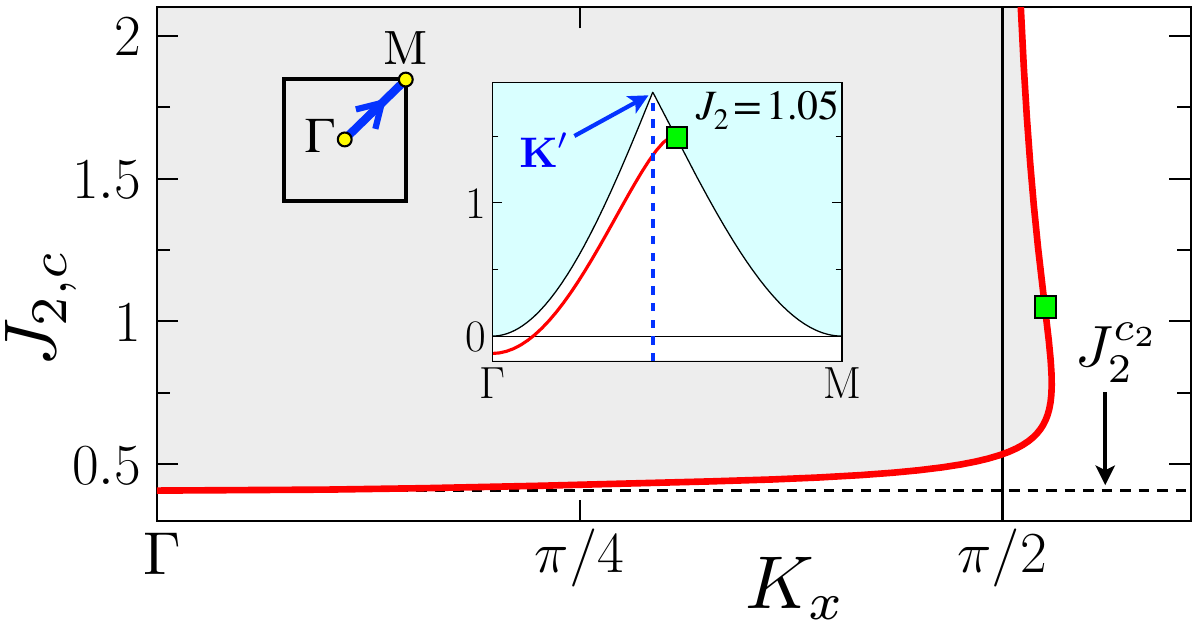}
\vskip -0.3cm
\caption{The threshold boundary $J_{2,c}$ as a function of $K_x$ along the $\Gamma$M direction for $S\!=\!1/2$ (solid line). The gray area marks the phase space for BS existence. Insets: the $\Gamma$M panel from Fig.~\ref{fig:EkJ1J2}(d) for $J_2\!=\!1.05$ where the dashed line marks ${\bf K}^{\prime}$ and the symbol highlights the termination point of the BS branch, mirrored in the main panel for the same $J_2$; the sketch of the BZ with the $\Gamma$M path.}
\label{fig:GM_Jc_vs_Kx_d_gs}
\vskip -0.4cm
\end{figure}

This analysis is accompanied by Fig.~\ref{fig:GM_Jc_vs_Kx_d_gs}, which shows the $J_2$--${\bf K}$ region of existence for the $d$-wave BS and its threshold boundary $J_{2,c}$ as a function of $K_x$. This boundary is obtained using the same approach as for Figs.~\ref{fig:MG_Jc_vs_Kx} and \ref{fig:MX_Jc_vs_Ky}; see Secs.~\ref{Sec:GammaM_dwaveJ1J2} and \ref{Sec:MX_symmetric} for technical details. The inset shows the $\Gamma$M panel from Fig.~\ref{fig:EkJ1J2}(d) for $J_2\!=\!1.05$ with the termination point highlighted by the solid square symbol, mirrored in the main panel for the same value of $J_2$. In Appendix~\ref{A:MmatrixGMJ1J2}, we combine Figs.~\ref{fig:MG_Jc_vs_Kx} and \ref{fig:GM_Jc_vs_Kx_d_gs} to emphasize the relative $J_2$--${\bf K}$ regions of existence for the $d$-wave BSs along the $\Gamma$M line. 

It is also important to note that before reaching its termination, the $d$-wave BS branch {\it must} go through a regime of enhanced pairing where the BS is 1D-like, owing to the degenerate lines of the two-magnon band minima at ${\bf K}\!=\!{\bf K}^{\prime}$ in Fig.~\ref{fig:GM_Ekq_largeJ2}(b). The value of ${\bf K}^{\prime}$ for $J_2\!=\!1.05$ is marked in the inset of Fig.~\ref{fig:GM_Jc_vs_Kx_d_gs} by the dashed line, where it is also clearly associated with the kink in the bottom of the two-magnon continuum energy, related to the switch of the two-magnon band minima described above. 

\subsubsection{$\Gamma$X direction}
\label{Sec:dwave_GX}

We make a few brief remarks on the BS spectrum evolution along the $\Gamma$X direction. The numerical phenomenology of Figs.~\ref{fig:EkJ1} and \ref{fig:EkJ1J2} and the SM~\cite{SM} shows the super-shallow BS branch continuously merging with the deep 1D-like BS at the X point for small $J_2$, followed by an emergence of the $d$-wave BS at the $\Gamma$ point at $J_2^{c_2}$ discussed above, with the $d$-wave branch also joining in and persisting as one continuous branch for all $J_2\!>\!J_2^{c_2}$. At $J_2^{c_3}\!=\!1.0$, the 1D-like BS at the X point is pushed out by the repulsive channel as discussed in Sec.~\ref{Sec:1DBSJ1J2}, while the BS branch continues to exist at least in some range of ${\bf K}$ at and away from the $\Gamma$ point along the $\Gamma$X line, similarly to the pure $d$-wave branch along the  $\Gamma$M direction, but with a different character of its termination, which is reminiscent of the cases of the super-shallow states considered in the previous sections. 

This phenomenology is entirely consistent with the symmetry considerations, which echo discussions of the mixing of the partial waves along the $\Gamma$X line in the $J_1$-only model, Sec.~\ref{Sec:XG_J1}, and at the X point for finite $J_2$, Sec.~\ref{Sec:1DBSJ1J2}. For ${\bf K}$ along the $\Gamma$X direction, the attractive $s$ and $d$ channels mix in some ${\bf K}$-dependent proportion, with no mixing at the $\Gamma$ point and fully symmetric mixing at the X point, while only one repulsive $s_{xy}$ channel adds to this mixture, with the $d_{xy}$ channel being fully orthogonal to it. All three channels are also affected by the hard-core terms in their interactions for the pair-momentum along the $\Gamma$X line.

This lack of orthogonality between the two attractive channels is responsible for the continuity of the transition from the $d$-wave-dominated BS branch near the $\Gamma$ point to the symmetric $s$- and $d$-wave mix of the 1D-like BS at the X-point via a shallow BS at the intermediate ${\bf K}$, shown in Fig.~\ref{fig:EkJ1J2}(c) for $J_2\!=\!0.45$ and also in the SM~\cite{SM} for other values of $J_2$. Since the BS at the X point is eventually eliminated by the admixture of the repulsive $s_{xy}$ channel, the same mechanism is expected to control the termination of this mixed branch away from the X point for $J_2\!>\!1.0$; see Fig.~\ref{fig:EkJ1J2}(d). 

The only intriguing question that remains for the $\Gamma$X direction is whether there is a second BS branch in the range of $J_2^{c_2}\!<\!J_2\!<\!0.5$, which is super-shallow and may be hard to detect numerically. This is because there are two orthogonal BS branches along the $\Gamma$M line in that same range of $J_2$, one $d$-wave and one super-shallow  $s$-$s_{xy}$-$d_{xy}$; see Secs.~\ref{Sec:GammaM_swaveJ1J2} and \ref{Sec:dwave_Kc_J1J2}. For the $\Gamma$X line, the partial-wave mixture is different, so the question remains whether there is a surviving second BS branch, or it is pushed into the continuum by the anti-bonding mixing. 

This question is addressed with the help of the analysis of the $3\times 3$ sector of the eigenvalue determinant $|\hat{\rm \bf M}_{\bf K}-\hat{\rm \bf I}|$ for the mixed $d$-$s$-$s_{xy}$ BS branch as a function of $\Delta$ for several representative values of pair-momentum ${\bf K}$ along the $\Gamma$X line and for $J_2$ from the $J_2^{c_2}\!<\!J_2\!<\!1.0$ range and also for $J_2\!>\!1.0$. The technical details of this analysis are outlined in Secs.~\ref{Sec:MX_asymmetric} and \ref{Sec:GammaM_swaveJ1J2}. With this analysis we confirm that there is at most one BS branch that is present at any $J_2$ for ${\bf K}$ along the $\Gamma$X line.

\subsection{Summary of the BSs in $J_1$--$J_2$ model}
\label{Sec:Summary_J1J2}

In this section, we have traced the evolution of the two-magnon BSs of the square-lattice ferromagnet upon introducing the antiferromagnetic $J_2$ interaction. The resulting transformation is considerably richer than a simple weakening of the $J_1$-driven attraction. The $J_2$ term acts in two distinct ways: it reshapes the single-magnon band and, consequently, the two-magnon continuum, removing many of the accidental degeneracies responsible for the anomalously deep BSs of the $J_1$-only model, while introducing the repulsive $s_{xy}$- and $d_{xy}$-wave interaction channels. Therefore, the evolution and eventual disappearance of the different BS branches can be understood as a competition between changes in the effective dimensionality of the pairing problem and changes of the effective interaction from attractive to repulsive.

The contrasting behaviors at the M and X points provide particularly transparent examples of these two mechanisms. At the M point, a finite $J_2$ lifts the complete flatness of the $J_1$-only continuum but preserves enough symmetry to keep the two $s$- and $d$-wave BSs degenerate. Their eventual disappearance at $J_2^{c_1}\!=\!1/\pi$ for $S\!=\!1/2$ is not caused by the repulsive $J_2$ channels. Instead, the new continuum minima lie precisely on the nodes of both partial waves, converting the problem from the 0D universality of the $J_1$-only limit to a dimensionally enhanced 2D problem with a finite threshold for binding. At the X point, by contrast, the 1D form of the two-magnon continuum is completely unaffected by $J_2$, making the corresponding BS remarkably robust. It survives up to $J_2^{c_3}\!=\!1$, where its disappearance is instead driven by mixing with the repulsive $s_{xy}$ channel, which reverses the sign of the $1/\sqrt{\Delta}$ pairing divergence.

The evolution along the high-symmetry directions combines these mechanisms in less trivial ways. Along the MX line, the lower branch continuously interpolates between the dimensionally enhanced M-point problem and the dimensionally reduced X-point one, with its suppression controlled by both the nodal crossing of the continuum minima and the increasing admixture of the repulsive channel. The upper branch is more subtle: because its relevant partial waves avoid the continuum minima, it retains the conventional 2D logarithmic pairing divergence and survives for $J_2$ above $J_2^{c_1}$, even outliving the lower branch over a finite momentum interval, although with a very small binding energy.

Along the $\Gamma$M direction, symmetry keeps the $d$-wave channel decoupled from the repulsive harmonics. However, its nodal crossing of the continuum minima preserves dimensional enhancement and ultimately causes this branch to disappear at $J_2^{c_1}$. The nonmonotonic boundary of its region of existence is a consequence of an anisotropic dimensional reduction associated with the pinching-off of the continuum minima and the resulting quartic dispersion in one momentum direction. The same effect strengthens the $s$-wave-like branch, locally changing its pairing divergence from the conventional logarithm to $1/\Delta^{1/4}$, before repulsive-channel mixing suppresses it completely at $J_2\!=\!0.5$.

A qualitatively new evolution is marked by the emergence of the $d$-wave BS at the $\Gamma$ point at $J_2^{c_2}\!\simeq\! 0.4078$, where it becomes the global energy minimum and provides the two-magnon instability relevant to the spin-nematic state. For $J_2\!<\!0.5$, this state is still subject to dimensional enhancement, but is stabilized by the increasing effective mass of the continuum. At $J_2\!=\!0.5$, the continuum acquires 1D lines of minima, while for $J_2\!>\!0.5$ its minima move away from the nodes of the $d$-wave harmonic. Then, from the viewpoint of its pairing kinematics, the nominally higher-partial-wave problem becomes equivalent to the conventional 2D $s$-wave problem, explaining the continual existence of the $d$-wave BS for arbitrarily large $J_2$ at the $\Gamma$ point. Away from the $\Gamma$ point, the $d$-wave branch eventually terminates when the evolving continuum minima cross its nodes again, restoring dimensional enhancement.

Finally, this analysis has provided a useful methodological tool for identifying BS phase boundaries even when the binding energies are far beyond direct numerical resolution. Rather than attempting to resolve a super-exponentially small $\Delta$, one can determine the existence of a BS from the leading small-$\Delta$ behavior of the eigenvalue determinant and, specifically, from the sign of its logarithmic or other lower-dimensional divergence. This approach unequivocally establishes that two BS branches can coexist, including over overlapping momentum regions, up to $J_2\!=\!0.5$, whereas for $J_2\!>\!0.5$, there is at most one BS branch at a given pair-momentum, and substantial regions of the BZ contain no BS at all. At large $J_2$, the only surviving branch is the one connected continuously to the $d$-wave paired ground state.

Thus, the seemingly complicated rearrangement of the BS spectrum with $J_2$ follows from the same basic ingredients identified in the $J_1$-only problem---partial-wave symmetry and dimensional reduction or enhancement---supplemented by the competing repulsive interaction channels generated by the AFM exchange.

\section{Summary}
\label{Sec:Summary}

The principal message of this work is that magnon bound states, despite the many peculiarities associated with spin algebra and lattice kinematics, do not require a conceptual framework that is fundamentally different from other two-particle pairing problems. Their phenomenology can be considerably more intricate, but the origins of this complexity can be traced systematically to familiar ingredients: the symmetry and partial-wave character of the pair wavefunction, the structure of the two-particle continuum, hard-core repulsion and its non-potential-like manifestations, and attractive or repulsive interactions in the relevant pairing channels.

Taken together, these ingredients produce a broad range of behaviors---from super-shallow to anomalously deep BSs, from conventional 2D pairing to effective 1D and 0D universalities, and from robust pairing to its complete suppression---while remaining fully consistent with the broader physics of two-particle bound states on a lattice. This resolves much of the apparent disconnect between magnon pairing and the conventional two-body problems in the existing literature.

An important part of this conclusion concerns the formal description of the problem itself. We have demonstrated that, for the fully polarized Heisenberg model on a Bravais lattice, the formally exact treatment in the non-orthogonal basis of spin flips and the standard $1/S$ spin-wave formulation in the orthogonal bosonic basis lead to the identical two-magnon Schr\"odinger equation. Thus, the overcompleteness of the bosonic Hilbert space has no bearing on the two-magnon BS problem, and the hard-core constraint does not invalidate the spin-wave approach. Instead, the effects of the spin algebra are encoded in the characteristic non-potential-like terms of the magnon interaction. Aside from providing a simpler and more transparent formulation, the bosonic approach has the important advantage of being naturally extendable to fluctuating states where an exact construction based on spin-flip states is generally unavailable.

This formulation also makes transparent the connection to conventional pairing. For a fixed total pair momentum $\mathbf{K}$, the two-magnon problem can be viewed as an effective one-particle problem for the relative motion, with the two-magnon continuum playing the role of its kinetic-energy band and the magnon-magnon vertex providing the effective interaction. On a lattice, neither ingredient needs to resemble its continuum counterpart: the effective band can develop accidental degeneracies and highly anisotropic minima, while the interaction depends not only on transferred momentum but also on the incident momenta. Nevertheless, the corresponding Schr\"odinger equation retains the familiar structure of a pairing problem. Its expansion in lattice partial waves provides a particularly natural language for separating symmetry channels, identifying their orthogonalities and mixings, and understanding whether the low-energy pairing falls into an $s$-wave-like universality or is suppressed by the nodal structure of a higher partial wave.

The nearest-neighbor square-lattice ferromagnet provides an especially instructive demonstration of this viewpoint. What superficially appears to be an assortment of unrelated anomalies is instead a convolution of several well-defined mechanisms. Near the $\Gamma$ point, the non-potential-like, hard-core terms suppress the effective attraction in the $s$-wave channel as $\mathbf{K}\!\rightarrow\!0$, producing the familiar but physically elusive super-exponentially shallow BS. The higher $d$-wave channel, on the other hand, is subject to dimensional enhancement and therefore requires a finite pairing threshold.

At large pair momenta, the situation is reversed by accidental degeneracies of the two-magnon continuum. Its complete flatness at the M point turns the pairing problem effectively zero-dimensional and produces deep, local, degenerate BSs, while its one-dimensional form along the MX line generates the corresponding 1D-like enhancement of binding. Thus, the anomalously deep states do not signal an exotic form of magnon attraction; they are consequences of the unusual effective kinetic energy of the pair. This also demonstrates why simple BS-counting rules based solely on the spatial dimensionality of the underlying lattice cannot be universal.

The addition of the antiferromagnetic $J_2$ coupling provides a stringent test of this interpretation, because it modifies both sides of the pairing problem. It makes the magnon band and the two-magnon continuum more generic, lifting many of the accidental degeneracies of the $J_1$-only model, while simultaneously introducing repulsive $s_{xy}$- and $d_{xy}$-wave interaction channels. The resulting evolution of the BS spectrum is controlled by two complementary mechanisms: dimensional reduction or enhancement originating from the structure of the continuum, and a change of the effective interaction from attractive to repulsive caused by the mixing of partial-wave channels.

Their interplay explains why some of the deep states inherited from the $J_1$-only model are rapidly destroyed whereas others remain remarkably resilient, and why substantial portions of momentum space eventually become completely free of BSs. It also explains the otherwise counterintuitive result that, although two BS branches can coexist up to $J_2\!=\!0.5$, for $J_2\!>\!0.5$ there is at most one BS branch at a given momentum.

Perhaps the most consequential example is the evolution of the $d$-wave BS into the true two-magnon ground state. Its appearance at the $\Gamma$ point is initially enabled by the increasing effective mass of the two-magnon continuum, despite the dimensional enhancement associated with the $d$-wave nodal structure. At $J_2\!=\!0.5$, the underlying magnon-band reconstruction changes the pairing kinematics qualitatively. For $J_2\!>\!0.5$, the continuum minima move to the antiferromagnetic wavevectors, away from the nodes of the $d$-wave harmonic. Consequently, the nominally higher-partial-wave pairing becomes effectively equivalent to the 2D $s$-wave problem and is guaranteed for arbitrarily weak attraction.

The same symmetry that keeps this channel free from the repulsive $J_2$ harmonics also eliminates the hard-core contribution to its interaction. These two properties account for the stability of the $d$-wave BS at large $J_2$ and provide a transparent two-particle origin for the instability associated with the spin-nematic state. Therefore, the dramatic transformation from two finite-energy BS branches in the ferromagnet to a single branch becoming the global energy minimum can be understood as a direct consequence of the restructuring of the pairing kinematics, not as a fundamentally new pairing mechanism.

Another outcome of this study is methodological. In two dimensions, exponentially shallow BSs can rapidly become inaccessible to a direct numerical determination of their binding energies, making the apparent disappearance of a branch an unreliable criterion for the actual loss of pairing. We have shown that the analysis of the small-$\Delta$ singularities of the eigenvalue determinant provides a substantially more powerful diagnostic.

The existence of a BS is controlled by the sign of the leading logarithmic divergence in the generic 2D case, while analogous criteria govern the $1/\sqrt{\Delta}$ and $1/\Delta^{1/4}$ divergences generated by dimensional reduction. Tracking the sign of these singular terms allows the BS boundaries to be determined even when the corresponding binding energies are many orders of magnitude below direct numerical resolution. This approach was essential for establishing the fate of several super-shallow branches and should be applicable more generally in lattice pairing problems near a BS threshold and for determining phase boundaries for various bound states.

The broader lesson is not that magnon BSs constitute an exceptional class of two-particle states, but rather that lattice pairing permits a substantially richer realization of otherwise familiar principles. Symmetry can project out hard-core terms or prevent mixing with repulsive channels; nodal structures can either suppress pairing through dimensional enhancement or become irrelevant when band minima move away from them; and accidental or symmetry-induced band degeneracies can lower the effective dimensionality of the problem and dramatically strengthen pairing.

The lattice partial-wave formulation brings these effects into a common framework and allows the complex evolution of magnon BSs throughout momentum space to be understood in terms of a small number of physical mechanisms. We expect this perspective to be useful well beyond the square-lattice $J_1$--$J_2$ model, particularly in frustrated magnets where competing exchanges naturally generate both unconventional magnon-band structures and higher-partial-wave pairing channels, and where the stability of multi-magnon bound states is central to the search for spin-multipolar phases, altogether paving the way to a realization of the robust spin nematic states.

\begin{acknowledgments}
Useful conversations with Mike Zhitomirsky and Karlo Penc are gratefully acknowledged.
The use of AI (Claude and ChatGPT) has been limited to proofreading the draft to improve grammar and style and taking a few particularly cumbersome integrals.
This work was supported by the U.S. Department of Energy, Office of Science, Basic Energy Sciences under Award No.~DE-SC0021221. 
C.~A.~G. was supported by the UCI-LANL-SoCal Hub Graduate Fellowship program and by the Eddleman Quantum Institute at UCI.
A.~L.~C.  would like to thank Aspen Center for Physics where some of this work was advanced. The Aspen Center for Physics is supported by National Science Foundation Grant No. PHY-2210452.
\end{acknowledgments}

\vskip -0.5cm
\section*{Data availability}
\vskip -0.5cm
The data that support the findings of this article are openly
available~\cite{dataset}.

\appendix

\section{BS periodicity} \label{A:BS_Periodicity}
As mentioned in the main text, the periodicity of the BS dispersion requires some clarification. This is because the center-of-mass coordinate of the BS, $({\bf r}_i+{\bf r}_j)/2$, can take values that do not belong to the original Bravais lattice. Therefore, one might naively expect the BS to have a different momentum periodicity. However, the BS is described by both its center-of-mass coordinate and the relative position of the pair, ${\bf r}_j-{\bf r}_i$. Consequently, the center-of-mass coordinate does not define an independent, finer lattice for the BS.

It is also intuitive to think about the momentum of each individual particle, ${\bf k}_{1(2)}\!=\!{\bf K}/2\pm {\bf q}$, so that the total and relative momenta are, respectively, ${\bf K}\!=\!{\bf k}_1+{\bf k}_2$ and ${\bf q}\!=\!({\bf k}_1\!-\!{\bf k}_2)/2$. A shift ${\bf k}_1\!\rightarrow\!{\bf k}_1+{\bf G}$ by a reciprocal vector ${\bf G}$ of the original lattice yields the simultaneous transformation $({\bf K},{\bf q})\!\rightarrow\!({\bf K}+{\bf G},{\bf q}+{\bf G}/2)$. Below, we show that the spectrum of the two-magnon Schr\"odinger equation, and therefore the BS energy, is invariant under this transformation.

The two-magnon SE~\eqref{eq:2MagSE} for a total momentum ${\bf K}\!+\!{\bf G}$ and a relative momentum ${\bf q}\!+\!{\bf G}/2$ reads
\begin{align}
\big(E &\!-\! \varepsilon_{\frac{{\bf K}}{2}+{\bf q}+{\bf G}} \!-\! \varepsilon_{\frac{{\bf K}}{2}-{\bf q}}\big) \psi_{{\bf K}+{\bf G}}({\bf q}+\tfrac{\bf G}{2})\!= \nonumber \\
&\qquad \qquad \frac{1}{2N}\! \sum_{\bf p}
V_{{\bf K}+{\bf G}}({\bf q}+\tfrac{\bf G}{2},{\bf p}) \psi_{{\bf K}+{\bf G}}({\bf p}).
\label{eqA:SE_EVPshift}
\end{align}
Using the periodicity of the single-magnon spectrum, $\varepsilon_{\frac{{\bf K}}{2}+{\bf q}+{\bf G}}\!=\!\varepsilon_{\frac{{\bf K}}{2}+{\bf q}}$, and introducing the change of variable ${\bf p}'\!=\!{\bf p}\!-\!{\bf G}/2$ in the sum on the right-hand side yields
\begin{align}
\big(E &\!-\! \varepsilon_{\frac{{\bf K}}{2}+{\bf q}} \!-\! \varepsilon_{\frac{{\bf K}}{2}-{\bf q}}\big) \psi_{{\bf K}+{\bf G}}({\bf q}\!+\!\tfrac{\bf G}{2})\!= \nonumber \\
&\quad \frac{1}{2N}\! \sum_{{\bf p}'}
V_{{\bf K}+{\bf G}}({\bf q}\!+\!\tfrac{\bf G}{2},{\bf p}'\!+\!\tfrac{\bf G}{2}) \psi_{{\bf K}+{\bf G}}({\bf p}'\!+\!\tfrac{\bf G}{2}).
\label{eqA:SE_VPshift}
\end{align}
The equation above can be simplified by noting that the interaction becomes
\begin{align}
V_{{\bf K}+{\bf G}}\left({\bf q}\!+\!\tfrac{\bf G}{2},
{\bf p}'\!+\!\tfrac{\bf G}{2}\right)=V_{\bf K}({\bf q},{\bf p}'),
\end{align}
where we have used $V_{\bf K}({\bf q},{\bf p})\! =\! {\cal J}_{{\bf q}-{\bf p}}\!+\!{\cal J}_{{\bf q}+{\bf p}}\!-\!{\cal J}_{\frac{{\bf K}}{2}-{\bf q}}\!-\!{\cal J}_{\frac{\bf K}{2}+{\bf q}}$ from Eq.~\eqref{eq:2MagVKqp} and ${\cal J}_{{\bf k}+{\bf G}}\!=\!{\cal J}_{\bf k}$. The latter follows from $e^{-i{\bf G}{\bm \delta}}\!=\!1$ for any lattice vector ${\bm \delta}$ of the Bravais lattice. Therefore, Eq.~\eqref{eqA:SE_VPshift} becomes
\begin{align}
\big(E &\!-\! \varepsilon_{\frac{{\bf K}}{2}+{\bf q}} \!-\! \varepsilon_{\frac{{\bf K}}{2}-{\bf q}}\big) \psi_{{\bf K}+{\bf G}}({\bf q}\!+\!\tfrac{\bf G}{2})\!= \nonumber \\
&\qquad \qquad\qquad \frac{1}{2N}\! \sum_{{\bf p}'}
V_{{\bf K}}({\bf q},{\bf p}') \psi_{{\bf K}+{\bf G}}({\bf p}'\!+\!\tfrac{\bf G}{2}).
\label{eqA:SE_Pshift}
\end{align}
Finally, the inverse of the Fourier transform of the wavefunction in Eq.~\eqref{eq:2mag_wfnFT} gives
\begin{align}
\psi_{{\bf K}+{\bf G}}({\bf q}\!+\!\tfrac{\bf G}{2})&\!=\!\frac{1}{N}\sum_{i,j}\!e^{-i\frac{{\bf K}+{\bf G}}{2}({\bf r}_i+{\bf r}_j)} e^{-i({\bf q}+\frac{\bf G}{2})({\bf r}_j-{\bf r}_i)} \psi_{i,j}\nonumber \\
&= \psi_{{\bf K}}({\bf q}),
\end{align}
where we used $e^{-i{\bf G}{\bf r}_j}\!=\!1$, since ${\bf r}_j$ is a lattice vector. Therefore, Eq.~\eqref{eqA:SE_Pshift} can be rewritten as
\begin{align}
\big(E \!-\! \varepsilon_{\frac{{\bf K}}{2}+{\bf q}} \!-\! \varepsilon_{\frac{{\bf K}}{2}-{\bf q}}\big) \psi_{{\bf K}}({\bf q})\!= \!\frac{1}{2N}\! \sum_{\bf p'}
V_{{\bf K}}({\bf q},{\bf p}') \psi_{{\bf K}}({\bf p}'),
\label{eqA:SE_rewrite}
\end{align}
which demonstrates that the BS energy $E$ is invariant under the transformation ${\bf K}\!\rightarrow\! {\bf K}\!+\!{\bf G}$ and is periodic in the same Brillouin zone as the energy of the single-magnon excitations.

\section{Magnon BSs in the square-lattice FM}

\subsection{$\Gamma$M direction: $s$-wave BS} \label{A:swave}
In this Section, we provide the details of the derivation for the binding energy of the $s$-wave BS along the $\Gamma$M high-symmetry line. We first simplify the 2D integral that appears in the matrix element $M_{ss,{\bf K}}$ and then use this approach to solve for the $s$-wave BS near the $\Gamma$ point.

\subsubsection{Effective density of states}
Two-dimensional integrals that depend only on the symmetric combination $z\!=\!\cos p_x\!+\!\cos p_y\!=\!2\gamma_{\bf p}$ can be written as one-dimensional integrals using
\begin{align}
\frac{1}{N}\sum_{\bf p} f(2\gamma_{\bf p})
\!=\!\int_{-2}^{2} f(z)D(z)\,dz,
\label{eqA:fzDz}
\end{align}
where $D(z)$ is an effective density of states (DOS),
\begin{align}
D(z) \!=
\! \frac{1}{(2\pi)^2}\int_{-\pi}^{\pi}\int_{-\pi}^{\pi}
\delta(z \!-\! \cos p_x \!-\! \cos p_y)dp_xdp_y.
\label{eqA:Dzoriginal_integral}
\end{align}
The effective DOS can be calculated by introducing the change of variables $(p_x,p_y)\!=\!(u+v,u-v)$ in Eq.~\eqref{eqA:Dzoriginal_integral}, followed by $t\!=\!\sin u$ and $t\!=\!\sqrt{1-z^2/4}\,\sin\theta$, yielding
\begin{align}
D(z)&= \frac{K(\kappa)}{\pi^2},
\quad \kappa=\sqrt{1-z^2/4},
\label{eqA:EllipticDoS}
\end{align}
where $K(\kappa)$ is the complete elliptic integral of the first kind with modulus $\kappa$~\cite{GradshteynRyzhik1994}.

\subsubsection{Exponentially shallow $s$-wave BS}
Along the $\Gamma$M line, the $s$- and $d$-wave harmonics are orthogonal. Therefore, the solution for the $s$-wave BS is obtained by solving the equation $M_{ss,{\bf K}}\!=\!1$, with the matrix element
\begin{align}
{M}_{ss,{\bf K}} &= -\frac{\alpha_{s}}{2N}\sum_{\bf p}
\frac{R_{s}(\mathbf{p}) \, \widetilde{R}_{s,{\bf K}}(\mathbf{p})}
{E-E_{\bf K}(\mathbf{p})},
\label{eqA:MssGammaM}
\end{align}
where $R_{s}({\bf p})\!=\!\gamma_{\bf p}$, $\widetilde{R}_{s,{\bf K}}({\bf p})\!=\!R_{s}({\bf p})\!-\!R_{s}({\bf K}/2)$, $\alpha_s\!=\!8$, and the energy denominator is given by
\begin{equation}
E-E_{\bf K}({\bf p})=-2\Delta-8S\big(
\gamma^{\phantom -}_{\frac{\bf K}2}
-\gamma^{\phantom -}_{\frac{\bf K}2}
\gamma_{\bf p}^{\phantom{-}}\big).
\label{eqA:DeltaE_GammaMJ1}
\end{equation}
The integrand of Eq.~\eqref{eqA:MssGammaM} depends only on the $s$-wave harmonic $\gamma_{\bf p}\!=\!\frac12(\cos p_x\!+\!\cos p_y)$. Therefore, using the effective DOS in Eqs.~\eqref{eqA:fzDz} and \eqref{eqA:EllipticDoS}, we can cast $M_{ss,{\bf K}}$ into a single-variable integral with $z\!=\!2\gamma_{\bf p}$,
\begin{align}
M_{ss,{\bf K}}\!&=\!\int_{-2}^2 f_{\bf K}(z)D(z)\,dz,
\label{eqA:M_ss-Gamma-DoS} \\
f_{\bf K}(z)&=\frac{z(z-2\gamma^{\phantom -}_{\frac{\bf K}2})}
{2\Delta_s-4S\gamma^{\phantom -}_{\frac{\bf K}2}(z-2)}.
\label{eqA:fzGammaM}
\end{align}
In the small-binding-energy limit, the integrand $f_{\bf K}(z)$ becomes singular at $z\!=\!2$, which corresponds to the minimum of the two-magnon continuum at ${\bf p}^*\!=\!0$ in the original variables. Therefore, the leading logarithmic divergence of Eq.~\eqref{eqA:M_ss-Gamma-DoS} can be obtained by evaluating the numerator of $f_{\bf K}(z)$ and $D(z)$ at $z\!=\!2$. Using $K(0)\!=\!\pi/2$~\cite{GradshteynRyzhik1994}, we obtain
\begin{align}
M_{ss,{\bf K}}&\approx
\frac{2(1-\gamma_{\frac{\bf K}2})}{\pi}
\int_{-2}^2
\frac{dz}{2\Delta_s-4S\gamma_{\frac{\bf K}2}(z-2)},
\nonumber \\
&\approx
-\frac{1-\gamma_{\frac{\bf K}2}}
{2S\pi\gamma_{\frac{\bf K}2}}
\ln\Big(\frac{\Delta_s}{8S\gamma_{\frac{\bf K}2}}\Big).
\end{align}
Therefore, setting $M_{ss,{\bf K}}\!=\!1$ gives
\begin{align}
\Delta_s =8
S\gamma_{\frac{\bf K}2}
\exp\left(
-\frac{2S\pi\gamma_{\frac{\bf K}2}}
{1-\gamma_{\frac{\bf K}2}}
\right),
\end{align}
which can be written in the familiar form
\begin{align}
\Delta_s = \frac{\alpha_s}{2m_{\bf K}^*}
\exp\Big(\!\!-\frac{8\pi}
{m_{\bf K}^*V_{\bf K}^{\rm eff}}\Big),
\label{eqA:Deltas_general}
\end{align}
where $\alpha_s\!=\!8$, $m_{\bf K}^*\!=\!(2S\gamma_{\frac{\bf K}2})^{-1}$ is the effective mass, and $V_{\bf K}^{\rm eff}\!=\!\alpha_s(1\!-\!\gamma_{\frac{\bf K}2})$ is the effective attractive potential.

In the small-$\bf K$ limit, near the $\Gamma$ point, Eq.~\eqref{eqA:Deltas_general} gives the leading dependence of the super-exponentially shallow $s$-wave BS,
\begin{align}
\Delta_s \propto S\exp\left(
-\frac{32\pi S}{{\bf K}^2}
\right),
\end{align}
which corresponds to Eq.~\eqref{eq:Delta_s_mag} in the main text.

\subsection{$\Gamma$M direction: $d$-wave BS} \label{A:dwaveBSJ1}
In this Section, we derive the critical momentum ${\bf K}_c$, corresponding to the boundary along the $\Gamma$M line above which the $d$-wave BS exists. Since the $s$- and $d$-wave harmonics are orthogonal along the $\Gamma$M line, the solution for the $d$-wave BS is obtained by solving the equation $M_{dd,{\bf K}}\!=\!1$, with the matrix element
\begin{align}
{M}_{dd,{\bf K}} &= -\frac{\alpha_{d}}{2N}\sum_{\bf p}
\frac{R_{d}(\mathbf{p}) \, \widetilde{R}_{d,{\bf K}}(\mathbf{p})}
{E-E_{\bf K}(\mathbf{p})},
\label{eqA:MddGammaM}
\end{align}
where $\widetilde{R}_{d,{\bf K}}({\bf p})\!=\!R_{d}({\bf p})\!=\!\gamma_{\bf p}^-$ because $R_{d}({\bf K}/2)\!=\!0$ along the $\Gamma$M line, $\alpha_d\!=\!8$, and the energy denominator is given by Eq.~\eqref{eqA:DeltaE_GammaMJ1}.

The critical momentum ${\bf K}_c$ at which the $d$-wave BS first appears corresponds to the vanishing binding energy, so it can be obtained by solving
\begin{equation}
M_{dd,{\bf K}_c}\left(\Delta\!=\!0\right)
=\frac{1}{2S\gamma_{\frac{{\bf K}_c}{2}}^{\phantom -}}
\frac{1}{N}\sum_{\bf p}
\frac{(\gamma_{\bf p}^-)^2}{1-\gamma_{\bf p}}
=1,
\label{eqA:Mdd_kc}
\end{equation}
where $\gamma^-_{\bf p}\!=\!\frac12(\cos p_x-\cos p_y)$.

To evaluate the expression above, we introduce the integrals
\begin{align}
I_0(a,b) &= \frac{1}{2\pi} \int_{-\pi}^{\pi}
\frac{dx}{a-b\cos x}
= \frac{1}{\sqrt{a^2-b^2}},
\label{eqA:I0}
\\
I_1(a,b) &= \frac{1}{2\pi} \int_{-\pi}^{\pi}
\frac{\cos x}{a-b\cos x}\,dx
= \frac{aI_0(a,b)-1}{b},
\label{eqA:I1}
\end{align}
valid for $a\!>\!|b|$~\cite{GradshteynRyzhik1994}. Using the integrals in Eqs.~\eqref{eqA:I0} and \eqref{eqA:I1}, Eq.~\eqref{eqA:Mdd_kc} can be written as
\begin{align}
M_{dd,{\bf K}_c}\!&=\!\frac{1}{2S\gamma_{\frac{{\bf K}_c}{2}}^{\phantom - }} \int_{-\pi}^\pi \!\frac{dp_x}{2\pi} \Big[\! \cos^2\!p_x I_0(a,b)\!-\!\cos p_x I_1(a,b) \Big],\nonumber \\
&=\!\frac{1}{2S\gamma_{\frac{{\bf K}_c}{2}}^{\phantom - }}\frac{4-\pi}{\pi},
\end{align}
where we used $a\!=\!2-\cos p_x$ and $b\!=\!1$. Setting $M_{dd,{\bf K}_c}\!=\!1$ yields
\begin{align}
|{\bf K}_c|=2\sqrt{2}\arccos\left(
\frac{4-\pi}{2S\pi}
\right),
\end{align}
corresponding to Eq.~\eqref{eq:kc} in the main text.

\subsection{MX direction: 1D BS} \label{A:1DBSJ1}
Along the MX line, ${\bf K}\!=\!(\pi,K_y)$, the eigenvalue matrix $\hat{\rm \bf M}_{\bf K}$ reduces to
\begin{align}
\hat{\rm \bf M}_{\bf K} =
\begin{pmatrix}
M_{{ss},{\bf K}} & M_{{sd},{\bf K}} \\
M_{{sd},{\bf K}} & M_{{ss},{\bf K}}
\end{pmatrix},
\label{eqA:MMtoXJ1}
\end{align}
with the matrix elements given by the general expression
\begin{align}
{M}_{\gamma \gamma',{\bf K}} &=
-\frac{\alpha_{\gamma'}}{2N}\sum_{\bf p}
\frac{R_{\gamma'}(\mathbf{p}) \,
\widetilde{R}_{\gamma,{\bf K}}(\mathbf{p})}
{E-E_{\bf K}(\mathbf{p})},
\label{eqA:M_gamma}
\end{align}
and the energy denominator given by
\begin{equation}
E-E_{\bf K}({\bf p})=-2\Delta
-4S\cos\frac{K_y}{2}\,\big(1-\cos p_y\big).
\label{eqA:2mag_MToX}
\end{equation}
The BS energies are obtained by setting the eigenvalues of Eq.~\eqref{eqA:MMtoXJ1}, $\lambda_{\pm}\!=\!M_{ss,{\bf K}}\pm M_{sd,{\bf K}}$, equal to one.

\subsubsection{Symmetric BS}
After some straightforward algebra, the eigenvalue of $\hat{\rm \bf M}_{\bf K}$ corresponding to the symmetric BS reads
\begin{align}
\lambda_+=\frac12 I_0\big(
\Delta+2S\cos({K_y}/2),
\,2S\cos({K_y}/2)\big),
\end{align}
with $I_0$ defined in Eq.~\eqref{eqA:I0} and $\lambda_+\!=\!1$ leading to
\begin{align}
\label{eqA:Delta_+_exact}
\Delta_+=\frac12 \sqrt{
16S^2\cos^2\frac{K_y}{2}+1}
-2S\cos \frac{K_y}{2},
\end{align}
which is given in Eq.~\eqref{eq:Delta_+_exact} in the main text.

\subsubsection{Antisymmetric BS}
Similarly, after straightforward algebra using the integrals in Eqs.~\eqref{eqA:I0} and \eqref{eqA:I1}, the eigenvalue corresponding to the antisymmetric BS is
\begin{align}
\lambda_-=
\frac{(1+\delta-\alpha)(1+\delta-c_y)}
{2S\alpha c_y},
\end{align}
where $c_y\!=\!\cos(K_y/2)$, $\delta\!=\!\Delta_-/2Sc_y$, and $\alpha\!=\!\sqrt{\delta(2\!+\!\delta)}$. Setting $\lambda_-=1$ yields the equation
\begin{align}
(1+\delta)(1+\delta-c_y)
-\big(1+\delta+(2S-1)c_y\big)\alpha=0.
\end{align}
While the equation above does not admit a compact analytical solution because of its nonlinear dependence on $\delta$, for $S\!=\!1/2$ it simplifies to
\begin{align}
1+\delta-c_y=\alpha,
\end{align}
from which one readily obtains
\begin{align}
\label{eqA:Delta_-_exact}
\Delta_-=\frac12\bigg(
1-\cos \frac{K_y}{2}
\bigg)^2,
\end{align}
corresponding to Eq.~\eqref{eq:Delta_-_exact} in the main text. 

\section{BSs in the square-lattice $J_1$--$J_2$ FM-AFM model}

\subsection{M point: degenerate BSs} \label{A:DeltaMJ1J2}
In this Section, we provide additional details of the derivation of the binding energy of the degenerate BSs at the M point and its asymptotes. As discussed in Sec.~\ref{Sec:M_point_J1J2}, all channels are orthogonal at ${\bf K}\!=\!{\rm M}$. Therefore, we only focus on the attractive $s$- and $d$-wave channels.
The two-magnon energy at the M point reduces the denominator of the two-particle SE in Eq.~\eqref{eqA:M_gamma} to
\begin{align}
E-E_{\rm M}({\bf p})=-2\Delta-8J_2S\big(1+\gamma_{\bf p}^{d_{xy}}\big),
\label{eqA:Ek2_MJ2}
\end{align}
which depends on the $d_{xy}$ harmonic, $\gamma_{\bf p}^{d_{xy}}\!=\!\sin p_x\sin p_y$, and has minima at the ${\bf Q}\!=\!\pm(\pi/2,-\pi/2)$ points.

It is convenient to write the matrix elements of the $s$- and $d$-wave channels in terms of the shifted momenta ${\bf x}\!=\!{\bf p}\!-\!{\bf Q}$, such that $(x,y)=(p_x\!-\!\pi/2,\,p_y\!+\!\pi/2)$. In these variables, one of the integrals is readily obtained using Eq.~\eqref{eqA:I0}, which yields
\begin{align}
M_{\gamma\gamma,{\rm M}}&=\frac{1}{4SJ_2}\int_{-\pi}^{\pi}
\frac{dx}{2\pi}\sin^2\!x\,I_0(1+\delta,\cos x),\nonumber\\
&=\frac{1}{2\pi SJ_2}\int_0^{\frac{\pi}{2}}
\frac{\sin^2\!x}{\sqrt{(1+\delta)^2-\cos^2\!x}}\,dx,
\end{align}
where $\gamma\!=\!\{s,d\}$ and $\delta=\Delta/4SJ_2$.
The remaining integral can be evaluated using the consecutive changes of variables $u=\cos x$ and $u=\sin\varphi$, leading to
\begin{align}
M_{\gamma\gamma,{\rm M}}&=\!\frac{1}{2\pi S \kappa J_2}
\!\big(E(\kappa)\!-\!(1-\kappa^2)K(\kappa)\big),
\label{eqA:MM_J1J2}
\end{align}
where $\kappa\!=\!1/(1\!+\!\delta)$, and $K(\kappa)$ and $E(\kappa)$ are the complete elliptic integrals of the first and second kind, respectively~\cite{GradshteynRyzhik1994}.

\subsubsection{Small-$J_2$ asymptote}
In the small-$J_2$ limit, $\delta\!\propto\!1/J_2S\!\gg\!1$. Expanding $M_{\gamma\gamma,{\rm M}}$ in Eq.~\eqref{eqA:MM_J1J2} yields
\begin{align}
M_{\gamma\gamma,{\rm M}}&\approx\frac{1}{2\Delta}
-\frac{2J_2S}{\Delta^2}
+\mathcal{O}\big((J_2S)^2\big),
\end{align}
and the asymptotic expansion of the binding energy
\begin{align}
\Delta=\frac{1}{2}-4J_2S +\mathcal{O}\big((J_2S)^2\big),
\label{eqA:DeltaMJ1J2_smallJ2}
\end{align}
where the linear correction for small $J_2$ follows from the linear $J_2$-dependence of the inverse mass at the M point, $1/m_{\rm M}^*\!\propto\!J_2S$, which in turn increases the threshold coupling for the existence of the BS by ${\cal O}(J_2S)$.

\subsubsection{Small-$\Delta$ asymptote}
In the $J_2\!\rightarrow\!J_2^{c_1}$ limit, for $J_2^{c_1}\!=\!1/2\pi S$, we can expand $M_{\gamma\gamma,{\rm M}}$ in Eq.~\eqref{eqA:MM_J1J2} in $\Delta$, yielding
\begin{align}
M_{\gamma\gamma,{\rm M}}& \approx \frac{J_2^{c_1}}{J_2} +
\frac{ \pi (J_2^{c_1})^2}{4J_2^2}\Delta\left(\ln\left(\frac{\Delta}{32SJ_2}\right)+1 \right).
\label{eqA:MssMapprox}
\end{align}
We note that the leading nonanalytic term $\Delta\ln\Delta$ differs from the standard $\ln\Delta$ of the 2D case due to the dimensional enhancement at the M point.

Using Eq.~\eqref{eqA:MssMapprox}, one arrives at the approximate equation for the binding energy
\begin{align}
-\alpha\Delta+\Delta\ln\Delta
=-\beta(J_2^{c_1}-J_2),
\label{eqA:SCeqDeltaM_smallDelta}
\end{align}
where $\alpha\!=\!\ln(32SJ_2)-1$ and $\beta\!=\!4J_2/\pi(J_2^{c_1})^2$. 

Eq.~\eqref{eqA:SCeqDeltaM_smallDelta} is solved using the change of variables $t\!=\!\ln\Delta$, $w\!=\!t-\alpha$, and $x\!=\!-e^{-\alpha}\beta(J_2^{c_1}\!-\!J_2)$, which yield
\begin{align}
we^w=x,
\label{eqA:Lambert}
\end{align}
with its solution given by the Lambert $W$ function, $w\!=\!W_k(x)$~\cite{Corless1996LambertW}.
Since $x\!<\!0$ for $J_2\!<\!J_2^{c_1}$, the relevant real branches are $k\!=\!0$ and $k\!=\!-1$ for $-1/e\!\leq\! x\!<\!0$. The physical solution with $\Delta\!\rightarrow\!0$ for $J_2\!\rightarrow\! J_2^{c_1}$ is on the $k\!=\!-1$ branch. Therefore, 
\begin{align}
\Delta\approx 32SJ_2
\exp\!\big(W_{-1}(x)-1\big).
\label{eqA:DeltaMLambertW_smallDelta}
\end{align}
In the limit $J_2\!\rightarrow \! J_2^{c_1}$, $W_{-1}(0^-)\!\rightarrow\!-\infty$, giving $\Delta\!\rightarrow\!0$, as expected. Eq.~\eqref{eqA:DeltaMLambertW_smallDelta} is simplified further using $W_{-1}(x)\!\approx \!\ln(-x)$ for $x\!\rightarrow\!0^-$~\cite{Corless1996LambertW}, which yields
\begin{align}
\Delta\approx
\frac{8S(J_2^{c_1}-J_2)}
{\big|\ln(J_2^{c_1}-J_2)\big|}.
\label{eqA:DeltaMJ1J2_smallDelta}
\end{align}
The asymptotic expressions in Eqs.~(\ref{eqA:DeltaMJ1J2_smallJ2}) and~(\ref{eqA:DeltaMJ1J2_smallDelta}) agree well with the behavior of $\Delta_{\rm M}$ in Fig.~\ref{fig:DeltasJ1J2} in the corresponding limits.

\subsection{MX line} \label{A:MmatrixMXJ1J2}
In this Section, we discuss the solution of the BS problem along the MX line. For ${\bf K}\!=\!(\pi,K_y)$ and finite $J_2$, the two-magnon continuum has the form
\begin{align}
\label{eqA:MX_2mag_E}
E-E_{\bf K}({\bf p})={}&-2\Delta-\Delta E_{\bf K}({\bf p}^*)\\
&+4S\Big(\cos\frac{K_y}{2}\cos p_y
-2J_2\sin\frac{K_y}{2}\,\gamma_{\bf p}^{d_{xy}}\Big),\nonumber
\end{align}
where $\Delta E_{\bf K}({\bf p}^*)\!=\!4S\sqrt{\cos^2(K_y/2)+4J_2^2\sin^2(K_y/2)}$ is a constant energy offset reached at the incommensurate momenta ${\bf p}^*\!=\!\pm(\pi/2,-q^*)$, with
\begin{align}
q^*=\arctan\left(2J_2\tan\frac{K_y}{2}\right).
\end{align}
The form of the two-magnon continuum in Eq.~\eqref{eqA:MX_2mag_E} enters in the denominator of the matrix elements of $\hat{\bf M}_{\bf K}$, given by ${M}_{\gamma \gamma',{\bf K}}$ in Eq.~\eqref{eqA:M_gamma}. The form of   ${M}_{\gamma \gamma',{\bf K}}$ is independent of the basis of the harmonics in which we choose to expand the interaction term.

\subsubsection{Elemental partial waves}
Along the MX line, it is convenient to replace the $s$- and $d$-wave harmonics by their symmetric and antisymmetric combinations
\begin{align}
R_S({\bf q})&=R_s({\bf q})+R_d({\bf q})=\cos q_x,
\label{eqA:RS_MX}\\
R_A({\bf q})&=R_s({\bf q})-R_d({\bf q})=\cos q_y.
\label{eqA:RA_MX}
\end{align}
Then the interaction is written in the separable form
\begin{align}
V_{\bf K}({\bf q},{\bf p})=-\sum_\gamma\alpha_\gamma
R_\gamma({\bf q})\widetilde{R}_{\gamma,{\bf K}}({\bf p}),
\label{eqA:Vkqp_expansion}
\end{align}
with $\gamma=\{S,A,s_{xy},d_{xy}\}$, where
\begin{align}
\alpha_{S(A)}=4,\qquad
R_{S(A)}({\bf q})=\cos q_x(\cos q_y),
\label{eqA:RSA_MX_harmonics}
\end{align}
and the $s_{xy}$ and $d_{xy}$ harmonics remain unchanged, with $\alpha_{s_{xy}}\!=\!\alpha_{d_{xy}}\!=\!-8J_2$, $R_{s_{xy}}({\bf q})\!=\!\gamma_{\bf q}^{(2)}$, and $R_{d_{xy}}({\bf q})\!=\!\gamma^{d_{xy}}_{\bf q}$.
Along the MX line, the ``hard-core'' terms vanish for the symmetric and $s_{xy}$ harmonics, so that $\widetilde{R}_{S,{\bf K}}({\bf p})\!=\!R_S({\bf p})$ and
$\widetilde{R}_{s_{xy},{\bf K}}({\bf p})\!=\!R_{s_{xy}}({\bf p})$. By contrast, such terms are nonzero for the antisymmetric and $d_{xy}$ harmonics.

As discussed in Sec.~\ref{Sec:MXJ1J2}, the mirror reflection about the $(\pi/2,p_y)$ line leaves the energy denominator in Eq.~\eqref{eqA:MX_2mag_E} invariant. Under this reflection, $R_A({\bf p})$ and the $d_{xy}$ partial waves are even, whereas $R_S({\bf p})\!=\!\cos p_x$ and the $s_{xy}$ partial waves  are odd. Therefore, in the ordered basis $\{S,s_{xy},A,d_{xy}\}$, the $4\times4$ eigenvalue matrix $\hat{\rm \bf M}_{\bf K}$ separates into two $2\times2$ blocks,
\begin{align}
\hat{\rm \bf M}_{\bf K}
=
\begin{pmatrix}
\hat{\rm \bf M}^{\{S,s_{xy}\}}_{\bf K} & {\bf 0}\\
{\bf 0} & \hat{\rm \bf M}^{\{A,d_{xy}\}}_{\bf K}
\end{pmatrix},
\label{eqA:MkMXJ1J2}
\end{align}
with
\begin{align}
\hat{\rm \bf M}^{\{S,s_{xy}\}}_{\bf K}
&=
\begin{pmatrix}
M_{SS,{\bf K}} & M_{Ss_{xy},{\bf K}}\\
M_{s_{xy}S,{\bf K}} & M_{s_{xy}s_{xy},{\bf K}}
\end{pmatrix},
\label{eqA:MS_MXJ1J2}\\
\hat{\rm \bf M}^{\{A,d_{xy}\}}_{\bf K}
&=
\begin{pmatrix}
M_{AA,{\bf K}} & M_{Ad_{xy},{\bf K}}\\
M_{d_{xy}A,{\bf K}} & M_{d_{xy}d_{xy},{\bf K}}
\end{pmatrix},
\end{align}
and ${\bf 0}$ is a $2\times2$ matrix of zeros.

\subsubsection{X point: 1D BS} \label{A:DeltaXJ1J2}

At the ${\rm X}\!=\!(\pi,0)$ point, the two-magnon continuum in Eq.~\eqref{eqA:MX_2mag_E} becomes one-dimensional even for finite $J_2$, with the energy denominator
\begin{align}
E-E_{\rm X}({\bf p})
=-2\Delta-4S(1-\cos p_y).
\label{eqA:DeltaE_XJ1J2}
\end{align}
The BS at the X point corresponds to the solution of the eigenvalue problem for the symmetric block $\hat{\rm \bf M}^{\{S,s_{xy}\}}_{\bf K}$ in Eq.~\eqref{eqA:MS_MXJ1J2}. This follows from continuity with the X-point BS of the $J_1$-only model, which is formed by the symmetric $R_S$ harmonic. Introducing $J_2$ mixes it with the repulsive $s_{xy}$-wave harmonic.

Because the energy denominator in Eq.~\eqref{eqA:DeltaE_XJ1J2} is $p_x$-independent, the $p_x$ integrations in the matrix elements of the symmetric block $\hat{\rm \bf M}^{\{S,s_{xy}\}}_{\bf K}$ are trivial. Therefore, all matrix elements can be written using the elementary integrals given in Eqs.~\eqref{eqA:I0} and~\eqref{eqA:I1}, together with
\begin{align}
I_2(a,b)&=\frac{1}{2\pi}\int_{-\pi}^{\pi}
\frac{\cos^2x}{a-b\cos x}\,dx
=\frac{aI_1(a,b)}{b}.
\label{eqA:I2}
\end{align}
The matrix elements for the symmetric block $\hat{\rm \bf M}^{\{S,s_{xy}\}}_{\bf K}$ are given by
\begin{align}
&M_{SS,{\rm X}}\!=\!\frac12 I_0(a,b),\qquad
M_{Ss_{xy},{\rm X}}\!=\!-J_2I_1(a,b),\nonumber\\
&M_{s_{xy}S,{\rm X}}\!=\!\frac12 I_1(a,b),\quad
M_{s_{xy}s_{xy},{\rm X}}\!=\!-J_2I_2(a,b),
\label{eqA:MmatrixXJ1J2}
\end{align}
where $a\!=\!\Delta+2S$ and $b\!=\!2S$. Therefore, the eigenvalue equation for the symmetric BS becomes
\begin{align}
\big(1\!-\!\tfrac12 I_0(a,b)\big)
\big(1\!+\!J_2I_2(a,b)\big)
+\frac{J_2}{2}I_1^2(a,b)=0.
\label{eqA:X_point_integrals}
\end{align}
Introducing $\delta\!=\!\Delta/2S$, one can rewrite Eq.~\eqref{eqA:X_point_integrals} as
\begin{align}
1\!-\!\frac{1}{4S}\frac{J_2}{J_2^{c_3}}f_1(\Delta) \!=\! \frac{1}{4S\sqrt{\delta(2\!+\!\delta)}}\left(1\!-\!\frac{J_2}{J_2^{c_3}}f_2(\Delta)\right),
\label{eq:X_point_exact_EV}
\end{align}
using $f_1(\Delta)\!=\!1\!+\!2\delta J_2^{c_3}$, $f_2(\Delta)\!=\!(1\!+\!\delta)f_1(\Delta)$, and
\begin{align}
J_2^{c_3}=\frac{1}{2(1-1/4S)},
\label{eqA:Jc3}
\end{align}
where $J_2^{c_3}\!=\!1$ for $S\!=\!1/2$, as expected. We note that Eq.~\eqref{eq:X_point_exact_EV} gives the exact relation between the binding energy $\Delta$ and $J_2$ for the BS at the X point. This equation further simplifies in the small-$\Delta$ limit, where $f_1(\Delta),f_2(\Delta)\!\rightarrow\!1$. Keeping the leading terms in Eq.~\eqref{eq:X_point_exact_EV} gives
\begin{align}
1-\frac{1}{4S}\frac{J_2}{J_2^{c_3}}
\approx\frac{1}{2\sqrt{4S\Delta}}
\left(1-\frac{J_2}{J_2^{c_3}}\right),
\label{eqA:smallDeltaXeq}
\end{align}
where the sign of the bracket on the right-hand side dictates whether the $1/\sqrt{\Delta}$ divergence results in a BS or not. Thus, the BS solution disappears at $J_2\geq J_2^{c_3}$.

Solving Eq.~\eqref{eqA:smallDeltaXeq} near $J_2^{c_3}$ gives the standard one-dimensional form
\begin{align}
\Delta\approx\frac{\alpha_{\rm eff}^2m_y^*}{32},
\label{eqA:DeltaXapproxJ1J2}
\end{align}
where $\alpha_{\rm eff}\!=\!4(J_2^{c_3}-J_2)$ and $m_y^*\!=\!(2S)^{-1}$; see Eq.~\eqref{eq:DeltaXapproxJ1J2}.

Therefore, the disappearance of the BS at the X point is controlled by the change of sign of the effective interaction due to the mixing of the attractive $R_S$ channel with the repulsive $s_{xy}$-wave channel.

\subsection{$\Gamma$M line} \label{A:MmatrixGMJ1J2}
In this Section, we consider the BSs along the  $\Gamma$M line. Along this line, the $d$-wave harmonic remains orthogonal to the other partial waves for any $J_2$. Therefore, the eigenvalue problem takes the simplified form
\begin{align}
\hat{\rm \bf M}_{\bf K}
=
\begin{pmatrix}
\hat{\rm \bf M}^{\{ssd\}}_{\bf K} & {\bf 0}\\
{\bf 0}^{T} & M_{dd,{\bf K}}
\end{pmatrix},
\label{eqA:MkGMJ1J2}
\end{align}
where ${\bf 0}$ denotes a $3\times1$ zero block and
\begin{align}
\hat{\rm \bf M}^{\{ssd\}}_{\bf K}
&\!=\!
\begin{pmatrix}
M_{ss,{\bf K}} & M_{ss_{xy},{\bf K}} & M_{sd_{xy},{\bf K}}\\
M_{s_{xy}s,{\bf K}} & M_{s_{xy}s_{xy},{\bf K}} & M_{s_{xy}d_{xy},{\bf K}}\\
M_{d_{xy}s,{\bf K}} & M_{d_{xy}s_{xy},{\bf K}} & M_{d_{xy}d_{xy},{\bf K}}
\end{pmatrix}\!.
\label{eqA:Mssd_GMJ1J2}
\end{align}
The matrix elements are given by Eq.~\eqref{eqA:M_gamma} with the $J_2$- and ${\bf K}$-dependent energy denominator $E\!-\!E_{\bf K}({\bf p})$ from Eq.~(\ref{eq:MG_2mag_E}) discussed in Sec.~\ref{Sec:GammaM}.

\subsubsection{$s$, $s_{xy}$, and $d_{xy}$ mixed BS} \label{A:ssdGammaMJ1J2}
\begin{figure}[t]
\includegraphics[width=\linewidth]{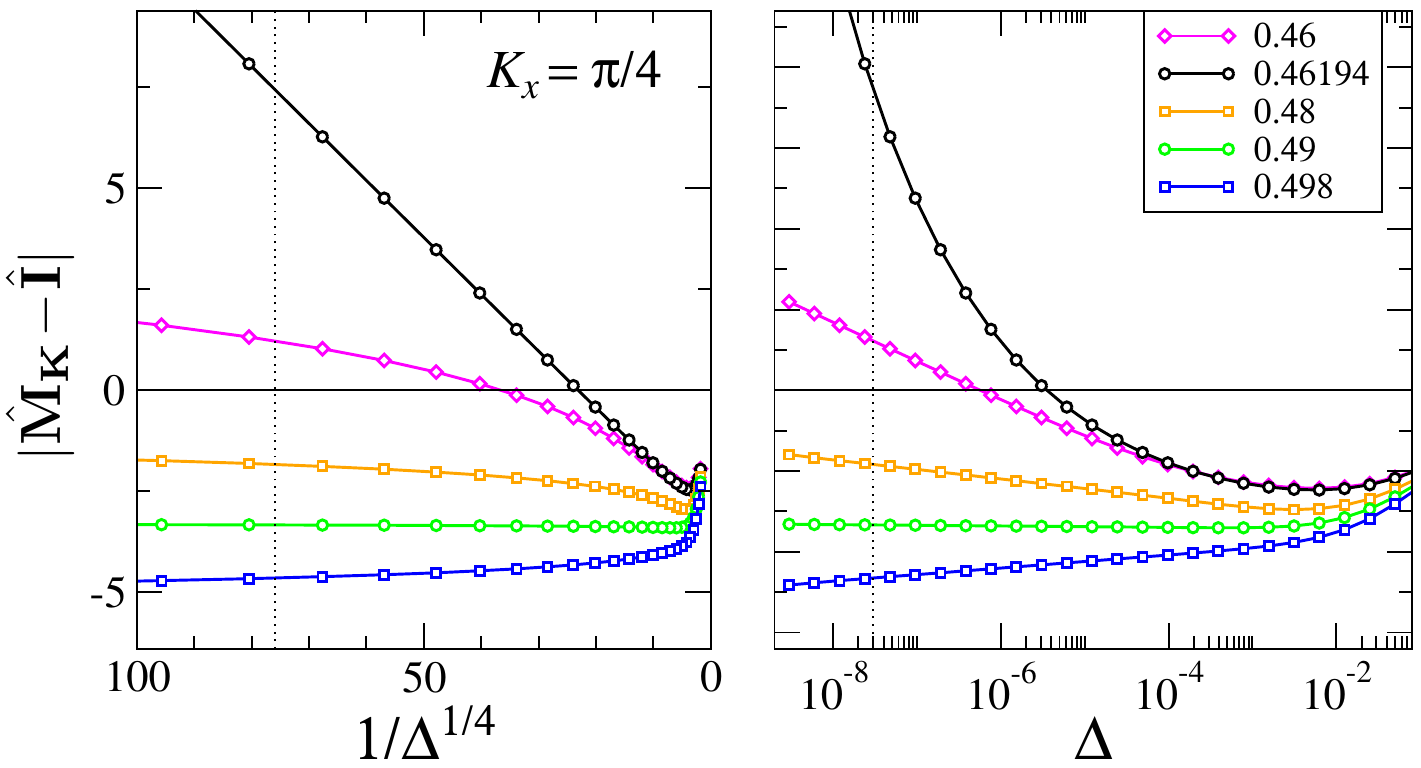}
\vskip -0.3cm
\caption{Same as in Fig.~\ref{fig:GM_ssd_MvsDelta_log}, for the $s$, $s_{xy}$, and $d_{xy}$ mixed channel along the $\Gamma$M line for $K_x\!=\!\pi/4$ and several $J_2$ near $J_2^*\!=\!0.5\cos(\pi/8)$. (a) and (b) $|\hat{\rm \bf M}_{\bf K}-\hat{\rm \bf I}|$ as a function of $\Delta$, plotted versus $1/\Delta^{1/4}$ and on a semi-log scale, respectively.}
\label{fig:GM_Kx_pi_4_spec}
\vskip -0.4cm
\end{figure}

Figure~\ref{fig:GM_Kx_pi_4_spec} illustrates the change in the small-$\Delta$ divergence of the eigenvalue determinant of the $3\times3$ matrix in Eq.~\eqref{eqA:Mssd_GMJ1J2} for $K_x\!=\!\pi/4$. At $J_2^*\!=\!\frac12\cos(K_x/2)\!=\!\frac12\cos(\pi/8)\!\approx\!0.46194$, the two-magnon band has a nonparabolic, $\propto p^4$, dispersion in one of the directions, which is associated with the pinching-off of its band minima at ${\bf K}\!=\!{\bf K}^*$; see Sec.~\ref{Sec:GammaM_E_J1J2}. Consequently, the determinant exhibits a $1/\Delta^{1/4}$ divergence at $J_2^*$ instead of the conventional logarithmic or $1/\sqrt{\Delta}$ behaviors.

\subsubsection{$d$-wave BS} \label{A:dwaveGammaMJ1J2}
Along the $\Gamma$M line, the $d$-wave BS is unaffected by the hard-core, non-potential-like terms since $\gamma^-_{{\bf K}/2}\!=\!0$, so that $\widetilde{R}_{d,{\bf K}}({\bf p})\!=\!R_d({\bf p})\!=\!\gamma^-_{\bf p}$.
Therefore, the $d$-wave BS is obtained by solving $M_{dd,{\bf K}}\!=\!1$, with the matrix element from Eq.~\eqref{eqA:M_gamma} given by
\begin{align}
M_{dd,{\bf K}}
&=-\frac{4}{N}\sum_{\bf p}
\frac{(\gamma^-_{\bf p})^2}{E-E_{\bf K}({\bf p})}.
\label{eqA:MddGammaMJ1J2}
\end{align}
We first consider the onset of the $d$-wave BS at the $\Gamma$ point, where it becomes the ground state for $J_2>J_2^{c_2}\approx0.408$. As discussed in Sec.~\ref{Sec:dwave_GammaJ1J2}, this transition occurs for $J_2\!<\!0.5$, so that the energy denominator of~\eqref{eqA:MddGammaMJ1J2} for ${\bf K}\!=\!\Gamma$ reads (\ref{eq:G_2mag_E})
\begin{align}
E\!-\!E_{\Gamma}({\bf p})
\!=-2\Delta \!-\!8S(1\!-\!\gamma_{\bf p})+8SJ_2(1\!-\!\gamma_{\bf p}^{(2)}).
\label{eqA:DeltaE_GammaJ1J2}
\end{align}
The critical value $J_2^{c_2}$ at which the $d$-wave BS first appears corresponds to vanishing binding energy. Therefore, setting $\Delta\!=\!0$ in $M_{dd,\Gamma}$ in Eq.~\eqref{eqA:MddGammaMJ1J2} gives
\begin{align}
M_{dd,\Gamma}(\Delta\!=\!0)=\frac{1}{4SN}\sum_{\bf p}
\frac{(\cos p_x-\cos p_y)^2}
{a-b\cos p_y},
\label{eqA:MddGammaDelta0}
\end{align}
where $a\!=\!2(1\!-\!J_2)\!-\!\cos p_x$ and $b\!=\!1\!-\!2J_2\cos p_x$. The integration in $p_y$ using Eqs.~\eqref{eqA:I0} and~\eqref{eqA:I1} yields
\begin{align}
M_{dd,\Gamma}
\!=\!\frac{1}{4S}\int_{-\pi}^{\pi}\frac{dp_x}{2\pi}
\Big(\!\cos^2\!p_x I_0\!+\!\left(\frac{a}{b}\!-\!2\cos p_x\right)I_1
\Big).
\label{eqA:MddGamma1D}
\end{align}
The remaining integral is evaluated analytically, giving
\begin{align}
M_{dd,{\Gamma}}&\!=\! 
\frac{1}{\pi S(1\!+\!2J_2)} +
\frac{1}{4SJ_2} \left(\frac{1}{\sqrt{1\!-\!2J_2}(1\!+\!2J_2)^{3/2}}\!-\!1 \right)
\nonumber\\
&+\frac{\arcsin(2J_2)} {2\pi SJ_2\sqrt{1\!-\!2J_2}(2J_2\!+\!1)^{3/2}}.
\label{eqA:MddGamma}
\end{align}
The transition point $J_2^{c_2}$ is found from
\begin{align}
M_{dd,\Gamma}(J_2^{c_2})=1.
\label{eqA:Jc2equation}
\end{align}
For $S\!=\!1/2$, its numerical solution gives
\begin{align}
J_2^{c_2}\approx0.4077593304754\dots,
\label{eqA:Jc2}
\end{align}
corresponding to Eq.~\eqref{eq:J2c2} in the main text.

\begin{figure}[t]
\includegraphics[width=\linewidth]{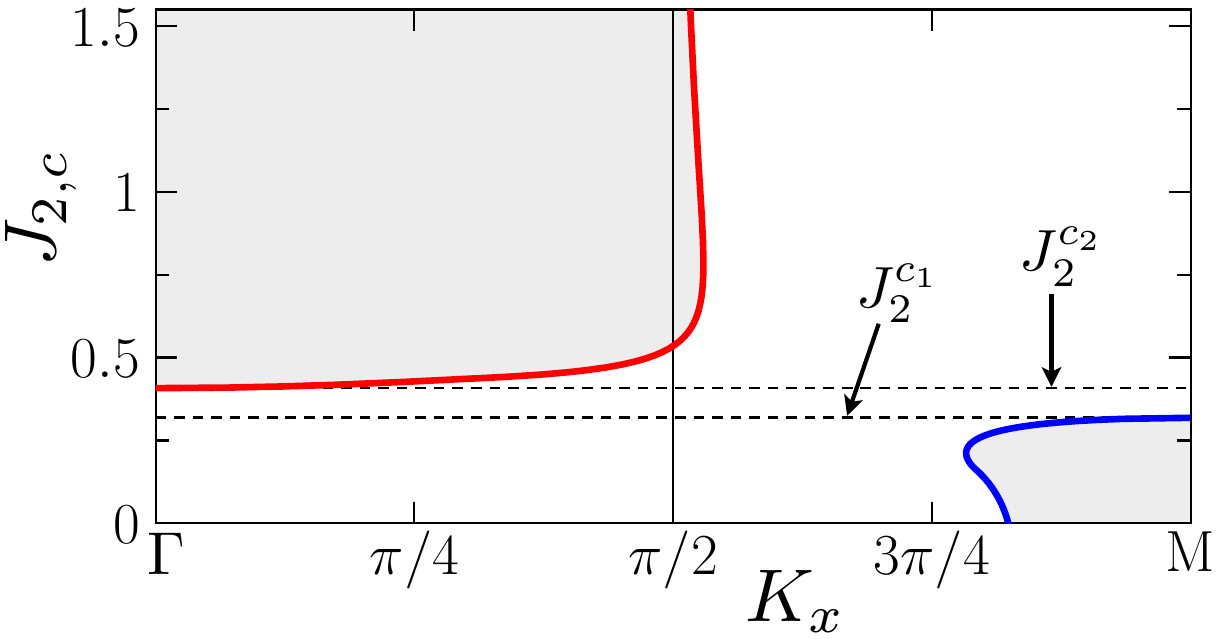}
\vskip -0.3cm
\caption{The threshold boundaries $J_{2,c}$ as a function of $K_x$ for the $\Gamma$M direction (lines) and regions of BS existence (shaded areas) for the two $d$-wave branches: near the M point in the $0\!\leq\!J_2\!<\!J_2^{c_1}$ range discussed in Sec.~\ref{Sec:GammaM_dwaveJ1J2}, and originating from the $\Gamma$ point and existing for $J_2\!>\!J_2^{c_2}$, discussed in Sec.~\ref{Sec:dwave_Kc_J1J2}; $S\!=\!1/2$. The results are combined from Figs.~\ref{fig:MG_Jc_vs_Kx} and~\ref{fig:GM_Jc_vs_Kx_d_gs}.}
\label{fig:GM_all_d_waves}
\vskip -0.4cm
\end{figure}

Figure~\ref{fig:GM_all_d_waves} combines Figs.~\ref{fig:MG_Jc_vs_Kx} and~\ref{fig:GM_Jc_vs_Kx_d_gs} to emphasize the relative $J_2$--${\bf K}$ regions of existence of the $d$-wave BSs along the $\Gamma$M line. It demonstrates that they do not overlap in either $J_2$ or ${\bf K}$. The boundaries of these regions are obtained by solving the eigenvalue problem $M_{dd,{\bf K}}\!=\!1$ from Eq.~\eqref{eqA:MddGammaMJ1J2} for $\Delta\!=\!0$; see Secs.~\ref{Sec:GammaM_dwaveJ1J2} and \ref{Sec:dwave_Kc_J1J2}.

\newpage

\bibliography{BoundStates}

\begin{thebibliography}{76}%
\makeatletter
\providecommand \@ifxundefined [1]{%
 \@ifx{#1\undefined}
}%
\providecommand \@ifnum [1]{%
 \ifnum #1\expandafter \@firstoftwo
 \else \expandafter \@secondoftwo
 \fi
}%
\providecommand \@ifx [1]{%
 \ifx #1\expandafter \@firstoftwo
 \else \expandafter \@secondoftwo
 \fi
}%
\providecommand \natexlab [1]{#1}%
\providecommand \enquote  [1]{``#1''}%
\providecommand \bibnamefont  [1]{#1}%
\providecommand \bibfnamefont [1]{#1}%
\providecommand \citenamefont [1]{#1}%
\providecommand \href@noop [0]{\@secondoftwo}%
\providecommand \href [0]{\begingroup \@sanitize@url \@href}%
\providecommand \@href[1]{\@@startlink{#1}\@@href}%
\providecommand \@@href[1]{\endgroup#1\@@endlink}%
\providecommand \@sanitize@url [0]{\catcode `\\12\catcode `\$12\catcode `\&12\catcode `\#12\catcode `\^12\catcode `\_12\catcode `\%12\relax}%
\providecommand \@@startlink[1]{}%
\providecommand \@@endlink[0]{}%
\providecommand \url  [0]{\begingroup\@sanitize@url \@url }%
\providecommand \@url [1]{\endgroup\@href {#1}{\urlprefix }}%
\providecommand \urlprefix  [0]{URL }%
\providecommand \Eprint [0]{\href }%
\providecommand \doibase [0]{https://doi.org/}%
\providecommand \selectlanguage [0]{\@gobble}%
\providecommand \bibinfo  [0]{\@secondoftwo}%
\providecommand \bibfield  [0]{\@secondoftwo}%
\providecommand \translation [1]{[#1]}%
\providecommand \BibitemOpen [0]{}%
\providecommand \bibitemStop [0]{}%
\providecommand \bibitemNoStop [0]{.\EOS\space}%
\providecommand \EOS [0]{\spacefactor3000\relax}%
\providecommand \BibitemShut  [1]{\csname bibitem#1\endcsname}%
\let\auto@bib@innerbib\@empty
\bibitem [{\citenamefont {Dean}\ and\ \citenamefont {Hjorth-Jensen}(2003)}]{PairingNuclearRMP2003}%
  \BibitemOpen
  \bibfield  {author} {\bibinfo {author} {\bibfnamefont {D.~J.}\ \bibnamefont {Dean}}\ and\ \bibinfo {author} {\bibfnamefont {M.}~\bibnamefont {Hjorth-Jensen}},\ }\bibfield  {title} {\bibinfo {title} {{Pairing in nuclear systems: from neutron stars to finite nuclei}},\ }\href {https://doi.org/10.1103/RevModPhys.75.607} {\bibfield  {journal} {\bibinfo  {journal} {Rev. Mod. Phys.}\ }\textbf {\bibinfo {volume} {75}},\ \bibinfo {pages} {607} (\bibinfo {year} {2003})}\BibitemShut {NoStop}%
\bibitem [{\citenamefont {Wentzel}(1950)}]{PairingParticlePR1950}%
  \BibitemOpen
  \bibfield  {author} {\bibinfo {author} {\bibfnamefont {G.}~\bibnamefont {Wentzel}},\ }\bibfield  {title} {\bibinfo {title} {{$\ensuremath{\mu}$-Pair Theories and the $\ensuremath{\pi}$-Meson}},\ }\href {https://doi.org/10.1103/PhysRev.79.710} {\bibfield  {journal} {\bibinfo  {journal} {Phys. Rev.}\ }\textbf {\bibinfo {volume} {79}},\ \bibinfo {pages} {710} (\bibinfo {year} {1950})}\BibitemShut {NoStop}%
\bibitem [{\citenamefont {Fermi}\ and\ \citenamefont {Yang}(1949)}]{Fermi1949}%
  \BibitemOpen
  \bibfield  {author} {\bibinfo {author} {\bibfnamefont {E.}~\bibnamefont {Fermi}}\ and\ \bibinfo {author} {\bibfnamefont {C.~N.}\ \bibnamefont {Yang}},\ }\bibfield  {title} {\bibinfo {title} {{Are Mesons Elementary Particles?}},\ }\href {https://doi.org/10.1103/PhysRev.76.1739} {\bibfield  {journal} {\bibinfo  {journal} {Phys. Rev.}\ }\textbf {\bibinfo {volume} {76}},\ \bibinfo {pages} {1739} (\bibinfo {year} {1949})}\BibitemShut {NoStop}%
\bibitem [{\citenamefont {Berestetskii}\ \emph {et~al.}(1982)\citenamefont {Berestetskii}, \citenamefont {Lifshitz},\ and\ \citenamefont {Pitaevskii}}]{QED}%
  \BibitemOpen
  \bibfield  {author} {\bibinfo {author} {\bibfnamefont {V.~B.}\ \bibnamefont {Berestetskii}}, \bibinfo {author} {\bibfnamefont {E.~M.}\ \bibnamefont {Lifshitz}},\ and\ \bibinfo {author} {\bibfnamefont {L.~P.}\ \bibnamefont {Pitaevskii}},\ }\bibfield  {title} {\bibinfo {title} {{C}hapter {XII} -- {R}adiative corrections, {T}he relativistic equation for bound states},\ }in\ \href {https://doi.org/https://doi.org/10.1016/B978-0-08-050346-2.50017-9} {\emph {\bibinfo {booktitle} {{Quantum Electrodynamics (Second Edition)}}}},\ \bibinfo {editor} {edited by\ \bibinfo {editor} {\bibfnamefont {L.~D.}\ \bibnamefont {Landau}}\ and\ \bibinfo {editor} {\bibfnamefont {E.~M.}\ \bibnamefont {Lifshitz}}}\ (\bibinfo  {publisher} {Pergamon Press},\ \bibinfo {address} {Oxford},\ \bibinfo {year} {1982})\ pp.\ \bibinfo {pages} {501--596}\BibitemShut {NoStop}%
\bibitem [{\citenamefont {Wannier}(1937)}]{Wannier1937}%
  \BibitemOpen
  \bibfield  {author} {\bibinfo {author} {\bibfnamefont {G.~H.}\ \bibnamefont {Wannier}},\ }\bibfield  {title} {\bibinfo {title} {{The Structure of Electronic Excitation Levels in Insulating Crystals}},\ }\href {https://doi.org/10.1103/PhysRev.52.191} {\bibfield  {journal} {\bibinfo  {journal} {Phys. Rev.}\ }\textbf {\bibinfo {volume} {52}},\ \bibinfo {pages} {191} (\bibinfo {year} {1937})}\BibitemShut {NoStop}%
\bibitem [{\citenamefont {Cooper}(1956)}]{Cooper1956}%
  \BibitemOpen
  \bibfield  {author} {\bibinfo {author} {\bibfnamefont {L.~N.}\ \bibnamefont {Cooper}},\ }\bibfield  {title} {\bibinfo {title} {{Bound Electron Pairs in a Degenerate Fermi Gas}},\ }\href {https://doi.org/10.1103/PhysRev.104.1189} {\bibfield  {journal} {\bibinfo  {journal} {Phys. Rev.}\ }\textbf {\bibinfo {volume} {104}},\ \bibinfo {pages} {1189} (\bibinfo {year} {1956})}\BibitemShut {NoStop}%
\bibitem [{\citenamefont {Combescot}\ and\ \citenamefont {Shiau}(2015)}]{Combescot2015}%
  \BibitemOpen
  \bibfield  {author} {\bibinfo {author} {\bibfnamefont {M.}~\bibnamefont {Combescot}}\ and\ \bibinfo {author} {\bibfnamefont {S.-Y.}\ \bibnamefont {Shiau}},\ }\href {https://doi.org/10.1093/acprof:oso/9780198753735.001.0001} {\emph {\bibinfo {title} {{Excitons and Cooper Pairs: Two Composite Bosons in Many-Body Physics}}}}\ (\bibinfo  {publisher} {Oxford University Press},\ \bibinfo {address} {New York},\ \bibinfo {year} {2015})\BibitemShut {NoStop}%
\bibitem [{\citenamefont {Hanus}(1963)}]{Hanus1963}%
  \BibitemOpen
  \bibfield  {author} {\bibinfo {author} {\bibfnamefont {J.}~\bibnamefont {Hanus}},\ }\bibfield  {title} {\bibinfo {title} {{Bound States in the Heisenberg Ferromagnet}},\ }\href {https://doi.org/10.1103/PhysRevLett.11.336} {\bibfield  {journal} {\bibinfo  {journal} {Phys. Rev. Lett.}\ }\textbf {\bibinfo {volume} {11}},\ \bibinfo {pages} {336} (\bibinfo {year} {1963})}\BibitemShut {NoStop}%
\bibitem [{\citenamefont {Wortis}(1963)}]{Wortis63}%
  \BibitemOpen
  \bibfield  {author} {\bibinfo {author} {\bibfnamefont {M.}~\bibnamefont {Wortis}},\ }\bibfield  {title} {\bibinfo {title} {{Bound States of Two Spin Waves in the Heisenberg Ferromagnet}},\ }\href {https://doi.org/10.1103/PhysRev.132.85} {\bibfield  {journal} {\bibinfo  {journal} {Phys. Rev.}\ }\textbf {\bibinfo {volume} {132}},\ \bibinfo {pages} {85} (\bibinfo {year} {1963})}\BibitemShut {NoStop}%
\bibitem [{\citenamefont {Rastelli}(2013)}]{Rastelli2013}%
  \BibitemOpen
  \bibfield  {author} {\bibinfo {author} {\bibfnamefont {E.}~\bibnamefont {Rastelli}},\ }\href {https://doi.org/10.1142/8189} {\emph {\bibinfo {title} {Statistical Mechanics of Magnetic Excitations}}}\ (\bibinfo  {publisher} {World Scientific},\ \bibinfo {address} {Singapore},\ \bibinfo {year} {2013})\BibitemShut {NoStop}%
\bibitem [{\citenamefont {Woodland}\ \emph {et~al.}(2023)\citenamefont {Woodland}, \citenamefont {Lovas}, \citenamefont {Telling}, \citenamefont {Prabhakaran}, \citenamefont {Balents},\ and\ \citenamefont {Coldea}}]{WoodlandLovas2023CoNb2O6}%
  \BibitemOpen
  \bibfield  {author} {\bibinfo {author} {\bibfnamefont {L.}~\bibnamefont {Woodland}}, \bibinfo {author} {\bibfnamefont {I.}~\bibnamefont {Lovas}}, \bibinfo {author} {\bibfnamefont {M.}~\bibnamefont {Telling}}, \bibinfo {author} {\bibfnamefont {D.}~\bibnamefont {Prabhakaran}}, \bibinfo {author} {\bibfnamefont {L.}~\bibnamefont {Balents}},\ and\ \bibinfo {author} {\bibfnamefont {R.}~\bibnamefont {Coldea}},\ }\bibfield  {title} {\bibinfo {title} {{Excitations of quantum Ising chain ${\text{CoNb}}_{2}{\text{O}}_{6}$ in low transverse field: Quantitative description of bound states stabilized by off-diagonal exchange and applied field}},\ }\href {https://doi.org/10.1103/PhysRevB.108.184417} {\bibfield  {journal} {\bibinfo  {journal} {Phys. Rev. B}\ }\textbf {\bibinfo {volume} {108}},\ \bibinfo {pages} {184417} (\bibinfo {year} {2023})}\BibitemShut {NoStop}%
\bibitem [{\citenamefont {Legros}\ \emph {et~al.}(2021)\citenamefont {Legros}, \citenamefont {Zhang}, \citenamefont {Bai}, \citenamefont {Zhang}, \citenamefont {Dun}, \citenamefont {Phelan}, \citenamefont {Batista}, \citenamefont {Mourigal},\ and\ \citenamefont {Armitage}}]{Armitage2021MagBoundStates}%
  \BibitemOpen
  \bibfield  {author} {\bibinfo {author} {\bibfnamefont {A.}~\bibnamefont {Legros}}, \bibinfo {author} {\bibfnamefont {S.-S.}\ \bibnamefont {Zhang}}, \bibinfo {author} {\bibfnamefont {X.}~\bibnamefont {Bai}}, \bibinfo {author} {\bibfnamefont {H.}~\bibnamefont {Zhang}}, \bibinfo {author} {\bibfnamefont {Z.}~\bibnamefont {Dun}}, \bibinfo {author} {\bibfnamefont {W.~A.}\ \bibnamefont {Phelan}}, \bibinfo {author} {\bibfnamefont {C.~D.}\ \bibnamefont {Batista}}, \bibinfo {author} {\bibfnamefont {M.}~\bibnamefont {Mourigal}},\ and\ \bibinfo {author} {\bibfnamefont {N.~P.}\ \bibnamefont {Armitage}},\ }\bibfield  {title} {\bibinfo {title} {{Observation of 4- and 6-Magnon Bound States in the Spin-Anisotropic Frustrated Antiferromagnet ${\mathrm{FeI}}_{2}$}},\ }\href {https://doi.org/10.1103/PhysRevLett.127.267201} {\bibfield  {journal} {\bibinfo  {journal} {Phys. Rev. Lett.}\ }\textbf {\bibinfo {volume} {127}},\ \bibinfo {pages} {267201} (\bibinfo {year} {2021})}\BibitemShut {NoStop}%
\bibitem [{\citenamefont {Ghioldi}\ \emph {et~al.}(2022)\citenamefont {Ghioldi}, \citenamefont {Zhang}, \citenamefont {Kamiya}, \citenamefont {Manuel}, \citenamefont {Trumper},\ and\ \citenamefont {Batista}}]{Ghioldi2022}%
  \BibitemOpen
  \bibfield  {author} {\bibinfo {author} {\bibfnamefont {E.~A.}\ \bibnamefont {Ghioldi}}, \bibinfo {author} {\bibfnamefont {S.-S.}\ \bibnamefont {Zhang}}, \bibinfo {author} {\bibfnamefont {Y.}~\bibnamefont {Kamiya}}, \bibinfo {author} {\bibfnamefont {L.~O.}\ \bibnamefont {Manuel}}, \bibinfo {author} {\bibfnamefont {A.~E.}\ \bibnamefont {Trumper}},\ and\ \bibinfo {author} {\bibfnamefont {C.~D.}\ \bibnamefont {Batista}},\ }\bibfield  {title} {\bibinfo {title} {{Evidence of two-spinon bound states in the magnetic spectrum of ${\mathrm{Ba}}_{3}{\mathrm{CoSb}}_{2}{\mathrm{O}}_{9}$}},\ }\href {https://doi.org/10.1103/PhysRevB.106.064418} {\bibfield  {journal} {\bibinfo  {journal} {Phys. Rev. B}\ }\textbf {\bibinfo {volume} {106}},\ \bibinfo {pages} {064418} (\bibinfo {year} {2022})}\BibitemShut {NoStop}%
\bibitem [{\citenamefont {El~Mendili}\ \emph {et~al.}(2025)\citenamefont {El~Mendili}, \citenamefont {Ziman},\ and\ \citenamefont {Zhitomirsky}}]{Zhitomisky2025}%
  \BibitemOpen
  \bibfield  {author} {\bibinfo {author} {\bibfnamefont {A.}~\bibnamefont {El~Mendili}}, \bibinfo {author} {\bibfnamefont {T.}~\bibnamefont {Ziman}},\ and\ \bibinfo {author} {\bibfnamefont {M.~E.}\ \bibnamefont {Zhitomirsky}},\ }\bibfield  {title} {\bibinfo {title} {{Longitudinal magnons in large-$S$ easy-axis magnets}},\ }\href {https://doi.org/10.1103/gy3k-c1w3} {\bibfield  {journal} {\bibinfo  {journal} {Phys. Rev. B}\ }\textbf {\bibinfo {volume} {112}},\ \bibinfo {pages} {174433} (\bibinfo {year} {2025})}\BibitemShut {NoStop}%
\bibitem [{\citenamefont {Mila}(2017)}]{Mila_17}%
  \BibitemOpen
  \bibfield  {author} {\bibinfo {author} {\bibfnamefont {F.}~\bibnamefont {Mila}},\ }\bibfield  {title} {\bibinfo {title} {{Closing in on a Magnetic Analog of Liquid Crystals}},\ }\href {https://doi.org/10.1103/Physics.10.64} {\bibfield  {journal} {\bibinfo  {journal} {Physics}\ }\textbf {\bibinfo {volume} {10}},\ \bibinfo {pages} {64} (\bibinfo {year} {2017})}\BibitemShut {NoStop}%
\bibitem [{\citenamefont {Orlova}\ \emph {et~al.}(2017)\citenamefont {Orlova}, \citenamefont {Green}, \citenamefont {Law}, \citenamefont {Gorbunov}, \citenamefont {Chanda}, \citenamefont {Kr\"amer}, \citenamefont {Horvati\ifmmode~\acute{c}\else \'{c}\fi{}}, \citenamefont {Kremer}, \citenamefont {Wosnitza},\ and\ \citenamefont {Rikken}}]{exp2017licuvo4}%
  \BibitemOpen
  \bibfield  {author} {\bibinfo {author} {\bibfnamefont {A.}~\bibnamefont {Orlova}}, \bibinfo {author} {\bibfnamefont {E.~L.}\ \bibnamefont {Green}}, \bibinfo {author} {\bibfnamefont {J.~M.}\ \bibnamefont {Law}}, \bibinfo {author} {\bibfnamefont {D.~I.}\ \bibnamefont {Gorbunov}}, \bibinfo {author} {\bibfnamefont {G.}~\bibnamefont {Chanda}}, \bibinfo {author} {\bibfnamefont {S.}~\bibnamefont {Kr\"amer}}, \bibinfo {author} {\bibfnamefont {M.}~\bibnamefont {Horvati\ifmmode~\acute{c}\else \'{c}\fi{}}}, \bibinfo {author} {\bibfnamefont {R.~K.}\ \bibnamefont {Kremer}}, \bibinfo {author} {\bibfnamefont {J.}~\bibnamefont {Wosnitza}},\ and\ \bibinfo {author} {\bibfnamefont {G.~L. J.~A.}\ \bibnamefont {Rikken}},\ }\bibfield  {title} {\bibinfo {title} {{Nuclear magnetic resonance signature of the spin-nematic phase in {L}i{C}u{VO}$_4$ at high magnetic fields}},\ }\href {https://doi.org/10.1103/PhysRevLett.118.247201} {\bibfield  {journal} {\bibinfo  {journal} {Phys. Rev. Lett.}\ }\textbf {\bibinfo {volume} {118}},\
  \bibinfo {pages} {247201} (\bibinfo {year} {2017})}\BibitemShut {NoStop}%
\bibitem [{\citenamefont {Kohama}\ \emph {et~al.}(2019)\citenamefont {Kohama}, \citenamefont {Ishikawa}, \citenamefont {Matsuo}, \citenamefont {Kindo}, \citenamefont {Shannon},\ and\ \citenamefont {Hiroi}}]{exp2019sq}%
  \BibitemOpen
  \bibfield  {author} {\bibinfo {author} {\bibfnamefont {Y.}~\bibnamefont {Kohama}}, \bibinfo {author} {\bibfnamefont {H.}~\bibnamefont {Ishikawa}}, \bibinfo {author} {\bibfnamefont {A.}~\bibnamefont {Matsuo}}, \bibinfo {author} {\bibfnamefont {K.}~\bibnamefont {Kindo}}, \bibinfo {author} {\bibfnamefont {N.}~\bibnamefont {Shannon}},\ and\ \bibinfo {author} {\bibfnamefont {Z.}~\bibnamefont {Hiroi}},\ }\bibfield  {title} {\bibinfo {title} {{Possible observation of quantum spin-nematic phase in a frustrated magnet}},\ }\href {https://doi.org/10.1073/pnas.1821969116} {\bibfield  {journal} {\bibinfo  {journal} {Proc. Natl. Acad. Sci. U.S.A.}\ }\textbf {\bibinfo {volume} {116}},\ \bibinfo {pages} {10686} (\bibinfo {year} {2019})}\BibitemShut {NoStop}%
\bibitem [{\citenamefont {Bhartiya}\ \emph {et~al.}(2019)\citenamefont {Bhartiya}, \citenamefont {Povarov}, \citenamefont {Blosser}, \citenamefont {Bettler}, \citenamefont {Yan}, \citenamefont {Gvasaliya}, \citenamefont {Raymond}, \citenamefont {Ressouche}, \citenamefont {Beauvois}, \citenamefont {Xu}, \citenamefont {Yokaichiya},\ and\ \citenamefont {Zheludev}}]{zheludev2}%
  \BibitemOpen
  \bibfield  {author} {\bibinfo {author} {\bibfnamefont {V.~K.}\ \bibnamefont {Bhartiya}}, \bibinfo {author} {\bibfnamefont {K.~Y.}\ \bibnamefont {Povarov}}, \bibinfo {author} {\bibfnamefont {D.}~\bibnamefont {Blosser}}, \bibinfo {author} {\bibfnamefont {S.}~\bibnamefont {Bettler}}, \bibinfo {author} {\bibfnamefont {Z.}~\bibnamefont {Yan}}, \bibinfo {author} {\bibfnamefont {S.}~\bibnamefont {Gvasaliya}}, \bibinfo {author} {\bibfnamefont {S.}~\bibnamefont {Raymond}}, \bibinfo {author} {\bibfnamefont {E.}~\bibnamefont {Ressouche}}, \bibinfo {author} {\bibfnamefont {K.}~\bibnamefont {Beauvois}}, \bibinfo {author} {\bibfnamefont {J.}~\bibnamefont {Xu}}, \bibinfo {author} {\bibfnamefont {F.}~\bibnamefont {Yokaichiya}},\ and\ \bibinfo {author} {\bibfnamefont {A.}~\bibnamefont {Zheludev}},\ }\bibfield  {title} {\bibinfo {title} {{Presaturation phase with no dipolar order in a quantum ferro-antiferromagnet}},\ }\href {https://doi.org/10.1103/PhysRevResearch.1.033078} {\bibfield  {journal} {\bibinfo  {journal} {Phys.
  Rev. Research}\ }\textbf {\bibinfo {volume} {1}},\ \bibinfo {pages} {033078} (\bibinfo {year} {2019})}\BibitemShut {NoStop}%
\bibitem [{\citenamefont {Blume}\ and\ \citenamefont {Hsieh}(1969)}]{1969nm}%
  \BibitemOpen
  \bibfield  {author} {\bibinfo {author} {\bibfnamefont {M.}~\bibnamefont {Blume}}\ and\ \bibinfo {author} {\bibfnamefont {Y.~Y.}\ \bibnamefont {Hsieh}},\ }\bibfield  {title} {\bibinfo {title} {{Biquadratic exchange and quadrupolar ordering}},\ }\href {https://doi.org/10.1063/1.1657616} {\bibfield  {journal} {\bibinfo  {journal} {J. Appl. Phys.}\ }\textbf {\bibinfo {volume} {40}},\ \bibinfo {pages} {1249} (\bibinfo {year} {1969})}\BibitemShut {NoStop}%
\bibitem [{\citenamefont {Andreev}\ and\ \citenamefont {Grishchuk}(1984)}]{1984nm}%
  \BibitemOpen
  \bibfield  {author} {\bibinfo {author} {\bibfnamefont {A.~F.}\ \bibnamefont {Andreev}}\ and\ \bibinfo {author} {\bibfnamefont {I.~A.}\ \bibnamefont {Grishchuk}},\ }\bibfield  {title} {\bibinfo {title} {{Spin nematics}},\ }\href {http://www.jetp.ras.ru/cgi-bin/dn/e_060_02_0267.pdf} {\bibfield  {journal} {\bibinfo  {journal} {Sov. Phys. JETP}\ }\textbf {\bibinfo {volume} {60}},\ \bibinfo {pages} {267} (\bibinfo {year} {1984})}\BibitemShut {NoStop}%
\bibitem [{\citenamefont {Papanicolaou}(1988)}]{Papanicolaou1988}%
  \BibitemOpen
  \bibfield  {author} {\bibinfo {author} {\bibfnamefont {N.}~\bibnamefont {Papanicolaou}},\ }\bibfield  {title} {\bibinfo {title} {{Unusual phases in quantum spin-1 systems}},\ }\href {https://doi.org/https://doi.org/10.1016/0550-3213(88)90073-9} {\bibfield  {journal} {\bibinfo  {journal} {Nucl. Phys. B}\ }\textbf {\bibinfo {volume} {305}},\ \bibinfo {pages} {367} (\bibinfo {year} {1988})}\BibitemShut {NoStop}%
\bibitem [{\citenamefont {Savary}\ and\ \citenamefont {Balents}(2017)}]{Savary2016QSL}%
  \BibitemOpen
  \bibfield  {author} {\bibinfo {author} {\bibfnamefont {L.}~\bibnamefont {Savary}}\ and\ \bibinfo {author} {\bibfnamefont {L.}~\bibnamefont {Balents}},\ }\bibfield  {title} {\bibinfo {title} {{Quantum spin liquids: a review}},\ }\href {https://doi.org/10.1088/0034-4885/80/1/016502} {\bibfield  {journal} {\bibinfo  {journal} {Rep. Prog. Phys.}\ }\textbf {\bibinfo {volume} {80}},\ \bibinfo {pages} {016502} (\bibinfo {year} {2017})}\BibitemShut {NoStop}%
\bibitem [{\citenamefont {Chubukov}(1991)}]{Chubukov1991Nematic}%
  \BibitemOpen
  \bibfield  {author} {\bibinfo {author} {\bibfnamefont {A.~V.}\ \bibnamefont {Chubukov}},\ }\bibfield  {title} {\bibinfo {title} {{Chiral, nematic, and dimer states in quantum spin chains}},\ }\href {https://doi.org/10.1103/PhysRevB.44.4693} {\bibfield  {journal} {\bibinfo  {journal} {Phys. Rev. B}\ }\textbf {\bibinfo {volume} {44}},\ \bibinfo {pages} {4693} (\bibinfo {year} {1991})}\BibitemShut {NoStop}%
\bibitem [{\citenamefont {Penc}\ and\ \citenamefont {L{\"a}uchli}(2011)}]{Penc2011Nematic}%
  \BibitemOpen
  \bibfield  {author} {\bibinfo {author} {\bibfnamefont {K.}~\bibnamefont {Penc}}\ and\ \bibinfo {author} {\bibfnamefont {A.~M.}\ \bibnamefont {L{\"a}uchli}},\ }\href {https://doi.org/10.1007/978-3-642-10589-0_13} {\emph {\bibinfo {title} {{Introduction to Frustrated Magnetism: Materials, Experiments, Theory}}}},\ edited by\ \bibinfo {editor} {\bibfnamefont {C.}~\bibnamefont {Lacroix}}, \bibinfo {editor} {\bibfnamefont {P.}~\bibnamefont {Mendels}},\ and\ \bibinfo {editor} {\bibfnamefont {F.}~\bibnamefont {Mila}}\ (\bibinfo  {publisher} {Springer},\ \bibinfo {address} {Berlin, Heidelberg},\ \bibinfo {year} {2011})\ pp.\ \bibinfo {pages} {331--362}\BibitemShut {NoStop}%
\bibitem [{\citenamefont {Jiang}\ \emph {et~al.}(2023)\citenamefont {Jiang}, \citenamefont {Romh\'anyi}, \citenamefont {White}, \citenamefont {Zhitomirsky},\ and\ \citenamefont {Chernyshev}}]{NematicShengtao2023}%
  \BibitemOpen
  \bibfield  {author} {\bibinfo {author} {\bibfnamefont {S.}~\bibnamefont {Jiang}}, \bibinfo {author} {\bibfnamefont {J.}~\bibnamefont {Romh\'anyi}}, \bibinfo {author} {\bibfnamefont {S.~R.}\ \bibnamefont {White}}, \bibinfo {author} {\bibfnamefont {M.~E.}\ \bibnamefont {Zhitomirsky}},\ and\ \bibinfo {author} {\bibfnamefont {A.~L.}\ \bibnamefont {Chernyshev}},\ }\bibfield  {title} {\bibinfo {title} {{Where is the Quantum Spin Nematic?}},\ }\href {https://doi.org/10.1103/PhysRevLett.130.116701} {\bibfield  {journal} {\bibinfo  {journal} {Phys. Rev. Lett.}\ }\textbf {\bibinfo {volume} {130}},\ \bibinfo {pages} {116701} (\bibinfo {year} {2023})}\BibitemShut {NoStop}%
\bibitem [{\citenamefont {Batyev}\ and\ \citenamefont {Braginskii}(1984)}]{Batyev}%
  \BibitemOpen
  \bibfield  {author} {\bibinfo {author} {\bibfnamefont {{\'{E}}.~G.}\ \bibnamefont {Batyev}}\ and\ \bibinfo {author} {\bibfnamefont {L.~S.}\ \bibnamefont {Braginskii}},\ }\bibfield  {title} {\bibinfo {title} {{Antiferrornagnet in a strong magnetic field: analogy with {Bose} gas}},\ }\href {https://jetp.ras.ru/cgi-bin/dn/e_069_05_1033.pdf} {\bibfield  {journal} {\bibinfo  {journal} {Sov. Phys. JETP}\ }\textbf {\bibinfo {volume} {60}},\ \bibinfo {pages} {781} (\bibinfo {year} {1984})}\BibitemShut {NoStop}%
\bibitem [{\citenamefont {Zapf}\ \emph {et~al.}(2014)\citenamefont {Zapf}, \citenamefont {Jaime},\ and\ \citenamefont {Batista}}]{Batista_14}%
  \BibitemOpen
  \bibfield  {author} {\bibinfo {author} {\bibfnamefont {V.}~\bibnamefont {Zapf}}, \bibinfo {author} {\bibfnamefont {M.}~\bibnamefont {Jaime}},\ and\ \bibinfo {author} {\bibfnamefont {C.~D.}\ \bibnamefont {Batista}},\ }\bibfield  {title} {\bibinfo {title} {{Bose-Einstein condensation in quantum magnets}},\ }\href {https://doi.org/10.1103/RevModPhys.86.563} {\bibfield  {journal} {\bibinfo  {journal} {Rev. Mod. Phys.}\ }\textbf {\bibinfo {volume} {86}},\ \bibinfo {pages} {563} (\bibinfo {year} {2014})}\BibitemShut {NoStop}%
\bibitem [{\citenamefont {Ueda}\ and\ \citenamefont {Momoi}(2013)}]{ueda_phasesep}%
  \BibitemOpen
  \bibfield  {author} {\bibinfo {author} {\bibfnamefont {H.~T.}\ \bibnamefont {Ueda}}\ and\ \bibinfo {author} {\bibfnamefont {T.}~\bibnamefont {Momoi}},\ }\bibfield  {title} {\bibinfo {title} {Nematic phase and phase separation near saturation field in frustrated ferromagnets},\ }\href {https://doi.org/10.1103/PhysRevB.87.144417} {\bibfield  {journal} {\bibinfo  {journal} {Phys. Rev. B}\ }\textbf {\bibinfo {volume} {87}},\ \bibinfo {pages} {144417} (\bibinfo {year} {2013})}\BibitemShut {NoStop}%
\bibitem [{\citenamefont {Mattis}(1986)}]{Mattis1986RMP}%
  \BibitemOpen
  \bibfield  {author} {\bibinfo {author} {\bibfnamefont {D.~C.}\ \bibnamefont {Mattis}},\ }\bibfield  {title} {\bibinfo {title} {{The few-body problem on a lattice}},\ }\href {https://doi.org/10.1103/RevModPhys.58.361} {\bibfield  {journal} {\bibinfo  {journal} {Rev. Mod. Phys.}\ }\textbf {\bibinfo {volume} {58}},\ \bibinfo {pages} {361} (\bibinfo {year} {1986})}\BibitemShut {NoStop}%
\bibitem [{\citenamefont {Kornilovitch}(2024)}]{Kornilovitch2024}%
  \BibitemOpen
  \bibfield  {author} {\bibinfo {author} {\bibfnamefont {P.~E.}\ \bibnamefont {Kornilovitch}},\ }\bibfield  {title} {\bibinfo {title} {{Two-particle bound states on a lattice}},\ }\href {https://doi.org/https://doi.org/10.1016/j.aop.2023.169574} {\bibfield  {journal} {\bibinfo  {journal} {Ann. Phys. (N. Y.)}\ }\textbf {\bibinfo {volume} {460}},\ \bibinfo {pages} {169574} (\bibinfo {year} {2024})}\BibitemShut {NoStop}%
\bibitem [{\citenamefont {Reiter}(1968)}]{Reiter1968}%
  \BibitemOpen
  \bibfield  {author} {\bibinfo {author} {\bibfnamefont {G.~F.}\ \bibnamefont {Reiter}},\ }\bibfield  {title} {\bibinfo {title} {{Magnon Density Fluctuations in the Heisenberg Ferromagnet}},\ }\href {https://doi.org/10.1103/PhysRev.175.631} {\bibfield  {journal} {\bibinfo  {journal} {Phys. Rev.}\ }\textbf {\bibinfo {volume} {175}},\ \bibinfo {pages} {631} (\bibinfo {year} {1968})}\BibitemShut {NoStop}%
\bibitem [{\citenamefont {Rodriguez-Nieva}\ \emph {et~al.}(2022)\citenamefont {Rodriguez-Nieva}, \citenamefont {Podolsky},\ and\ \citenamefont {Demler}}]{RodriguezDemlerPRB2022}%
  \BibitemOpen
  \bibfield  {author} {\bibinfo {author} {\bibfnamefont {J.~F.}\ \bibnamefont {Rodriguez-Nieva}}, \bibinfo {author} {\bibfnamefont {D.}~\bibnamefont {Podolsky}},\ and\ \bibinfo {author} {\bibfnamefont {E.}~\bibnamefont {Demler}},\ }\bibfield  {title} {\bibinfo {title} {{Probing hydrodynamic sound modes in magnon fluids using spin magnetometers}},\ }\href {https://doi.org/10.1103/PhysRevB.105.174412} {\bibfield  {journal} {\bibinfo  {journal} {Phys. Rev. B}\ }\textbf {\bibinfo {volume} {105}},\ \bibinfo {pages} {174412} (\bibinfo {year} {2022})}\BibitemShut {NoStop}%
\bibitem [{\citenamefont {Bethe}(1931)}]{Bethe1931}%
  \BibitemOpen
  \bibfield  {author} {\bibinfo {author} {\bibfnamefont {H.~A.}\ \bibnamefont {Bethe}},\ }\bibfield  {title} {\bibinfo {title} {{Zur Theorie der Metalle. I. Eigenwerte und Eigenfunktionen der linearen Atomkette}},\ }\href {https://doi.org/10.1007/BF01341708} {\bibfield  {journal} {\bibinfo  {journal} {Zeitschrift f{\"u}r Physik}\ }\textbf {\bibinfo {volume} {71}},\ \bibinfo {pages} {205} (\bibinfo {year} {1931})}\BibitemShut {NoStop}%
\bibitem [{\citenamefont {Dyson}(1956)}]{Dyson1956}%
  \BibitemOpen
  \bibfield  {author} {\bibinfo {author} {\bibfnamefont {F.~J.}\ \bibnamefont {Dyson}},\ }\bibfield  {title} {\bibinfo {title} {{General Theory of Spin-Wave Interactions}},\ }\href {https://doi.org/10.1103/PhysRev.102.1217} {\bibfield  {journal} {\bibinfo  {journal} {Phys. Rev.}\ }\textbf {\bibinfo {volume} {102}},\ \bibinfo {pages} {1217} (\bibinfo {year} {1956})}\BibitemShut {NoStop}%
\bibitem [{\citenamefont {Feynman}(1998)}]{Feynman1998SM}%
  \BibitemOpen
  \bibfield  {author} {\bibinfo {author} {\bibfnamefont {R.~P.}\ \bibnamefont {Feynman}},\ }\href {https://doi.org/10.1201/9780429493034} {\emph {\bibinfo {title} {{Statistical Mechanics: A Set of Lectures}}}}\ (\bibinfo  {publisher} {CRC Press},\ \bibinfo {address} {Boca Raton},\ \bibinfo {year} {1998})\ pp.\ \bibinfo {pages} {198--220}\BibitemShut {NoStop}%
\bibitem [{\citenamefont {Akhiezer}\ \emph {et~al.}(1968)\citenamefont {Akhiezer}, \citenamefont {Bar'yakhtar},\ and\ \citenamefont {Peletminskii}}]{Akhiezer1968SW}%
  \BibitemOpen
  \bibfield  {author} {\bibinfo {author} {\bibfnamefont {A.~I.}\ \bibnamefont {Akhiezer}}, \bibinfo {author} {\bibfnamefont {V.~G.}\ \bibnamefont {Bar'yakhtar}},\ and\ \bibinfo {author} {\bibfnamefont {S.~V.}\ \bibnamefont {Peletminskii}},\ }\href@noop {} {\emph {\bibinfo {title} {{Spin Waves}}}},\ edited by\ \bibinfo {editor} {\bibfnamefont {S.}~\bibnamefont {Doniach}}\ (\bibinfo  {publisher} {North-Holland Publishing Company},\ \bibinfo {address} {Amsterdam},\ \bibinfo {year} {1968})\BibitemShut {NoStop}%
\bibitem [{\citenamefont {Mattis}(2006)}]{Mattis2006book}%
  \BibitemOpen
  \bibfield  {author} {\bibinfo {author} {\bibfnamefont {D.~C.}\ \bibnamefont {Mattis}},\ }\href {https://doi.org/10.1142/5372} {\emph {\bibinfo {title} {{The Theory of Magnetism Made Simple}}}}\ (\bibinfo  {publisher} {World Scientific},\ \bibinfo {address} {Singapore},\ \bibinfo {year} {2006})\BibitemShut {NoStop}%
\bibitem [{\citenamefont {Fukuda}\ and\ \citenamefont {Wortis}(1963)}]{Fukuda1963}%
  \BibitemOpen
  \bibfield  {author} {\bibinfo {author} {\bibfnamefont {N.}~\bibnamefont {Fukuda}}\ and\ \bibinfo {author} {\bibfnamefont {M.}~\bibnamefont {Wortis}},\ }\bibfield  {title} {\bibinfo {title} {Bound states in the spin wave problem},\ }\href {https://doi.org/https://doi.org/10.1016/0022-3697(63)90115-X} {\bibfield  {journal} {\bibinfo  {journal} {J. Phys. Chem. Solids}\ }\textbf {\bibinfo {volume} {24}},\ \bibinfo {pages} {1675} (\bibinfo {year} {1963})}\BibitemShut {NoStop}%
\bibitem [{\citenamefont {Boyd}\ and\ \citenamefont {Callaway}(1965)}]{Callaway1965}%
  \BibitemOpen
  \bibfield  {author} {\bibinfo {author} {\bibfnamefont {R.~G.}\ \bibnamefont {Boyd}}\ and\ \bibinfo {author} {\bibfnamefont {J.}~\bibnamefont {Callaway}},\ }\bibfield  {title} {\bibinfo {title} {{Spin-Wave--Spin-Wave Scattering in a Heisenberg Ferromagnet}},\ }\href {https://doi.org/10.1103/PhysRev.138.A1621} {\bibfield  {journal} {\bibinfo  {journal} {Phys. Rev.}\ }\textbf {\bibinfo {volume} {138}},\ \bibinfo {pages} {A1621} (\bibinfo {year} {1965})}\BibitemShut {NoStop}%
\bibitem [{\citenamefont {Oguchi}(1971)}]{Oguchi1971}%
  \BibitemOpen
  \bibfield  {author} {\bibinfo {author} {\bibfnamefont {T.}~\bibnamefont {Oguchi}},\ }\bibfield  {title} {\bibinfo {title} {{Theory of Two-Magnon Bound States in the Heisenberg Ferro- and Antiferromagnet}},\ }\href {https://doi.org/10.1143/JPSJ.31.394} {\bibfield  {journal} {\bibinfo  {journal} {J. Phys. Soc. Jpn}\ }\textbf {\bibinfo {volume} {31}},\ \bibinfo {pages} {394} (\bibinfo {year} {1971})}\BibitemShut {NoStop}%
\bibitem [{\citenamefont {Zhitomirsky}\ and\ \citenamefont {Tsunetsugu}(2010)}]{Zhitomirsky2010}%
  \BibitemOpen
  \bibfield  {author} {\bibinfo {author} {\bibfnamefont {M.~E.}\ \bibnamefont {Zhitomirsky}}\ and\ \bibinfo {author} {\bibfnamefont {H.}~\bibnamefont {Tsunetsugu}},\ }\bibfield  {title} {\bibinfo {title} {{Magnon pairing in quantum spin nematic}},\ }\href {https://doi.org/10.1209/0295-5075/92/37001} {\bibfield  {journal} {\bibinfo  {journal} {EPL}\ }\textbf {\bibinfo {volume} {92}},\ \bibinfo {pages} {37001} (\bibinfo {year} {2010})}\BibitemShut {NoStop}%
\bibitem [{\citenamefont {Chiu-Tsao}\ \emph {et~al.}(1975)\citenamefont {Chiu-Tsao}, \citenamefont {Levy},\ and\ \citenamefont {Paulson}}]{Paulson1975}%
  \BibitemOpen
  \bibfield  {author} {\bibinfo {author} {\bibfnamefont {S.~T.}\ \bibnamefont {Chiu-Tsao}}, \bibinfo {author} {\bibfnamefont {P.~M.}\ \bibnamefont {Levy}},\ and\ \bibinfo {author} {\bibfnamefont {C.}~\bibnamefont {Paulson}},\ }\bibfield  {title} {\bibinfo {title} {{Elementary excitations of high-degree pair interactions: The two-spin-deviation spectra for a spin-1 ferromagnet}},\ }\href {https://doi.org/10.1103/PhysRevB.12.1819} {\bibfield  {journal} {\bibinfo  {journal} {Phys. Rev. B}\ }\textbf {\bibinfo {volume} {12}},\ \bibinfo {pages} {1819} (\bibinfo {year} {1975})}\BibitemShut {NoStop}%
\bibitem [{\citenamefont {Loly}\ and\ \citenamefont {Choudhury}(1976)}]{Loly1976}%
  \BibitemOpen
  \bibfield  {author} {\bibinfo {author} {\bibfnamefont {P.~D.}\ \bibnamefont {Loly}}\ and\ \bibinfo {author} {\bibfnamefont {B.~J.}\ \bibnamefont {Choudhury}},\ }\bibfield  {title} {\bibinfo {title} {{Two-magnon spectra and Ising anisotropy: The relationship between resonances and bound states}},\ }\href {https://doi.org/10.1103/PhysRevB.13.4019} {\bibfield  {journal} {\bibinfo  {journal} {Phys. Rev. B}\ }\textbf {\bibinfo {volume} {13}},\ \bibinfo {pages} {4019} (\bibinfo {year} {1976})}\BibitemShut {NoStop}%
\bibitem [{\citenamefont {Wada}\ \emph {et~al.}(1975)\citenamefont {Wada}, \citenamefont {Ishikawa},\ and\ \citenamefont {Oguchi}}]{Oguchi1975}%
  \BibitemOpen
  \bibfield  {author} {\bibinfo {author} {\bibfnamefont {K.}~\bibnamefont {Wada}}, \bibinfo {author} {\bibfnamefont {T.}~\bibnamefont {Ishikawa}},\ and\ \bibinfo {author} {\bibfnamefont {T.}~\bibnamefont {Oguchi}},\ }\bibfield  {title} {\bibinfo {title} {{Two-Magnon Bound States in the Triangular and Honeycomb Heisenberg Ferromagnets}},\ }\href {https://doi.org/10.1143/PTP.54.1589} {\bibfield  {journal} {\bibinfo  {journal} {Prog. Theor. Phys.}\ }\textbf {\bibinfo {volume} {54}},\ \bibinfo {pages} {1589} (\bibinfo {year} {1975})}\BibitemShut {NoStop}%
\bibitem [{\citenamefont {Tonegawa}(1970)}]{Tonegawa1970}%
  \BibitemOpen
  \bibfield  {author} {\bibinfo {author} {\bibfnamefont {T.}~\bibnamefont {Tonegawa}},\ }\bibfield  {title} {\bibinfo {title} {{Two-Magnon Bound States in the Heisenberg Ferromagnet with Anisotropic Exchange and Uniaxial Anisotropy Energies}},\ }\href {https://doi.org/10.1143/PTPS.46.61} {\bibfield  {journal} {\bibinfo  {journal} {Prog. Theor. Phys. Suppl.}\ }\textbf {\bibinfo {volume} {46}},\ \bibinfo {pages} {61} (\bibinfo {year} {1970})}\BibitemShut {NoStop}%
\bibitem [{\citenamefont {Silberglitt}\ and\ \citenamefont {Torrance}(1970)}]{Torrance1970}%
  \BibitemOpen
  \bibfield  {author} {\bibinfo {author} {\bibfnamefont {R.}~\bibnamefont {Silberglitt}}\ and\ \bibinfo {author} {\bibfnamefont {J.~B.}\ \bibnamefont {Torrance}},\ }\bibfield  {title} {\bibinfo {title} {{Effect of Single-Ion Anisotropy on Two-Spin-Wave Bound State in a Heisenberg Ferromagnet}},\ }\href {https://doi.org/10.1103/PhysRevB.2.772} {\bibfield  {journal} {\bibinfo  {journal} {Phys. Rev. B}\ }\textbf {\bibinfo {volume} {2}},\ \bibinfo {pages} {772} (\bibinfo {year} {1970})}\BibitemShut {NoStop}%
\bibitem [{\citenamefont {Hood}\ and\ \citenamefont {Loly}(1986)}]{Loly1986}%
  \BibitemOpen
  \bibfield  {author} {\bibinfo {author} {\bibfnamefont {M.}~\bibnamefont {Hood}}\ and\ \bibinfo {author} {\bibfnamefont {P.~D.}\ \bibnamefont {Loly}},\ }\bibfield  {title} {\bibinfo {title} {{The two-magnon spectrum for the Heisenberg ferromagnet with NN interactions on a square lattice}},\ }\href {https://doi.org/10.1088/0022-3719/19/24/015} {\bibfield  {journal} {\bibinfo  {journal} {J. Phys. C: Solid State Phys.}\ }\textbf {\bibinfo {volume} {19}},\ \bibinfo {pages} {4729} (\bibinfo {year} {1986})}\BibitemShut {NoStop}%
\bibitem [{\citenamefont {Mogil'ner}(1989)}]{Mogilner89}%
  \BibitemOpen
  \bibfield  {author} {\bibinfo {author} {\bibfnamefont {A.~I.}\ \bibnamefont {Mogil'ner}},\ }\bibfield  {title} {\bibinfo {title} {{Magnon bound states in an easy-axis Heisenberg ferromagnet of arbitrary dimensionality. Relation to magnetic solitons}},\ }\href {https://jetp.ras.ru/cgi-bin/dn/e_069_05_1033.pdf} {\bibfield  {journal} {\bibinfo  {journal} {Sov. Phys. JETP}\ }\textbf {\bibinfo {volume} {69}},\ \bibinfo {pages} {1033} (\bibinfo {year} {1989})}\BibitemShut {NoStop}%
\bibitem [{\citenamefont {Rastelli}\ \emph {et~al.}(1991)\citenamefont {Rastelli}, \citenamefont {Sedazzari},\ and\ \citenamefont {Tassi}}]{Rastelli1991}%
  \BibitemOpen
  \bibfield  {author} {\bibinfo {author} {\bibfnamefont {E.}~\bibnamefont {Rastelli}}, \bibinfo {author} {\bibfnamefont {S.}~\bibnamefont {Sedazzari}},\ and\ \bibinfo {author} {\bibfnamefont {A.}~\bibnamefont {Tassi}},\ }\bibfield  {title} {\bibinfo {title} {{Long-range order by crucial non-linear effects in Heisenberg models}},\ }\href {https://doi.org/10.1088/0953-8984/3/31/009} {\bibfield  {journal} {\bibinfo  {journal} {J. Phys.: Condens. Matter}\ }\textbf {\bibinfo {volume} {3}},\ \bibinfo {pages} {5861} (\bibinfo {year} {1991})}\BibitemShut {NoStop}%
\bibitem [{\citenamefont {Rastelli}\ \emph {et~al.}(1992)\citenamefont {Rastelli}, \citenamefont {Sedazzari},\ and\ \citenamefont {Tassi}}]{Rastelli1992}%
  \BibitemOpen
  \bibfield  {author} {\bibinfo {author} {\bibfnamefont {E.}~\bibnamefont {Rastelli}}, \bibinfo {author} {\bibfnamefont {S.}~\bibnamefont {Sedazzari}},\ and\ \bibinfo {author} {\bibfnamefont {A.}~\bibnamefont {Tassi}},\ }\bibfield  {title} {\bibinfo {title} {{Two-magnon bound states in the triangular ferromagnet}},\ }\href {https://doi.org/10.1088/0953-8984/4/29/012} {\bibfield  {journal} {\bibinfo  {journal} {J. Phys.: Condens. Matter}\ }\textbf {\bibinfo {volume} {4}},\ \bibinfo {pages} {6283} (\bibinfo {year} {1992})}\BibitemShut {NoStop}%
\bibitem [{\citenamefont {Akaki}\ \emph {et~al.}(2017)\citenamefont {Akaki}, \citenamefont {Yoshizawa}, \citenamefont {Okutani}, \citenamefont {Kida}, \citenamefont {Romh\'anyi}, \citenamefont {Penc},\ and\ \citenamefont {Hagiwara}}]{Penc_17}%
  \BibitemOpen
  \bibfield  {author} {\bibinfo {author} {\bibfnamefont {M.}~\bibnamefont {Akaki}}, \bibinfo {author} {\bibfnamefont {D.}~\bibnamefont {Yoshizawa}}, \bibinfo {author} {\bibfnamefont {A.}~\bibnamefont {Okutani}}, \bibinfo {author} {\bibfnamefont {T.}~\bibnamefont {Kida}}, \bibinfo {author} {\bibfnamefont {J.}~\bibnamefont {Romh\'anyi}}, \bibinfo {author} {\bibfnamefont {K.}~\bibnamefont {Penc}},\ and\ \bibinfo {author} {\bibfnamefont {M.}~\bibnamefont {Hagiwara}},\ }\bibfield  {title} {\bibinfo {title} {{Direct observation of spin-quadrupolar excitations in ${\mathrm{Sr}}_{2}{\mathrm{CoGe}}_{2}{\mathrm{O}}_{7}$ by high-field electron spin resonance}},\ }\href {https://doi.org/10.1103/PhysRevB.96.214406} {\bibfield  {journal} {\bibinfo  {journal} {Phys. Rev. B}\ }\textbf {\bibinfo {volume} {96}},\ \bibinfo {pages} {214406} (\bibinfo {year} {2017})}\BibitemShut {NoStop}%
\bibitem [{\citenamefont {Mook}\ \emph {et~al.}(2023)\citenamefont {Mook}, \citenamefont {Hoyer}, \citenamefont {Klinovaja},\ and\ \citenamefont {Loss}}]{Mook2023}%
  \BibitemOpen
  \bibfield  {author} {\bibinfo {author} {\bibfnamefont {A.}~\bibnamefont {Mook}}, \bibinfo {author} {\bibfnamefont {R.}~\bibnamefont {Hoyer}}, \bibinfo {author} {\bibfnamefont {J.}~\bibnamefont {Klinovaja}},\ and\ \bibinfo {author} {\bibfnamefont {D.}~\bibnamefont {Loss}},\ }\bibfield  {title} {\bibinfo {title} {{Magnons, magnon bound pairs, and their hybrid spin-multipolar topology}},\ }\href {https://doi.org/10.1103/PhysRevB.107.064429} {\bibfield  {journal} {\bibinfo  {journal} {Phys. Rev. B}\ }\textbf {\bibinfo {volume} {107}},\ \bibinfo {pages} {064429} (\bibinfo {year} {2023})}\BibitemShut {NoStop}%
\bibitem [{\citenamefont {H\"ohler}(1950)}]{firstBEC_FM1950}%
  \BibitemOpen
  \bibfield  {author} {\bibinfo {author} {\bibfnamefont {G.}~\bibnamefont {H\"ohler}},\ }\bibfield  {title} {\bibinfo {title} {{Ferromagnetismus als Einstein-Kondensation der Blochschen Spinwellen}},\ }\href {https://doi.org/https://doi.org/10.1002/andp.19504420110} {\bibfield  {journal} {\bibinfo  {journal} {Ann. Phys.}\ }\textbf {\bibinfo {volume} {442}},\ \bibinfo {pages} {93} (\bibinfo {year} {1950})}\BibitemShut {NoStop}%
\bibitem [{\citenamefont {Sudan}\ \emph {et~al.}(2009)\citenamefont {Sudan}, \citenamefont {L\"uscher},\ and\ \citenamefont {L\"auchli}}]{Lauchli09}%
  \BibitemOpen
  \bibfield  {author} {\bibinfo {author} {\bibfnamefont {J.}~\bibnamefont {Sudan}}, \bibinfo {author} {\bibfnamefont {A.}~\bibnamefont {L\"uscher}},\ and\ \bibinfo {author} {\bibfnamefont {A.~M.}\ \bibnamefont {L\"auchli}},\ }\bibfield  {title} {\bibinfo {title} {{Emergent multipolar spin correlations in a fluctuating spiral: The frustrated ferromagnetic spin-$\frac{1}{2}$ {H}eisenberg chain in a magnetic field}},\ }\href {https://doi.org/10.1103/PhysRevB.80.140402} {\bibfield  {journal} {\bibinfo  {journal} {Phys. Rev. B}\ }\textbf {\bibinfo {volume} {80}},\ \bibinfo {pages} {140402(R)} (\bibinfo {year} {2009})}\BibitemShut {NoStop}%
\bibitem [{\citenamefont {Rudner}\ and\ \citenamefont {Penc}(2026)}]{Penc_26}%
  \BibitemOpen
  \bibfield  {author} {\bibinfo {author} {\bibfnamefont {L.}~\bibnamefont {Rudner}}\ and\ \bibinfo {author} {\bibfnamefont {K.}~\bibnamefont {Penc}},\ }\href@noop {} {\bibinfo {title} {{Chiral enhancement of two-magnon bound states in an $S=1/2$ triangular-lattice magnet}}} (\bibinfo {year} {2026}),\ \Eprint {https://arxiv.org/abs/2607.01062} {arXiv:2607.01062} \BibitemShut {NoStop}%
\bibitem [{\citenamefont {Shannon}\ \emph {et~al.}(2004)\citenamefont {Shannon}, \citenamefont {Schmidt}, \citenamefont {Penc},\ and\ \citenamefont {Thalmeier}}]{Shannon_2004}%
  \BibitemOpen
  \bibfield  {author} {\bibinfo {author} {\bibfnamefont {N.}~\bibnamefont {Shannon}}, \bibinfo {author} {\bibfnamefont {B.}~\bibnamefont {Schmidt}}, \bibinfo {author} {\bibfnamefont {K.}~\bibnamefont {Penc}},\ and\ \bibinfo {author} {\bibfnamefont {P.}~\bibnamefont {Thalmeier}},\ }\bibfield  {title} {\bibinfo {title} {Finite temperature properties and frustrated ferromagnetism in a square lattice heisenberg model},\ }\href {https://doi.org/10.1140/epjb/e2004-00156-3} {\bibfield  {journal} {\bibinfo  {journal} {Eur. Phys. J. B}\ }\textbf {\bibinfo {volume} {38}},\ \bibinfo {pages} {599} (\bibinfo {year} {2004})}\BibitemShut {NoStop}%
\bibitem [{\citenamefont {Richter}\ \emph {et~al.}(2010)\citenamefont {Richter}, \citenamefont {Darradi}, \citenamefont {Schulenburg}, \citenamefont {Farnell},\ and\ \citenamefont {Rosner}}]{Richter2010}%
  \BibitemOpen
  \bibfield  {author} {\bibinfo {author} {\bibfnamefont {J.}~\bibnamefont {Richter}}, \bibinfo {author} {\bibfnamefont {R.}~\bibnamefont {Darradi}}, \bibinfo {author} {\bibfnamefont {J.}~\bibnamefont {Schulenburg}}, \bibinfo {author} {\bibfnamefont {D.~J.~J.}\ \bibnamefont {Farnell}},\ and\ \bibinfo {author} {\bibfnamefont {H.}~\bibnamefont {Rosner}},\ }\bibfield  {title} {\bibinfo {title} {{Frustrated spin-$\frac{1}{2}$ ${J}_{1}\text{\ensuremath{-}}{J}_{2}$ Heisenberg ferromagnet on the square lattice studied via exact diagonalization and coupled-cluster method}},\ }\href {https://doi.org/10.1103/PhysRevB.81.174429} {\bibfield  {journal} {\bibinfo  {journal} {Phys. Rev. B}\ }\textbf {\bibinfo {volume} {81}},\ \bibinfo {pages} {174429} (\bibinfo {year} {2010})}\BibitemShut {NoStop}%
\bibitem [{\citenamefont {Iqbal}\ \emph {et~al.}(2016)\citenamefont {Iqbal}, \citenamefont {Ghosh}, \citenamefont {Narayanan}, \citenamefont {Kumar}, \citenamefont {Reuther},\ and\ \citenamefont {Thomale}}]{Iqbal2016}%
  \BibitemOpen
  \bibfield  {author} {\bibinfo {author} {\bibfnamefont {Y.}~\bibnamefont {Iqbal}}, \bibinfo {author} {\bibfnamefont {P.}~\bibnamefont {Ghosh}}, \bibinfo {author} {\bibfnamefont {R.}~\bibnamefont {Narayanan}}, \bibinfo {author} {\bibfnamefont {B.}~\bibnamefont {Kumar}}, \bibinfo {author} {\bibfnamefont {J.}~\bibnamefont {Reuther}},\ and\ \bibinfo {author} {\bibfnamefont {R.}~\bibnamefont {Thomale}},\ }\bibfield  {title} {\bibinfo {title} {{Intertwined nematic orders in a frustrated ferromagnet}},\ }\href {https://doi.org/10.1103/PhysRevB.94.224403} {\bibfield  {journal} {\bibinfo  {journal} {Phys. Rev. B}\ }\textbf {\bibinfo {volume} {94}},\ \bibinfo {pages} {224403} (\bibinfo {year} {2016})}\BibitemShut {NoStop}%
\bibitem [{\citenamefont {Shannon}\ \emph {et~al.}(2006)\citenamefont {Shannon}, \citenamefont {Momoi},\ and\ \citenamefont {Sindzingre}}]{Shannon2006}%
  \BibitemOpen
  \bibfield  {author} {\bibinfo {author} {\bibfnamefont {N.}~\bibnamefont {Shannon}}, \bibinfo {author} {\bibfnamefont {T.}~\bibnamefont {Momoi}},\ and\ \bibinfo {author} {\bibfnamefont {P.}~\bibnamefont {Sindzingre}},\ }\bibfield  {title} {\bibinfo {title} {{Nematic Order in Square Lattice Frustrated Ferromagnets}},\ }\href {https://doi.org/10.1103/PhysRevLett.96.027213} {\bibfield  {journal} {\bibinfo  {journal} {Phys. Rev. Lett.}\ }\textbf {\bibinfo {volume} {96}},\ \bibinfo {pages} {027213} (\bibinfo {year} {2006})}\BibitemShut {NoStop}%
\bibitem [{\citenamefont {Smerald}\ \emph {et~al.}(2015)\citenamefont {Smerald}, \citenamefont {Ueda},\ and\ \citenamefont {Shannon}}]{Shannon2015}%
  \BibitemOpen
  \bibfield  {author} {\bibinfo {author} {\bibfnamefont {A.}~\bibnamefont {Smerald}}, \bibinfo {author} {\bibfnamefont {H.~T.}\ \bibnamefont {Ueda}},\ and\ \bibinfo {author} {\bibfnamefont {N.}~\bibnamefont {Shannon}},\ }\bibfield  {title} {\bibinfo {title} {{Theory of inelastic neutron scattering in a field-induced spin-nematic state}},\ }\href {https://doi.org/10.1103/PhysRevB.91.174402} {\bibfield  {journal} {\bibinfo  {journal} {Phys. Rev. B}\ }\textbf {\bibinfo {volume} {91}},\ \bibinfo {pages} {174402} (\bibinfo {year} {2015})}\BibitemShut {NoStop}%
\bibitem [{SM()}]{SM}%
  \BibitemOpen
  \href@noop {} {}\bibinfo {note} {{{See Supplemental Material at http://link.aps.org/supplemental/..., for an animation of the evolution of the bound states vs $J_2$ along the high-symmetry contour in the BZ.}}}\BibitemShut {Stop}%
\bibitem [{\citenamefont {Fetter}\ and\ \citenamefont {Walecka}(2012)}]{FetterWalecka2012}%
  \BibitemOpen
  \bibfield  {author} {\bibinfo {author} {\bibfnamefont {A.}~\bibnamefont {Fetter}}\ and\ \bibinfo {author} {\bibfnamefont {J.}~\bibnamefont {Walecka}},\ }\href {https://books.google.com/books?id=t5_DAgAAQBAJ} {\emph {\bibinfo {title} {{Quantum Theory of Many-Particle Systems}}}}\ (\bibinfo  {publisher} {Dover Publications},\ \bibinfo {address} {New York},\ \bibinfo {year} {2012})\BibitemShut {NoStop}%
\bibitem [{Note1()}]{Note1}%
  \BibitemOpen
  \bibinfo {note} {We also note that a reduced form of the Bethe-Salpeter equation has been utilized in the related hard-core boson problems under the same name~\cite {ueda_phasesep, JackeliZhitomirsky2004}.}\BibitemShut {Stop}%
\bibitem [{\citenamefont {Izyumov}\ and\ \citenamefont {Skryabin}(1988)}]{Izyumov1988book}%
  \BibitemOpen
  \bibfield  {author} {\bibinfo {author} {\bibfnamefont {Y.~A.}\ \bibnamefont {Izyumov}}\ and\ \bibinfo {author} {\bibfnamefont {Y.~N.}\ \bibnamefont {Skryabin}},\ }\href@noop {} {\emph {\bibinfo {title} {Statistical Mechanics of Magnetically Ordered Systems}}}\ (\bibinfo  {publisher} {Springer},\ \bibinfo {address} {New York},\ \bibinfo {year} {1988})\BibitemShut {NoStop}%
\bibitem [{\citenamefont {Chernyshev}\ \emph {et~al.}(1994)\citenamefont {Chernyshev}, \citenamefont {Dotsenko},\ and\ \citenamefont {Sushkov}}]{tJ94}%
  \BibitemOpen
  \bibfield  {author} {\bibinfo {author} {\bibfnamefont {A.~L.}\ \bibnamefont {Chernyshev}}, \bibinfo {author} {\bibfnamefont {A.~V.}\ \bibnamefont {Dotsenko}},\ and\ \bibinfo {author} {\bibfnamefont {O.~P.}\ \bibnamefont {Sushkov}},\ }\bibfield  {title} {\bibinfo {title} {{Hole-hole contact interaction in the $t$--$J$ model}},\ }\href {https://doi.org/10.1103/PhysRevB.49.6197} {\bibfield  {journal} {\bibinfo  {journal} {Phys. Rev. B}\ }\textbf {\bibinfo {volume} {49}},\ \bibinfo {pages} {6197} (\bibinfo {year} {1994})}\BibitemShut {NoStop}%
\bibitem [{\citenamefont {Callaway}(1964)}]{Callaway1964}%
  \BibitemOpen
  \bibfield  {author} {\bibinfo {author} {\bibfnamefont {J.}~\bibnamefont {Callaway}},\ }\bibfield  {title} {\bibinfo {title} {{Theory of Scattering in Solids}},\ }\href {https://doi.org/10.1063/1.1704180} {\bibfield  {journal} {\bibinfo  {journal} {J. Math. Phys.}\ }\textbf {\bibinfo {volume} {5}},\ \bibinfo {pages} {783} (\bibinfo {year} {1964})}\BibitemShut {NoStop}%
\bibitem [{\citenamefont {Maleev}(1958)}]{Maleev1958}%
  \BibitemOpen
  \bibfield  {author} {\bibinfo {author} {\bibfnamefont {S.~V.}\ \bibnamefont {Maleev}},\ }\bibfield  {title} {\bibinfo {title} {Scattering of slow neutrons in ferromagnets},\ }\href@noop {} {\bibfield  {journal} {\bibinfo  {journal} {Sov. Phys. JETP}\ }\textbf {\bibinfo {volume} {6}},\ \bibinfo {pages} {776} (\bibinfo {year} {1958})}\BibitemShut {NoStop}%
\bibitem [{\citenamefont {Holstein}\ and\ \citenamefont {Primakoff}(1940)}]{hp1940}%
  \BibitemOpen
  \bibfield  {author} {\bibinfo {author} {\bibfnamefont {T.}~\bibnamefont {Holstein}}\ and\ \bibinfo {author} {\bibfnamefont {H.}~\bibnamefont {Primakoff}},\ }\bibfield  {title} {\bibinfo {title} {Field dependence of the intrinsic domain magnetization of a ferromagnet},\ }\href {https://doi.org/10.1103/PhysRev.58.1098} {\bibfield  {journal} {\bibinfo  {journal} {Phys. Rev.}\ }\textbf {\bibinfo {volume} {58}},\ \bibinfo {pages} {1098} (\bibinfo {year} {1940})}\BibitemShut {NoStop}%
\bibitem [{\citenamefont {Zhitomirsky}\ and\ \citenamefont {Chernyshev}(2013)}]{RMP_13}%
  \BibitemOpen
  \bibfield  {author} {\bibinfo {author} {\bibfnamefont {M.~E.}\ \bibnamefont {Zhitomirsky}}\ and\ \bibinfo {author} {\bibfnamefont {A.~L.}\ \bibnamefont {Chernyshev}},\ }\bibfield  {title} {\bibinfo {title} {Colloquium: Spontaneous magnon decays},\ }\href {https://doi.org/10.1103/RevModPhys.85.219} {\bibfield  {journal} {\bibinfo  {journal} {Rev. Mod. Phys.}\ }\textbf {\bibinfo {volume} {85}},\ \bibinfo {pages} {219} (\bibinfo {year} {2013})}\BibitemShut {NoStop}%
\bibitem [{Note2()}]{Note2}%
  \BibitemOpen
  \bibinfo {note} {For the relevant derivations, see Ref.~\cite {Galitskii}: Problem 2.17 (see also 2.7) for 1D, Problem 4.38 for 2D, and Problems 4.1 and 4.10 for 3D.}\BibitemShut {Stop}%
\bibitem [{\citenamefont {Galitski}\ \emph {et~al.}(2013)\citenamefont {Galitski}, \citenamefont {Karnakov}, \citenamefont {Galitski},\ and\ \citenamefont {Kogan}}]{Galitskii}%
  \BibitemOpen
  \bibfield  {author} {\bibinfo {author} {\bibfnamefont {V.~M.}\ \bibnamefont {Galitski}}, \bibinfo {author} {\bibfnamefont {B.}~\bibnamefont {Karnakov}}, \bibinfo {author} {\bibfnamefont {V.}~\bibnamefont {Galitski}},\ and\ \bibinfo {author} {\bibfnamefont {V.~I.}\ \bibnamefont {Kogan}},\ }\href {https://books.google.com/books?id=NwrdQHswYwkC} {\emph {\bibinfo {title} {{Exploring Quantum Mechanics: A Collection of 700+ Solved Problems for Students, Lecturers, and Researchers}}}}\ (\bibinfo  {publisher} {OUP Oxford},\ \bibinfo {year} {2013})\BibitemShut {NoStop}%
\bibitem [{\citenamefont {Chernyshev}\ and\ \citenamefont {Maksimov}(2016)}]{Kagome_FM}%
  \BibitemOpen
  \bibfield  {author} {\bibinfo {author} {\bibfnamefont {A.~L.}\ \bibnamefont {Chernyshev}}\ and\ \bibinfo {author} {\bibfnamefont {P.~A.}\ \bibnamefont {Maksimov}},\ }\bibfield  {title} {\bibinfo {title} {{Damped Topological Magnons in the Kagome-Lattice Ferromagnets}},\ }\href {https://doi.org/10.1103/PhysRevLett.117.187203} {\bibfield  {journal} {\bibinfo  {journal} {Phys. Rev. Lett.}\ }\textbf {\bibinfo {volume} {117}},\ \bibinfo {pages} {187203} (\bibinfo {year} {2016})}\BibitemShut {NoStop}%
\bibitem [{\citenamefont {Gallegos}\ and\ \citenamefont {Chernyshev}(2026)}]{dataset}%
  \BibitemOpen
  \bibfield  {author} {\bibinfo {author} {\bibfnamefont {C.~A.}\ \bibnamefont {Gallegos}}\ and\ \bibinfo {author} {\bibfnamefont {A.~L.}\ \bibnamefont {Chernyshev}},\ }\bibfield  {title} {\bibinfo {title} {{Revisiting magnon bound states: ferro-antiferromagnetic $J_1$–$J_2$ square-lattice model}},\ }\href {https://doi.org/10.5281/zenodo.21859286} {10.5281/zenodo.21859286} (\bibinfo {year} {2026})\BibitemShut {NoStop}%
\bibitem [{\citenamefont {Gradshteyn}\ and\ \citenamefont {Ryzhik}(1994)}]{GradshteynRyzhik1994}%
  \BibitemOpen
  \bibfield  {author} {\bibinfo {author} {\bibfnamefont {I.~S.}\ \bibnamefont {Gradshteyn}}\ and\ \bibinfo {author} {\bibfnamefont {I.~M.}\ \bibnamefont {Ryzhik}},\ }\href@noop {} {\emph {\bibinfo {title} {{Table of Integrals, Series, and Products}}}},\ \bibinfo {edition} {5th}\ ed.,\ edited by\ \bibinfo {editor} {\bibfnamefont {A.}~\bibnamefont {Jeffrey}}\ (\bibinfo  {publisher} {Academic Press},\ \bibinfo {address} {San Diego},\ \bibinfo {year} {1994})\BibitemShut {NoStop}%
\bibitem [{\citenamefont {Corless}\ \emph {et~al.}(1996)\citenamefont {Corless}, \citenamefont {Gonnet}, \citenamefont {Hare}, \citenamefont {Jeffrey},\ and\ \citenamefont {Knuth}}]{Corless1996LambertW}%
  \BibitemOpen
  \bibfield  {author} {\bibinfo {author} {\bibfnamefont {R.~M.}\ \bibnamefont {Corless}}, \bibinfo {author} {\bibfnamefont {G.~H.}\ \bibnamefont {Gonnet}}, \bibinfo {author} {\bibfnamefont {D.~E.~G.}\ \bibnamefont {Hare}}, \bibinfo {author} {\bibfnamefont {D.~J.}\ \bibnamefont {Jeffrey}},\ and\ \bibinfo {author} {\bibfnamefont {D.~E.}\ \bibnamefont {Knuth}},\ }\bibfield  {title} {\bibinfo {title} {{On the Lambert$W$ function}},\ }\href {https://doi.org/10.1007/BF02124750} {\bibfield  {journal} {\bibinfo  {journal} {Adv Comput Math}\ }\textbf {\bibinfo {volume} {5}},\ \bibinfo {pages} {329} (\bibinfo {year} {1996})}\BibitemShut {NoStop}%
\bibitem [{\citenamefont {Jackeli}\ and\ \citenamefont {Zhitomirsky}(2004)}]{JackeliZhitomirsky2004}%
  \BibitemOpen
  \bibfield  {author} {\bibinfo {author} {\bibfnamefont {G.}~\bibnamefont {Jackeli}}\ and\ \bibinfo {author} {\bibfnamefont {M.~E.}\ \bibnamefont {Zhitomirsky}},\ }\bibfield  {title} {\bibinfo {title} {{Frustrated Antiferromagnets at High Fields: Bose-Einstein Condensation in Degenerate Spectra}},\ }\href {https://doi.org/10.1103/PhysRevLett.93.017201} {\bibfield  {journal} {\bibinfo  {journal} {Phys. Rev. Lett.}\ }\textbf {\bibinfo {volume} {93}},\ \bibinfo {pages} {017201} (\bibinfo {year} {2004})}\BibitemShut {NoStop}%
\end{thebibliography}%


\end{document}